\documentclass[12pt]{article}
\usepackage{jheppub, bm, wrapfig, slashed}
\numberwithin{equation}{section}
\renewcommand\afterTocRuleSpace{\bigskip}

\graphicspath{{./figures/}}

\usepackage{tikz}
\usetikzlibrary{arrows}
\usetikzlibrary{shapes.geometric,calc, arrows, positioning,shapes.misc}

\newcommand{\arrowpath}[3]{
	\draw[line width=1pt] #1--#2;
	\draw[->, >=stealth, line width=1pt, black] #1--($#1!#3!#2$);
}

\newcommand{\arrowpathdouble}[3]{
	\draw[double] #1--#2;
	\draw[double,->, >=stealth] #1--($#1!#3!#2$);
}

\newcommand{\paddedline}[3]{
	\draw[white, fill=white] ($#1-#3$)--($#1+#3$)--($#2+#3$)--($#2-#3$)--cycle;
}

\newcommand{\bea}{\begin{eqnarray}}
\newcommand{\eea}{\end{eqnarray}}
\newcommand{\be}{\begin{equation}}
\newcommand{\ee}{\end{equation}}
\newcommand{\bse}{\begin{subequations}}
\newcommand{\ese}{\end{subequations}}

\newcommand{\ben}{\begin{enumerate}}
\newcommand{\een}{\end{enumerate}}
\newcommand{\bit}{\begin{itemize}}
\newcommand{\eit}{\end{itemize}}

\newcommand{\wt}{\widetilde}

\newcommand{\ie}{\emph{i.e.}}

\newcommand{\Z}{{\mathbb Z}}
\newcommand{\R}{{\mathbb R}}
\newcommand{\C}{{\mathbb C}}

\newcommand{\CC}{\mathcal{C}}

\newcommand{\CG}{\mathcal{G}}

\newcommand{\CI}{\mathcal{I}}

\newcommand{\CL}{\mathcal{L}}
\newcommand{\CM}{\mathcal{M}}

\newcommand{\CO}{\mathcal{O}}

\newcommand{\CW}{\mathcal{W}}

\newcommand{\fB}{\mathfrak{B}}

\newcommand{\fT}{\mathfrak{T}}

\newcommand{\PP}{Pin$^+$}
\newcommand{\PM}{Pin$^-$}

\title{Unoriented 3d TFTs}

\author{Lakshya Bhardwaj\footnote{email: lbhardwaj@pitp.ca}}

\affiliation{Perimeter Institute for Theoretical Physics, Waterloo, Ontario, Canada N2L 2Y5}

\abstract{This paper generalizes two facts about oriented 3d TFTs to the unoriented case. On one hand, it is known that oriented 3d TFTs having a topological boundary condition admit a state-sum construction known as the Turaev-Viro construction. This is related to the string-net construction of fermionic phases of matter. We show how Turaev-Viro construction can be generalized to unoriented 3d TFTs. On the other hand, it is known that the ``fermionic" versions of oriented TFTs, known as Spin-TFTs, can be constructed in terms of ``shadow" TFTs which are ordinary oriented TFTs with an anomalous $\Z_2$ 1-form symmetry. We generalize this correspondence to \PP-TFTs by showing that they can be constructed in terms of ordinary unoriented TFTs with anomalous $\Z_2$ 1-form symmetry having a mixed anomaly with time-reversal symmetry. The corresponding \PP-TFT does not have any anomaly for time-reversal symmetry however and hence it can be unambiguously defined on a non-orientable manifold. In case a \PP-TFT admits a topological boundary condition, one can combine the above two statements to obtain a Turaev-Viro-like construction of \PP-TFTs. As an application of these ideas, we construct a large class of \PP-SPT phases.
}

\begin{document}

\maketitle

\section{Introduction}
It is an age-old problem to provide a complete definition of quantum field theories. A part of the problem is to understand on what kinds of manifolds can we put a quantum field theory. For instance, we can ask the following question: Given a theory that can be defined on orientable manifolds, what sort of extra data do we need in order to extend the definition of the theory to non-orientable manifolds? First of all, such an extension may not be possible. For instance, if the theory has a framing anomaly, then it will not be well-defined on non-orientable manifolds. This was recently explained in a footnote of \cite{Tachikawa:2016cha}. Second, if such an extension is possible, then it need not be unique. That is, there can be different unoriented theories which reduce to the same oriented theory on orientable manifolds. We will see plenty of examples like this in this paper.

On a non-orientable manifold, we can choose a consistent orientation everywhere if we remove a locus homologous to the Poincare dual of first Stiefel-Whitney class $w_1$. The induced local orientation flips as we cross this locus. In order to be able to define an unoriented theory in terms of the data of the oriented theory, we need the existence of orientation reversing codimension one defects which we place along this locus. These orientation reversing defects implement orientation reversing symmetries akin to the orientation preserving codimension one defects which implement a global symmetry transformation \cite{Gaiotto:2014kfa}. These defects can be placed on top of each other forming the structure of a group $G$ with a homomorphism $\rho: G\to\Z_2$ whose kernel $G_0$ is the global symmetry group of the theory. The set $G_1=G-G_0$ parametrizes the orientation reversing symmetries.

In this paper, we explore the consequences of the existence of such orientation reversing defects in the context of 3d TFTs which admit a topological boundary condition. We restrict ourselves to the case in which the structure group of the TFT can be decomposed as $O(3)\times G$ for a finite global symmetry group $G$. In such cases, the properties of orientation reversing defects allow us to propose a generalization of Turaev-Viro state-sum construction of 3d TFTs \cite{turaev1992state,2010arXiv1004.1533K} to the unoriented case.\footnote{Historically, the original construction due to Turaev and Viro was based on a modular tensor category treated as a spherical fusion category. This construction produced theories which could be defined on an unoriented manifold. This construction was generalized to arbitrary spherical fusion categories but such theories could only be defined on oriented manifolds. It is this latter construction that we call ``oriented Turaev-Viro construction" in this paper. This paper presents a further generalization of this setup which we call ``unoriented Turaev-Viro construction". Our construction can be used to construct any unoriented 3d TFT with a topological boundary condition.} We check that this proposal indeed defines a 3d unoriented TFT. From now on, whenever we say ``unoriented TFT", we mean this particular structure group.

For an oriented 3d TFT $\fT$ with global symmetry $G$, the input data for the construction is a $G$-graded spherical fusion category $\CC$. We will find that an unoriented 3d TFT $\wt\fT$ extending $\fT$ is constructed in terms of a $G$-graded ``twisted" spherical fusion category $\wt\CC$ where $\CC$ is embedded as a subcategory of $\wt\CC$. In terms of the data of $\wt\CC$, we give a prescription to construct the partition function of $\wt\fT$ on any (possibly non-orientable) 3-manifold.

We also apply these ideas to 3d \PP-TFTs ({\ie} TFTs with structure group \PP$(3)\times G$) which are a generalization of 3d Spin-TFTs. Spin-TFTs are ``fermionic" analogs of ordinary oriented TFTs as they are sensitive to the spin structure of the underlying orientable manifold. To define fermions on an unorientable manifold, we need to choose either a \PP-structure or a \PM-structure on the manifold. Correspondingly, the natural unoriented generalizations of Spin-TFTs are \PP-TFTs and \PM-TFTs. 

In \cite{Bhardwaj:2016clt}, a recipe was given to construct a 3d Spin-TFT from an ordinary 3d TFT with an anomalous $\Z_2$ 1-form symmetry. This ordinary TFT $\fT_f$ was called the shadow of the corresponding Spin-TFT $\fT_s$. The idea was to use a kernel TFT $\fT_k$ to connect the shadow theories with their spin counterparts. $\fT_k$ is a Spin-TFT with an anomalous $\Z_2$ 1-form symmetry. The diagonal $\Z_2$ 1-form symmetry in the product theory $\fT_f\times\fT_k$ is non-anomalous. This non-anomalous 1-form symmetry is then gauged to obtain the spin TFT $\fT_s$.

We extend their recipe by constructing shadows for \PP-TFTs. The \PP-shadows correspond to theories with anomalous $\Z_2$ 1-form symmetry and a certain time-reversal anomaly in the presence of a background 2-connection for the $\Z_2$ 1-form symmetry. The \PP-kernel TFT $\fT_k^+$ has a corresponding time-reversal anomaly which cancels the anomaly of the shadow. Hence, the resulting \PP-TFTs are time-reversal invariant and can be put on non-orientable manifolds without any ambiguity \cite{Witten:2016cio}.

As an application, we construct a large class of \PP-SPT phases with global symmetry $G$. SPT phases are TFTs which are invertible under the product operation on TFTs. In the condensed matter literature, these are referred to as fermionic SPT phases protected by $G\times\Z_2^T$ with $T^2=(-1)^F$. In the case when $G$ is trivial, cobordism hypothesis predicts two \PP-SPT phases forming a $\Z_2$ group structure \cite{Kapustin:2014dxa}. Our construction reproduces both of these SPT phases along with the $\Z_2$ structure.

This paper is organized as follows. In section \ref{2}, we propose a Turaev-Viro construction for unoriented 3d TFTs. In section \ref{3}, we provide a construction of \PP-TFTs in terms of ordinary unoriented TFTs with a $\Z_2$ 1-form symmetry which is anomalous and has a mixed anomaly with time-reversal symmetry. In section \ref{4}, we construct a large class of \PP-SPT phases with global symmetry $G$ and reproduce the $\Z_2$ group of \PP-SPT phases in the case of trivial $G$. In section \ref{5}, we present our conclusions and comment on future directions which include a strategy to classify all \PP-SPT phases with global symmetry $G$.

\section{Turaev-Viro construction} \label{2}
For an exhaustive review and physical understanding of the Turaev-Viro state-sum construction of oriented 3d TFTs, the reader is referred to \cite{Bhardwaj:2016clt}. In this section, we first review relevant aspects of this construction. Then, we propose a generalization of the construction to the unoriented case. We also provide a physical understanding of our proposal in terms of orientation reversing defects. We close the section by discussing invertible unoriented TFTs with global symmetry $G$, or in other words bosonic SPT phases protected by $G\times\Z_2^T$.  

A reader only interested in the Turaev-Viro construction of unoriented 3d TFTs is referred to subsections \ref{2.3} and \ref{2.4}.

\subsection{Boundary line defects and spherical fusion category}

In general, a boundary condition $\fB$ allows one to define a TFT $\fT$ on a manifold $M$ with boundary $B$ by placing $\fB$ on the boundary. The boundary condition is called topological if topological deformations of $M$ (including topological deformations of $B$) leave the partition function of $\fT$ on $M$ invariant.

Turaev-Viro procedure constructs an oriented 3d (unitary) TFT $\fT$ from the knowledge of a topological boundary condition $\fB$ of $\fT$ \cite{Fuchs:2012dt}. $\fT$ can be recovered from any one of its topological boundary conditions. For simplicity, we will assume that $\fT$ has a one-dimensional Hilbert space on $S^2$. The Turaev-Viro construction for such a TFT $\fT$ is phrased in terms of a (unitary) spherical fusion category $\CC$.

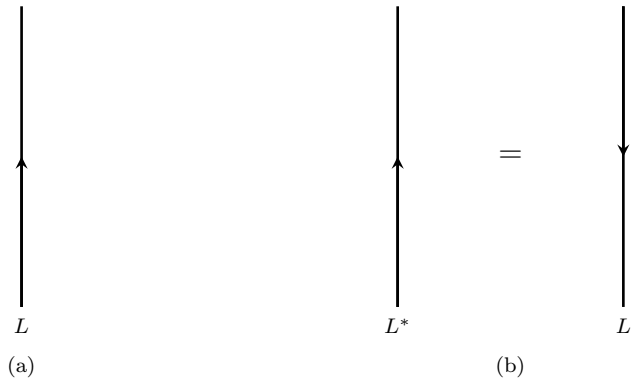
\begin{figure}
\centering
\begin{tikzpicture}[line width=1pt]
\begin{scope}[every node/.style={sloped,allow upside down}]
\coordinate (lowA) at (0,0);
\coordinate (highA) at (0,4);
\arrowpath{(lowA)}{(highA)}{0.5};
\node[below] at (lowA) {\scriptsize{$L$}};
\node[below] at ($(lowA) - (0, 0.5)$) {\scriptsize{(a)}};

\coordinate (lowB1) at ($(lowA)+(5,0)$);
\coordinate (highB1) at ($(lowB1) + (0,4)$);
\arrowpath{(lowB1)}{(highB1)}{0.5};
\node[below] at (lowB1) {\scriptsize{$L^*$}};

\coordinate (eqB) at ($(lowB1)!0.5!(highB1)+(1.5,0)$);
\node at (eqB) {$=$};

\coordinate (lowB2) at ($(lowB1) + (3,0)$);
\coordinate (highB2) at ($(lowB2) + (0,4)$);
\arrowpath{(highB2)}{(lowB2)}{0.5};
\node[below] at (lowB2) {\scriptsize{$L$}};
\node[below] at ($(lowB1) - (-1.5, 0.5)$) {\scriptsize{(b)}};
\end{scope}
\end{tikzpicture}
\caption{(a) A boundary line $L$. (b) The \emph{dual} line $L^*$ is defined by reversing the orientation of $L$.}
\label{fig:dual}
\end{figure}

The objects of $\CC$ are line defects living on $\fB$. Such line defects are specified by a label $L$ and an orientation along the line corresponding to $L$. If a line defect with a certain orientation is denoted as an object $L$ in $\CC$, the same line defect with opposite choice of orientation is denoted as the dual object $L^*$. See Figure \ref{fig:dual}.

\begin{figure}
\centering
\begin{tikzpicture}[line width=1pt]
\begin{scope}[every node/.style={sloped,allow upside down}]
\coordinate (C) at (0,0);
\coordinate (mid) at ($(C)+(0,2)$);
\coordinate (A) at ($(mid)+(-1.2,2)$);
\coordinate (B) at ($(mid)+(1.2,2)$);

\arrowpath{(mid)}{(C)}{0.5};
\arrowpath{(mid)}{(A)}{0.5};
\arrowpath{(mid)}{(B)}{0.5};
\draw[fill=black] (mid) circle(2pt);

\node[below] at (C) {\scriptsize{$A_3$}};
\node[above] at (A) {\scriptsize{$A_1$}};
\node[above] at (B) {\scriptsize{$A_2$}};
\node[right] at (mid) {\scriptsize{$m$}};
\node[below] at ($(C)-(0,0.5)$) {\scriptsize{(a)}};

\coordinate (H) at ($(mid)+(8,0)$);
\coordinate (A1) at ($(H)+({1.75*cos(135)},{1.75*sin(135)})$);
\coordinate (A2) at ($(H)+({1.75*cos(45)},{1.75*sin(45)})$);
\coordinate (A3) at ($(H)+({1.75*cos(270)},{1.75*sin(270)})$);
\draw (H) circle(1.75cm);
\draw[fill=black] (A1) circle(2pt);
\draw[fill=black] (A2) circle(2pt);
\draw[fill=black] (A3) circle(2pt);
\node[above left] at (A1) {\scriptsize{$A_1$}};
\node[above right] at (A2) {\scriptsize{$A_2$}};
\node[below] at (A3) {\scriptsize{$A_3$}};
\node at (H) {\scriptsize{$m$}};
\node[below] at ($(A3)-(0,0.75)$) {\scriptsize{(b)}};
\end{scope}
\end{tikzpicture}
\caption{A morphism $m$ between outgoing lines $A_1$, $A_2$ and $A_3$ corresponds to a state $m$ in the Hilbert space on a disk with boundary punctures $A_1$, $A_2$ and $A_3$. Consider on a hemisphere geometry with a boundary on the spherical part and the disk shown in (b) being the cross-section. The state shown in (b) is produced on the cross-section if the boundary has the graph shown on (a) inserted on it such that $A_i$ end on their respective punctures.}
\label{fig:Hilbert}
\end{figure}
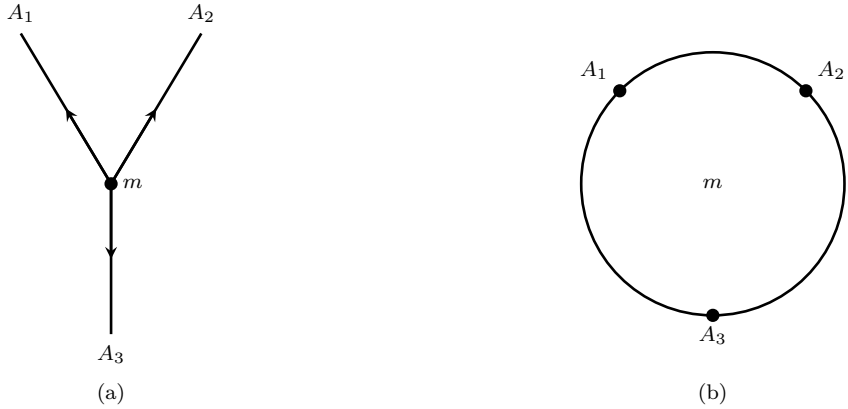

The morphisms $m_{AB}$ from $A$ to $B$ in $\CC$ are local operators living between two boundary lines. Thus, $m_{AB}$ form a vector space. This vector space can also be identified with the Hilbert space of states on the disk with boundary punctures corresponding to $A^*$ and $B$. Similarly, the local operators living at the junction of multiple outgoing lines $A_i$ is the space of states on disk with boundary punctures corresponding to $A_i$. The space of states can be generated by placing a hemispherical cap on which the lines $A_i$ emanate from a point on the boundary of the cap and go to their respective punctures. See Figure \ref{fig:Hilbert}.

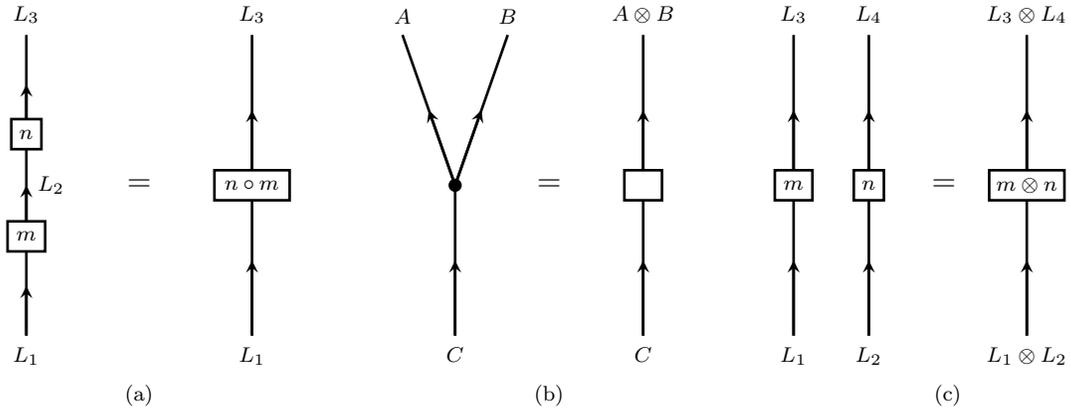
\begin{figure}
\centering
\begin{tikzpicture}[line width=1pt]
\begin{scope}[every node/.style={sloped,allow upside down}]
\coordinate (lowA1) at (0,0);
\coordinate (highA1) at ($(lowA1) + (0,4)$);
\coordinate (thirdA1) at ($(lowA1)!0.33!(highA1)$);
\coordinate (midA1) at ($(lowA1)!0.5!(highA1)$);
\coordinate (twothirdA1) at ($(lowA1)!0.67!(highA1)$);
\arrowpath{(lowA1)}{(thirdA1)}{0.5};
\arrowpath{(thirdA1)}{(twothirdA1)}{0.5};
\arrowpath{(twothirdA1)}{(highA1)}{0.5};
\coordinate (halfboxA11) at (0.25, 0.2);
\coordinate (halfboxA12) at (0.2, 0.2);
\draw[fill=white] ($(thirdA1) - (halfboxA11)$) rectangle ($(thirdA1) + (halfboxA11)$);
\draw[fill=white] ($(twothirdA1) - (halfboxA12)$) rectangle ($(twothirdA1) + (halfboxA12)$);
\node at (thirdA1) {\scriptsize{$m$}};
\node at (twothirdA1) {\scriptsize{$n$}};
\node[below] at (lowA1) {\scriptsize{$L_1$}};
\node[right] at (midA1) {\scriptsize{$L_2$}};
\node[above] at (highA1) {\scriptsize{$L_3$}};

\coordinate (eqA) at ($(midA1)+(1.5,0)$);
\node at (eqA) {$=$};

\coordinate (lowA2) at ($(lowA1)+(3,0)$);
\coordinate (highA2) at ($(lowA2) + (0,4)$);
\coordinate (midA2) at ($(lowA2)!0.5!(highA2)$);
\arrowpath{(lowA2)}{(midA2)}{0.5};
\arrowpath{(midA2)}{(highA2)}{0.5};
\coordinate (halfboxA2) at (0.5, 0.2);
\draw[fill=white] ($(midA2) - (halfboxA2)$) rectangle ($(midA2) + (halfboxA2)$);
\node at (midA2) {\scriptsize{$n \circ m$}};
\node[below] at (lowA2) {\scriptsize{$L_1$}};
\node[above] at (highA2) {\scriptsize{$L_3$}};

\node[below] at ($(lowA1)!0.5!(lowA2) + (0,-0.5)$) {\scriptsize{(a)}};

\coordinate (C) at ($(lowA2)+(2.7,0)$);
\coordinate (mid) at ($(C)+(0,2)$);
\coordinate (A) at ($(mid)+(-0.7,2)$);
\coordinate (B) at ($(mid)+(0.7,2)$);

\arrowpath{(C)}{(mid)}{0.5};
\arrowpath{(mid)}{(A)}{0.5};
\arrowpath{(mid)}{(B)}{0.5};
\draw[fill=black] (mid) circle(2pt);

\node[below] at (C) {\scriptsize{$C$}};
\node[above] at (A) {\scriptsize{$A$}};
\node[above] at (B) {\scriptsize{$B$}};

\coordinate (eq2) at ($(mid)+(1.25,0)$);
\node at (eq2) {$=$};

\coordinate (C2) at ($(C)+(2.5,0)$);
\coordinate (mid2) at ($(C2)+(0,2)$);
\coordinate (AB) at ($(mid2)+(0,2)$);

\arrowpath{(C2)}{(mid2)}{0.5};
\arrowpath{(mid2)}{(AB)}{0.5};
\coordinate (halfbox) at (0.25, 0.2);
\draw[fill=white] ($(mid2) - (halfbox)$) rectangle ($(mid2) + (halfbox)$);

\node[below] at (C2) {\scriptsize{$C$}};
\node[above] at (AB) {\scriptsize{$A\otimes B$}};

\node[below] at ($(eq2)-(0,2.5)$) {\scriptsize{(b)}};

\coordinate (lowC1) at ($(C2)+(2,0)$);
\coordinate (highC1) at ($(lowC1) + (0,4)$);
\coordinate (midC1) at ($(lowC1)!0.5!(highC1)$);
\arrowpath{(lowC1)}{(midC1)}{0.5};
\arrowpath{(midC1)}{(highC1)}{0.5};
\coordinate (halfboxC1) at (0.25, 0.2);
\draw[fill=white] ($(midC1) - (halfboxC1)$) rectangle ($(midC1) + (halfboxC1)$);
\node at (midC1) {\scriptsize{$m$}};
\node[below] at (lowC1) {\scriptsize{$L_1$}};
\node[above] at (highC1) {\scriptsize{$L_3$}};

\coordinate (lowC2) at ($(lowC1)+(1,0)$);
\coordinate (highC2) at ($(lowC2) + (0,4)$);
\coordinate (midC2) at ($(lowC2)!0.5!(highC2)$);
\arrowpath{(lowC2)}{(midC2)}{0.5};
\arrowpath{(midC2)}{(highC2)}{0.5};
\coordinate (halfboxC2) at (0.2, 0.2);
\draw[fill=white] ($(midC2) - (halfboxC2)$) rectangle ($(midC2) + (halfboxC2)$);
\node at (midC2) {\scriptsize{$n$}};
\node[below] at (lowC2) {\scriptsize{$L_2$}};
\node[above] at (highC2) {\scriptsize{$L_4$}};

\coordinate (eqC) at ($(lowC2)!0.5!(highC2)+(1,0)$);
\node at (eqC) {$=$};

\coordinate (lowC3) at ($(lowC2)+(2.1,0)$);
\coordinate (highC3) at ($(lowC3) + (0,4)$);
\coordinate (midC3) at ($(lowC3)!0.5!(highC3)$);
\arrowpath{(lowC3)}{(midC3)}{0.5};
\arrowpath{(midC3)}{(highC3)}{0.5};
\coordinate (halfboxC3) at (0.5, 0.2);
\draw[fill=white] ($(midC3) - (halfboxC3)$) rectangle ($(midC3) + (halfboxC3)$);
\node at (midC3) {\scriptsize{$m \otimes n$}};
\node[below] at (lowC3) {\scriptsize{$L_1 \otimes L_2$}};
\node[above] at (highC3) {\scriptsize{$L_3 \otimes L_4$}};

\node[below] at ($(lowC2)!0.5!(lowC3) - (0, 0.5)$) {\scriptsize{(c)}};
\end{scope}
\end{tikzpicture}
\caption{(a) Composition of morphisms. The box is our alternative notation for a morphism. (b,c) Tensor product of objects and morphisms.}
\label{fig:fusion}
\end{figure}

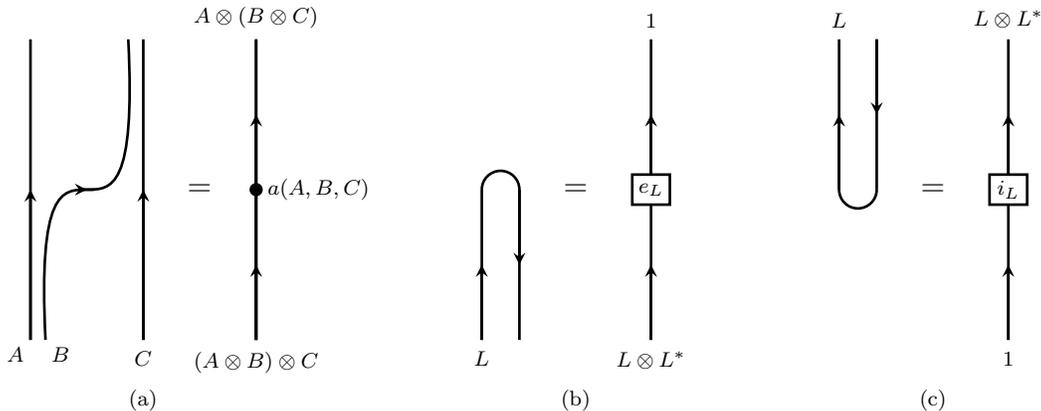
\begin{figure}
\centering
\begin{tikzpicture}[line width=1pt]
\begin{scope}[every node/.style={sloped,allow upside down}]
\coordinate (lowA) at (0,0);
\coordinate (highA) at ($(lowA)+(0,4)$);
\coordinate (lowB) at ($(lowA)+(0.2,0)$);
\coordinate (lowC) at ($(lowA)+(1.5,0)$);
\coordinate (highC) at ($(lowC)+(0,4)$);
\coordinate (highB) at ($(highC)-(0.2,0)$);
\coordinate (midBl) at ($(lowB)!0.5!(highB) - (0.1,0)$);
\coordinate (midBr) at ($(lowB)!0.5!(highB) + (0.1,0)$);
\arrowpath{(lowA)}{(highA)}{0.5};
\node at ($(lowA)-(0.2,0.2)$) {\scriptsize{$A$}};
\draw (lowB) .. controls (highA) and (lowC) .. (highB);
\node at ($(lowB)+(0.2,-0.2)$) {\scriptsize{$B$}};
\arrowpath{(midBl)}{(midBr)}{0.5};
\arrowpath{(lowC)}{(highC)}{0.5};
\node[below] at (lowC) {\scriptsize{$C$}};

\coordinate (eq) at ($(lowC)!0.5!(highC) + (0.75,0)$);
\node at (eq) {$=$};

\coordinate (lowT) at ($(lowC) + (1.5,0)$);
\coordinate (highT) at ($(lowT) + (0,4)$);
\coordinate (midT) at ($(lowT)!0.5!(highT)$);
\draw[fill=black] (midT) circle(2pt);
\arrowpath{(lowT)}{(highT)}{0.25};
\arrowpath{(lowT)}{(highT)}{0.75};
\node[below] at (lowT) {\scriptsize{$(A \otimes B)\otimes C$}};
\node[above] at (highT) {\scriptsize{$A \otimes (B\otimes C)$}};
\node[right] at (midT) {\scriptsize{$a(A,B,C)$}};

\node[below] at ($(lowC) - (0,0.5)$) {\scriptsize{(a)}};

\coordinate (lowA1) at ($(lowT)+(3,0)$);
\coordinate (highA1) at ($(lowA1) + (0,2)$);
\coordinate (lowA2) at ($(lowA1)+(0.5,0)$);
\coordinate (highA2) at ($(lowA2) + (0,2)$);
\arrowpath{(lowA1)}{(highA1)}{0.5};
\arrowpath{(highA2)}{(lowA2)}{0.5};
\draw (highA1) arc[radius=0.25, start angle=180, end angle=0];
\node[below] at (lowA1) {\scriptsize{$L$}};

\coordinate (eqA) at ($(highA2)+(0.75,0)$);
\node at (eqA) {$=$};

\coordinate (lowA3) at ($(lowA2)+(1.75,0)$);
\coordinate (highA3) at ($(lowA3) + (0,4)$);
\coordinate (midA3) at ($(lowA3)!0.5!(highA3)$);
\arrowpath{(lowA3)}{(midA3)}{0.5};
\arrowpath{(midA3)}{(highA3)}{0.5};
\coordinate (halfboxA3) at (0.25, 0.2);
\draw[fill=white] ($(midA3) - (halfboxA3)$) rectangle ($(midA3) + (halfboxA3)$);
\node at (midA3) {\scriptsize{$e_L$}};
\node[below] at (lowA3) {\scriptsize{$L \otimes L^*$}};
\node[above] at (highA3) {\scriptsize{$1$}};

\node[below] at ($(lowA2) + (0.75,-0.5)$) {\scriptsize{(b)}};

\coordinate (lowB1) at ($(midA3)+(2.5,0)$);
\coordinate (highB1) at ($(lowB1) + (0,2)$);
\coordinate (lowB2) at ($(lowB1)+(0.5,0)$);
\coordinate (highB2) at ($(lowB2) + (0,2)$);
\arrowpath{(highB2)}{(lowB2)}{0.5};
\arrowpath{(lowB1)}{(highB1)}{0.5};
\draw (lowB1) arc[radius=0.25, start angle=180, end angle=360];
\node[above] at (highB1) {\scriptsize{$L$}};

\coordinate (eqB) at ($(lowB2)+(0.75,0)$);
\node at (eqB) {$=$};

\coordinate (lowB3) at ($(lowB2)+(1.75,-2)$);
\coordinate (highB3) at ($(lowB3) + (0,4)$);
\coordinate (midB3) at ($(lowB3)!0.5!(highB3)$);
\arrowpath{(lowB3)}{(midB3)}{0.5};
\arrowpath{(midB3)}{(highB3)}{0.5};
\coordinate (halfboxB3) at (0.25, 0.2);
\draw[fill=white] ($(midB3) - (halfboxB3)$) rectangle ($(midB3) + (halfboxB3)$);
\node at (midB3) {\scriptsize{$i_L$}};
\node[above] at (highB3) {\scriptsize{$L \otimes L^*$}};
\node[below] at (lowB3) {\scriptsize{$1$}};

\node[below] at ($(eqB) + (0,-2.5)$) {\scriptsize{(c)}};
\end{scope}
\end{tikzpicture}
\caption{Canonical maps: (a) Associator $a(A,B,C)$, (b) Evaluation $e_L$, and (c) Co-evaluation $i_L$.}\label{fig:canonical}
\end{figure}

The composition of morphisms corresponds to fusion of local operators along the line. There is a tensor product corresponding to fusion of lines as they are brought together. There are also canonical associator, evaluation and coevaluation maps which physically correspond to placing the lines in a certain fashion and fusing them. See Figures \ref{fig:fusion} and \ref{fig:canonical}. Using these canonical morphisms, we can assign a morphism $m$ from $\otimes_i A_i$ to $\otimes_j B_j$ to any planar graph $\Gamma$ of boundary line defects (with local operators at their junctions) such that $\Gamma$ has incoming lines $A_i$ and outgoing lines $B_j$. The canonical morphisms satisfy certain identities which guarantee that a topologically equivalent graph $\Gamma'$ evaluates to the same morphism $m$.

Consider vacua $i$ of $\fB$ which can be characterized by the expectation value of a line $L_i$. Such lines are called simple lines. Morphism space from $L_i$ to $L_j$ is empty for $i\ne j$ and is one-dimensional for $i=j$. The space of local operators living on $L_i$ can be identified as $\C$ because there is a canonical identity operator living on $L_i$. Every line $L$ can be written as a sum of simple lines $L=\oplus n_i L_i$ where $n_i$ denotes the multiplicity of the simple line $L_i$ in the sum. The identity line $1$ can be treated as a special simple line which can be inserted anywhere without changing any answers. The duals of simple lines are simple as well.

Turaev-Viro construction uses $\CC$ as an input and produces the partition of $\fT$ on any oriented manifold $M$ as the output. We will describe the construction in a very hands-on fashion in the next subsection. We will see that the basic object in the construction is a graph $\Gamma$ in $\CC$ drawn on the sphere. $\Gamma$ can be projected down to a closed graph $\Gamma_p$ drawn on the plane. $\Gamma_p$ constructs a morphism from identity line to itself which evaluates to a definite number. This number is the partition function $Z_\Gamma$ of $\fT$ on a 3-ball along with a network of boundary lines (and local operators at their junctions) $\Gamma$ inserted at the boundary 2-sphere. The word ``spherical" in spherical fusion category corresponds to certain axioms which guarantee that different projections to the plane evaluate to the same number.

This construction can be easily generalized to TFTs with a global symmetry group $G$. The symmetry manifests itself in the existence of codimension one topological defects $U_g$ labeled by $g\in G$. Going across the locus of $U_g$ implements a symmetry transformation on the system by $g$. These defects fuse according to the group law and can end on $\fB$ giving rise to new lines at the junction. Thus the category of boundary lines living on $\fB$ becomes graded by $G$, {\ie} $\mathcal{C}=\oplus_{g}\,\mathcal{C}_{g}$.

%

\subsection{Oriented Turaev-Viro}\label{2.2}

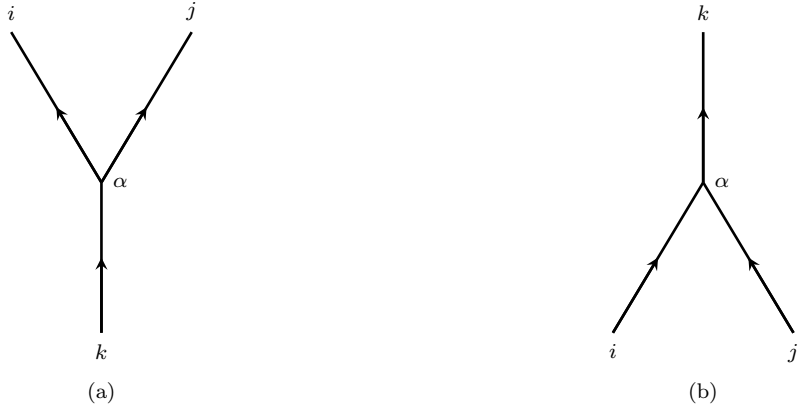
\begin{figure}
\centering
\begin{tikzpicture}[line width=1pt]
\begin{scope}[every node/.style={sloped,allow upside down}]
\coordinate (C) at (0,0);
\coordinate (mid) at ($(C)+(0,2)$);
\coordinate (A) at ($(mid)+(-1.2,2)$);
\coordinate (B) at ($(mid)+(1.2,2)$);

\arrowpath{(C)}{(mid)}{0.5};
\arrowpath{(mid)}{(A)}{0.5};
\arrowpath{(mid)}{(B)}{0.5};

\node[below] at (C) {\scriptsize{$k$}};
\node[above] at (A) {\scriptsize{$i$}};
\node[above] at (B) {\scriptsize{$j$}};
\node[right] at (mid) {\scriptsize{$\alpha$}};
\node[below] at ($(C)-(0,0.5)$) {\scriptsize{(a)}};

\coordinate (C1) at ($(mid)+(8,2)$);
\coordinate (mid1) at ($(C1)-(0,2)$);
\coordinate (A1) at ($(mid1)+(-1.2,-2)$);
\coordinate (B1) at ($(mid1)+(1.2,-2)$);

\arrowpath{(mid1)}{(C1)}{0.5};
\arrowpath{(A1)}{(mid1)}{0.5};
\arrowpath{(B1)}{(mid1)}{0.5};

\node[above] at (C1) {\scriptsize{$k$}};
\node[below] at (A1) {\scriptsize{$i$}};
\node[below] at (B1) {\scriptsize{$j$}};
\node[right] at (mid1) {\scriptsize{$\alpha$}};
\node[below] at ($(C1)-(0,4.5)$) {\scriptsize{(b)}};
\end{scope}
\end{tikzpicture}
\caption{(a) Graphical representation of the chosen basis. (b) Graphical representation of the dual basis.}\label{fig:basis}
\end{figure}

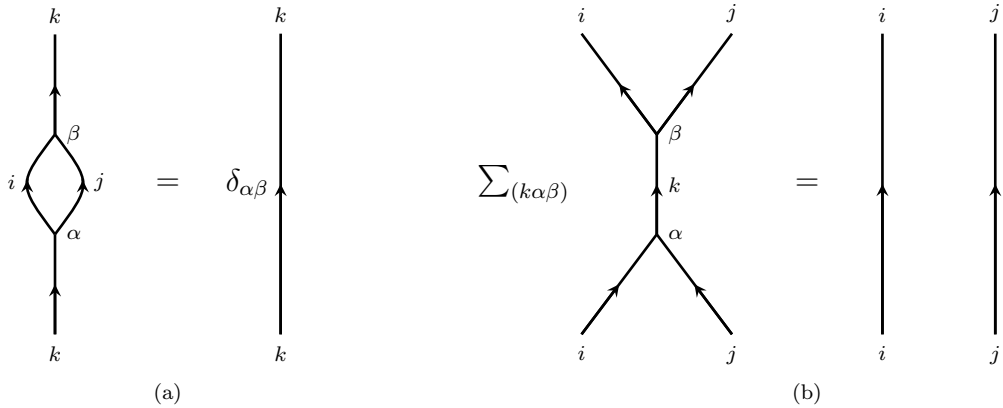
\begin{figure}
\centering
\begin{tikzpicture}[line width=1pt]
\begin{scope}[every node/.style={sloped,allow upside down}]
\coordinate (C2) at (0,0);
\coordinate (mid2) at ($(C2)+(0,1.33)$);
\coordinate (mid3) at ($(mid2)+(0,1.33)$);
\coordinate (C3) at ($(mid3)+(0,1.33)$);

\arrowpath{(C2)}{(mid2)}{0.5};
\arrowpath{(mid3)}{(C3)}{0.5};
\draw (mid2) .. controls ($(mid2)+(-0.5,0.67)$) .. (mid3);
\draw (mid2) .. controls ($(mid2)+(0.5,0.67)$) .. (mid3);
\arrowpath{($(mid2)+(-0.37,0.7)$)}{($(mid2)+(-0.37,0.74)$)}{0.5};
\arrowpath{($(mid2)+(0.37,0.7)$)}{($(mid2)+(0.37,0.74)$)}{0.5};

\node[below] at (C2) {\scriptsize{$k$}};
\node[above] at (C3) {\scriptsize{$k$}};
\node[left] at ($(mid2)+(-0.37,0.7)$) {\scriptsize{$i$}};
\node[right] at ($(mid2)+(0.37,0.7)$) {\scriptsize{$j$}};
\node[right] at (mid2) {\scriptsize{$\alpha$}};
\node[right] at (mid3) {\scriptsize{$\beta$}};

\coordinate (eq) at ($(mid2)+(1.5,0.67)$);
\node at (eq) {$=$};

\coordinate (C4) at ($(C2)+(3,0)$);
\coordinate (C5) at ($(C4)+(0,4)$);
\arrowpath{(C4)}{(C5)}{0.5};

\node[left] at ($(C4)+(0,2)$) {$\delta_{\alpha\beta}$};
\node[below] at (C4) {\scriptsize{$k$}};
\node[above] at (C5) {\scriptsize{$k$}};
\node[below] at ($(C4)-(1.5,0.5)$) {\scriptsize{(a)}};

\coordinate (i0) at ($(C4)+(4,0)$);
\coordinate (j0) at ($(i0)+(2,0)$);
\coordinate (k0) at ($(i0)+(1,1.33)$);
\coordinate (k1) at ($(k0)+(0,1.33)$);
\coordinate (i1) at ($(i0)+(0,4)$);
\coordinate (j1) at ($(j0)+(0,4)$);
\coordinate (sum) at ($(i0)+(0,2)$);

\arrowpath{(i0)}{(k0)}{0.5};
\arrowpath{(j0)}{(k0)}{0.5};
\arrowpath{(k0)}{(k1)}{0.5};
\arrowpath{(k1)}{(i1)}{0.5};
\arrowpath{(k1)}{(j1)}{0.5};

\node[above] at (i1) {\scriptsize{$i$}};
\node[above] at (j1) {\scriptsize{$j$}};
\node[below] at (i0) {\scriptsize{$i$}};
\node[below] at (j0) {\scriptsize{$j$}};
\node[right] at (k0) {\scriptsize{$\alpha$}};
\node[right] at (k1) {\scriptsize{$\beta$}};
\node[right] at ($(k0)!0.5!(k1)$) {\scriptsize{$k$}};
\node[left] at (sum) {$\sum_{(k\alpha\beta)}$};

\coordinate (eq2) at ($(j0)+(1,2)$);
\node at (eq2) {$=$};

\coordinate (I0) at ($(j0)+(2,0)$);
\coordinate (J0) at ($(I0)+(1.5,0)$);
\coordinate (I1) at ($(I0)+(0,4)$);
\coordinate (J1) at ($(J0)+(0,4)$);

\arrowpath{(I0)}{(I1)}{0.5};
\arrowpath{(J0)}{(J1)}{0.5};

\node[above] at (I1) {\scriptsize{$i$}};
\node[above] at (J1) {\scriptsize{$j$}};
\node[below] at (I0) {\scriptsize{$i$}};
\node[below] at (J0) {\scriptsize{$j$}};

\node[below] at ($(eq2)-(0,2.5)$) {\scriptsize{(b)}};
\end{scope}
\end{tikzpicture}
\caption{(a) Graphical representation of the fact that the two basis are dual to each other. (b) Completeness of the basis. Here, sum over $(k\alpha\beta)$ represents a sum over all such consistent triples.}\label{fig:duality}
\end{figure}

\begin{figure}
\centering
\begin{tikzpicture}[line width=1pt]
\begin{scope}[every node/.style={sloped,allow upside down}]
\coordinate (C) at (0,0);
\coordinate (mid) at ($(C)+(0,2)$);
\coordinate (A) at ($(mid)+(-1.2,2)$);
\coordinate (B) at ($(mid)+(1.2,2)$);

\arrowpath{(C)}{(mid)}{0.5};
\arrowpath{(mid)}{(A)}{0.5};
\arrowpath{(mid)}{(B)}{0.5};

\node[below] at (C) {\scriptsize{$k$}};
\node[above] at (A) {\scriptsize{$i$}};
\node[above] at (B) {\scriptsize{$j$}};
\node[right] at (mid) {\scriptsize{$\alpha'$}};

\coordinate (to) at ($(mid)+(3,0)$);
\node at (to) {$=$};
\node[right] at ($(to)+(0.5,0)$) {$\sum_\alpha\:\:\:(U^{ij}_k)_{\alpha'\alpha}$};

\coordinate (mid2) at ($(to)+(4.5,0)$);
\coordinate (C2) at ($(mid2)-(0,2)$);
\coordinate (A2) at ($(mid2)+(-1.2,2)$);
\coordinate (B2) at ($(mid2)+(1.2,2)$);

\arrowpath{(C2)}{(mid2)}{0.5};
\arrowpath{(mid2)}{(A2)}{0.5};
\arrowpath{(mid2)}{(B2)}{0.5};

\node[below] at (C2) {\scriptsize{$k$}};
\node[above] at (A2) {\scriptsize{$i$}};
\node[above] at (B2) {\scriptsize{$j$}};
\node[right] at (mid2) {\scriptsize{$\alpha$}};
\end{scope}
\end{tikzpicture}
\caption{A change of basis via a unitary matrix.}\label{fig:gauge}
\end{figure}
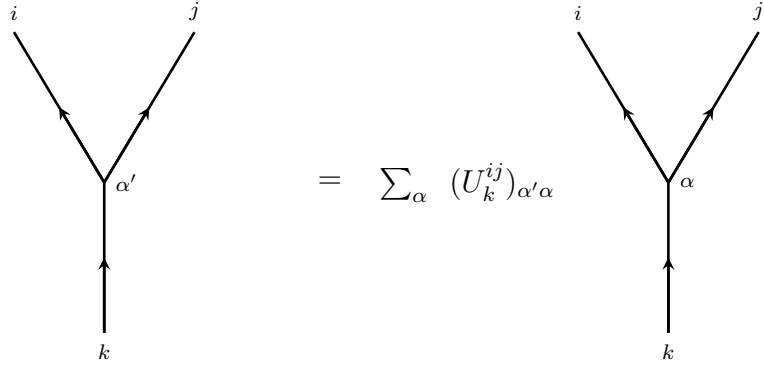

Let's look at the decomposition of the tensor product of two simple lines $L_i\otimes L_j=\oplus n_{ij}^k L_k$. This means that there is a $n_{ij}^k$ dimensional space of morphisms from $L_k$ to $L_i\otimes L_j$. We pick a basis of this space labeled by $\alpha$. Similarly we pick a dual basis for the space of morphisms from $L_i\otimes L_j$ to $L_k$ which we also label by $\alpha$. See Figure \ref{fig:basis}. The completeness of the basis can be written graphically as in Figure \ref{fig:duality}(b). \footnote{We are assuming that the quantum dimensions of all $L_i$ is 1 for simplicity. For generic quantum dimensions, we have to normalize these morphisms appropriately so that any graph of line defects and its topological deformations define the same morphism.} We can transform to a basis labeled by $\alpha'$. We denote the unitary matrix corresponding to the transformation as $(U^{ij}_k)_{\alpha'\alpha}$. See Figure \ref{fig:gauge}.

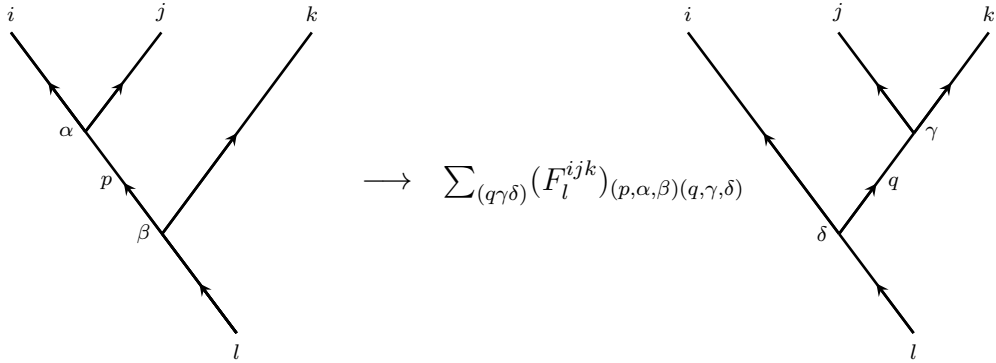
\begin{figure}
\centering
\begin{tikzpicture}[line width=1pt]
\begin{scope}[every node/.style={sloped,allow upside down}]
\coordinate (l) at (3,0);
\coordinate (i) at (0,4);
\coordinate (j) at (2,4);
\coordinate (k) at (4,4);
\coordinate (ij) at ($(l)!0.67!(i)$);
\coordinate (jk) at ($(l)!0.33!(i)$);

\arrowpath{(l)}{(i)}{0.5};
\arrowpath{(l)}{(jk)}{0.5};
\arrowpath{(ij)}{(i)}{0.5};
\arrowpath{(jk)}{(k)}{0.5};
\arrowpath{(ij)}{(j)}{0.5};

\node[above] at (i) {\scriptsize{$i$}};
\node[above] at (j) {\scriptsize{$j$}};
\node[above] at (k) {\scriptsize{$k$}};
\node[below] at (l) {\scriptsize{$l$}};
\node[left] at (ij) {\scriptsize{$\alpha$}};
\node[left] at (jk) {\scriptsize{$\beta$}};
\node[left] at ($(ij)!0.5!(jk)$) {\scriptsize{$p$}};

\coordinate (to) at ($(l)+(2,2)$);
\node at (to) {$\longrightarrow$};
\node at ($(to)+(2.75,0)$) {$\sum_{(q\gamma\delta)}(F^{ijk}_l)_{(p,\alpha,\beta)(q,\gamma,\delta)}$};

\coordinate (L) at ($(l)+(9,0)$);
\coordinate (I) at ($(i)+(9,0)$);
\coordinate (J) at ($(j)+(9,0)$);
\coordinate (K) at ($(k)+(9,0)$);
\coordinate (JK) at ($(jk)+(9,0)$);
\coordinate (IJ) at ($(JK)!0.5!(K)$);

\arrowpath{(JK)}{(IJ)}{0.5};
\arrowpath{(L)}{(JK)}{0.5};
\arrowpath{(IJ)}{(K)}{0.5};
\arrowpath{(JK)}{(I)}{0.5};
\arrowpath{(IJ)}{(J)}{0.5};

\node[above] at (I) {\scriptsize{$i$}};
\node[above] at (J) {\scriptsize{$j$}};
\node[above] at (K) {\scriptsize{$k$}};
\node[below] at (L) {\scriptsize{$l$}};
\node[right] at (IJ) {\scriptsize{$\gamma$}};
\node[left] at (JK) {\scriptsize{$\delta$}};
\node[right] at ($(IJ)!0.5!(JK)$) {\scriptsize{$q$}};

\end{scope}
\end{tikzpicture}
\caption{Definition of $F$-symbols.}\label{fig:F}
\end{figure}

The associator induces an isomorphism between the morphism space from $L_l$ to $(L_i\otimes L_j)\otimes L_k$ and the morphism space from $L_l$ to $L_i\otimes (L_j\otimes L_k)$. In terms of our chosen basis, this isomorphism can be captured in terms of $F$-symbols $(F^{ijk}_l)_{(p,\alpha,\beta)(q,\gamma,\delta)}$ which are defined in Figure \ref{fig:F}. Under a change of basis, $F$-symbols transform as
\be
(F^{ijk}_l)_{(p,\alpha,\beta)(q,\gamma,\delta)}\to(F^{ijk}_l)_{(p,\alpha,\beta)(q,\gamma,\delta)}(U^{jk}_q)^*_{\gamma'\gamma}(U^{iq}_l)^*_{\delta'\delta}(U^{ij}_p)_{\alpha'\alpha}(U^{pk}_l)_{\beta'\beta} \label{gauge}
\ee

We are now ready to describe Turaev-Viro prescription for the partition function of $\fT$ on a manifold $M$. Pick a branched triangulation $T$ of $M$. A branched triangulation requires an ordering $>$ of the vertices of the triangulation. To an edge $e$ between vertices $a$ and $b$, a branched triangulation assigns a direction $a\to b$ if $a>b$. The $G$-connection $\alpha_1$ on $M$ assigns an element $g_e$ of the group $G$ to each directed edge $e$. We now label each directed edge $e$ by a simple element living in $\CC_{g_e}$. Pick a face $f$ of $T$. Rotating it and flipping it, $f$ looks like as shown in Figure \ref{fig:face}(a). Then, we label $f$ by some $\alpha$ corresponding to a morphism as shown in the Figure \ref{fig:face}(b). Thus we have a labeling of edges and faces of a branched triangulation. Call one such labeling as $\tilde l$.

\begin{figure}
\centering
\begin{tikzpicture}[line width=1pt]
\begin{scope}[every node/.style={sloped,allow upside down}]
\coordinate (a) at (0,0);
\coordinate (b) at (4,0);
\coordinate (c) at (2,2.5);

\arrowpath{(a)}{(b)}{0.5};
\arrowpath{(a)}{(c)}{0.5};
\arrowpath{(c)}{(b)}{0.5};

\node[below] at ($(a)!0.5!(b)$) {\scriptsize{$k$}};
\node[above left] at ($(a)!0.5!(c)$) {\scriptsize{$i$}};
\node[above right] at ($(c)!0.5!(b)$) {\scriptsize{$j$}};
\node[below] at ($(a)!0.5!(b)+(0,-0.5)$) {\scriptsize{$(a)$}};

\coordinate (C) at ($(b)+(6,0)$);
\coordinate (mid) at ($(C)+(0,2)$);
\coordinate (A) at ($(mid)+(-1.2,2)$);
\coordinate (B) at ($(mid)+(1.2,2)$);

\arrowpath{(C)}{(mid)}{0.5};
\arrowpath{(mid)}{(A)}{0.5};
\arrowpath{(mid)}{(B)}{0.5};

\node[below] at (C) {\scriptsize{$k$}};
\node[above] at (A) {\scriptsize{$i$}};
\node[above] at (B) {\scriptsize{$j$}};
\node[right] at (mid) {\scriptsize{$\alpha$}};
\node[below] at ($(C)-(0,0.5)$) {\scriptsize{(b)}};
\end{scope}
\end{tikzpicture}
\caption{Labeling of faces.}\label{fig:face}
\end{figure}
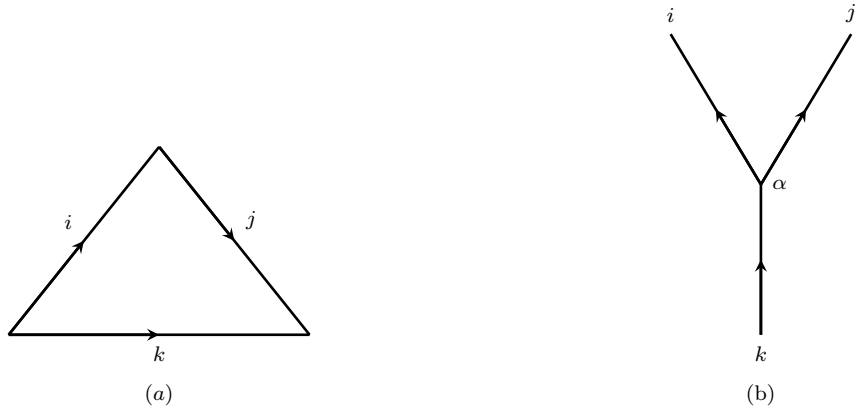

\begin{figure}
\begin{tikzpicture}[line width=1pt]
\begin{scope}[every node/.style={sloped,allow upside down}]
\coordinate (2) at (0,0);
\coordinate (1) at ($(2)+(3,-0.5)$);
\coordinate (0) at ($(2)+(2.1,3.25)$);
\coordinate (3) at ($(1)+(1.5,1.5)$);

\arrowpath{(2)}{(3)}{0.5};
\paddedline{(0)}{(1)}{(0.1,0)};
\arrowpath{(0)}{(1)}{0.5};
\arrowpath{(0)}{(2)}{0.5};
\arrowpath{(1)}{(2)}{0.5};
\arrowpath{(0)}{(3)}{0.5};
\arrowpath{(1)}{(3)}{0.5};

\node[above] at ($(2)-(1.5,0)$) {};
\node[above] at (0) {\scriptsize{$A$}};
\node[below] at (1) {\scriptsize{$B$}};
\node[left] at (2) {\scriptsize{$C$}};
\node[right] at (3) {\scriptsize{$D$}};

\node[right] at ($(0)!0.5!(1)$) {\scriptsize{$i$}};
\node[below] at ($(1)!0.5!(2)$) {\scriptsize{$j$}};
\node[above left] at ($(2)!0.5!(3)$) {\scriptsize{$k$}};
\node[left] at ($(0)!0.5!(2)$) {\scriptsize{$p$}};
\node[below right] at ($(1)!0.5!(3)$) {\scriptsize{$q$}};
\node[right] at ($(0)!0.5!(3)$) {\scriptsize{$l$}};

\node (a) at ($(0)+(0.25,-4.5)$) {\scriptsize{(a)}};

\coordinate (1') at ($(2)+(8.5,0)$);
\coordinate (2') at ($(1')+(3,-0.5)$);
\coordinate (0') at ($(1')+(2.1,3.25)$);
\coordinate (3') at ($(2')+(1.5,1.5)$);

\arrowpath{(1')}{(3')}{0.5};
\paddedline{(0')}{(2')}{(0.1,0)};
\arrowpath{(0')}{(2')}{0.5};
\arrowpath{(0')}{(1')}{0.5};
\arrowpath{(1')}{(2')}{0.5};
\arrowpath{(0')}{(3')}{0.5};
\arrowpath{(2')}{(3')}{0.5};

\node[above] at (0') {\scriptsize{$A$}};
\node[left] at (1') {\scriptsize{$B$}};
\node[below] at (2') {\scriptsize{$C$}};
\node[right] at (3') {\scriptsize{$D$}};

\node[left] at ($(0')!0.5!(1')$) {\scriptsize{$i$}};
\node[below] at ($(1')!0.5!(2')$) {\scriptsize{$j$}};
\node[below right] at ($(2')!0.5!(3')$) {\scriptsize{$k$}};
\node[right] at ($(0')!0.5!(2')$) {\scriptsize{$p$}};
\node[below left] at ($(1')!0.5!(3')$) {\scriptsize{$q$}};
\node[right] at ($(0')!0.5!(3')$) {\scriptsize{$l$}};

\node (b) at ($(0')+(0.25,-4.5)$) {\scriptsize{(b)}};
\end{scope}
\end{tikzpicture}
\caption{A tetrahedron can have two chiralities - (a) positive and (b) negative. There is a label attached to every face but we don't show it in the figure for brevity.} \label{fig:chirality}
\end{figure}
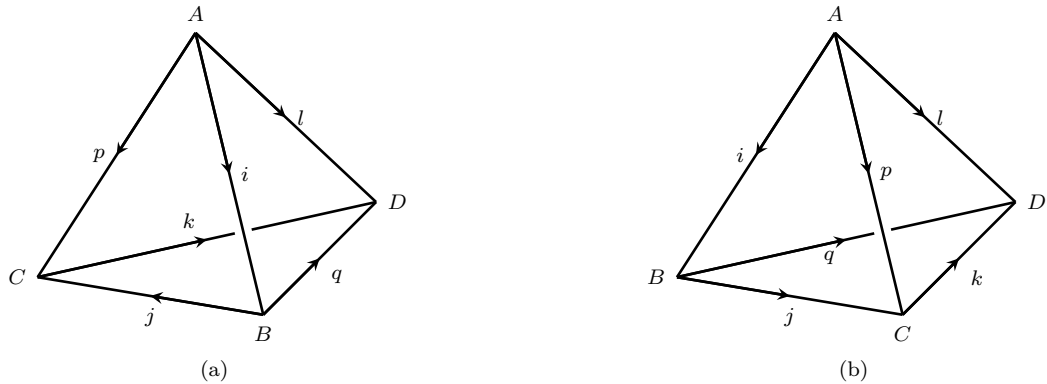

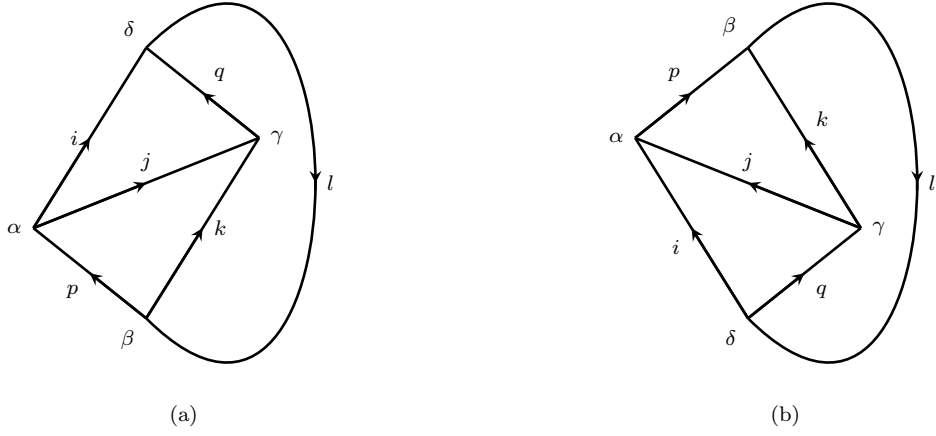
\begin{figure}
\centering
\begin{tikzpicture}[line width=1pt]
\begin{scope}[every node/.style={sloped,allow upside down}]
\coordinate (023) at (0,0);
\coordinate (012) at ($(023)+(-1.5,1.2)$);
\coordinate (123) at ($(023)+(1.5,2.4)$);
\coordinate (013) at ($(023)+(0,3.6)$);
\coordinate (controlu) at ($(013)+(3,3)$);
\coordinate (controld) at ($(023)+(3,-3)$);
\coordinate (mid) at ($(controlu)!0.5!(controld)-(0.75,0)$);

\arrowpath{(023)}{(012)}{0.5};
\arrowpath{(023)}{(123)}{0.5};
\arrowpath{(012)}{(123)}{0.5};
\arrowpath{(012)}{(013)}{0.5};
\arrowpath{(123)}{(013)}{0.5};
\draw (013)..controls(controlu) and (controld)..(023);
\arrowpath{($(mid)+(0,0.1)$)}{($(mid)-(0,0.1)$)}{0.5};

\node[left] at (012) {\scriptsize{$\alpha$}};
\node[below left] at (023) {\scriptsize{$\beta$}};
\node[above left] at (013) {\scriptsize{$\delta$}};
\node[right] at (123) {\scriptsize{$\gamma$}};

\node[below left] at ($(023)!0.5!(012)$) {\scriptsize{$p$}};
\node[right] at ($(023)!0.5!(123)$) {\scriptsize{$k$}};
\node[above] at ($(012)!0.5!(123)$) {\scriptsize{$j$}};
\node[left] at ($(012)!0.5!(013)$) {\scriptsize{$i$}};
\node[above right] at ($(123)!0.5!(013)$) {\scriptsize{$q$}};
\node[right] at (mid) {\scriptsize{$l$}};


\node[below] at ($(023) + (0.5,-1)$) {\scriptsize{(a)}};

\coordinate (023') at ($(023)+(8,0)$);
\coordinate (012') at ($(023')+(-1.5,2.4)$);
\coordinate (123') at ($(023')+(1.5,1.2)$);
\coordinate (013') at ($(023')+(0,3.6)$);
\coordinate (controlu') at ($(013')+(3,3)$);
\coordinate (controld') at ($(023')+(3,-3)$);
\coordinate (mid') at ($(controlu')!0.5!(controld')-(0.75,0)$);

\arrowpath{(023')}{(012')}{0.5};
\arrowpath{(023')}{(123')}{0.5};
\arrowpath{(123')}{(012')}{0.5};
\arrowpath{(012')}{(013')}{0.5};
\arrowpath{(123')}{(013')}{0.5};
\draw (013')..controls(controlu') and (controld')..(023');
\arrowpath{($(mid')+(0,0.1)$)}{($(mid')-(0,0.1)$)}{0.5};

\node[left] at (012') {\scriptsize{$\alpha$}};
\node[below left] at (023') {\scriptsize{$\delta$}};
\node[above left] at (013') {\scriptsize{$\beta$}};
\node[right] at (123') {\scriptsize{$\gamma$}};

\node[below left] at ($(023')!0.5!(012')$) {\scriptsize{$i$}};
\node[below right] at ($(023')!0.5!(123')$) {\scriptsize{$q$}};
\node[above] at ($(012')!0.5!(123')$) {\scriptsize{$j$}};
\node[above left] at ($(012')!0.5!(013')$) {\scriptsize{$p$}};
\node[above right] at ($(123')!0.5!(013')$) {\scriptsize{$k$}};
\node[right] at (mid') {\scriptsize{$l$}};


\node[below] at ($(023') + (0.5,-1)$) {\scriptsize{(b)}};
\end{scope}
\end{tikzpicture}
\caption{Graph attached to a tetrahedron: (a) positive chirality and (b) negative chirality.} \label{fig:graph}
\end{figure}

Pick a tetrahedron $t$ in $\tilde l$. To each $t$ we assign a planar graph $\Gamma_t$ in $\CC$ and we call such a graph as a \emph{tetrahedron graph}. Notice that $t$ can have two chiralities - positive and negative as shown in Figure \ref{fig:chirality}. $\Gamma_t$ for a positive chirality $t$ and a negative chirality $t$ are shown in Figure \ref{fig:graph}. The first one evaluates to $(F^{ijk}_l)_{(p,\alpha,\beta)(q,\gamma,\delta)}$ and the second one evaluates to $(F^{ijk}_l)^*_{(p,\alpha,\beta)(q,\gamma,\delta)}$. Let's call this number as $n_t(\tilde l)$ and define $N(l)=\prod_t n_t(\tilde l)$. To each edge $e$ of $T$, we can associate a number $d_e(\tilde l)$ which is the quantum dimension of the simple line assigned to $e$ in $\tilde l$. Define $d(\tilde l)=\prod_e d_e(\tilde l)$. The partition function $Z(M)$ is then given by
\be
Z(M)=D^{-2v}\sum_{\tilde l} N(\tilde l)d(\tilde l)
\ee
where $D=\sqrt{\sum_i d_i^2}$ is the total quantum dimension of $\CC$ (where $d_i$ is the quantum dimension of simple line $L_i$) and $v$ is the number of vertices in $T$.

The invariance of $Z(M)$ under Pachner moves is guaranteed by the pentagon equation satisfied by the associators in $\CC$. The pentagon equation says that the following morphism made by composing associators
\be
((A\otimes B)\otimes C)\otimes D\to(A\otimes (B\otimes C))\otimes D\to A\otimes ((B\otimes C)\otimes D)\to A\otimes (B\otimes (C\otimes D))
\ee
and the following morphism made by composing associators
\be
((A\otimes B)\otimes C)\otimes D\to (A\otimes B)\otimes (C\otimes D)\to A\otimes (B\otimes (C\otimes D))
\ee
are equal. In terms of $F$-symbols, this means that
\be
\sum_{r,\delta,\epsilon,\mu} (F^{ijk}_q)_{(p,\alpha,\beta)(q,\gamma,\delta)}(F^{irl}_m)_{(q,\epsilon,\gamma)(s,\mu,\nu)}(F^{jkl}_s)_{(r,\delta,\mu)(t,\rho,\sigma)}=\sum_\tau(F^{pkl}_m)_{(q,\beta,\gamma)(t,\rho,\tau)}(F^{ijt}_m)_{(p,\alpha,\tau)(s,\sigma,\nu)} \label{pentagon}
\ee
One can check that (\ref{pentagon}) is invariant under an arbitrary gauge transformation (\ref{gauge}).

\subsection{Twisted spherical fusion category and orientation reversing defects} \label{2.3}
Consider an oriented theory defined by $\CC$. We propose that an unoriented parent theory can be constructed in terms of a larger ``twisted" spherical fusion category $\tilde\CC$. This larger category is assembled from four pieces $\tilde\CC=\tilde\CC_{0,0}\oplus\tilde\CC_{0,1}\oplus\tilde\CC_{1,0}\oplus\tilde\CC_{1,1}$ such that each of the subcategories $\tilde\CC_{\epsilon,\epsilon'}$ is $G$-graded. $\tilde\CC_{0,1}$ is a bimodule on which $\tilde\CC_{0,0}$ acts from left and $\tilde\CC_{1,1}$ acts from right. Similarly, $\tilde\CC_{1,0}$ is a (non-empty) bimodule with $\tilde\CC_{1,1}$ acting from the left and $\tilde\CC_{0,0}$ acting from the right. An object in $\tilde\CC_{0,1}$ fuses with an object in $\tilde\CC_{1,0}$ to give an object in $\tilde\CC_{0,0}$. Similarly, an object in $\tilde\CC_{1,0}$ fuses with an object in $\tilde\CC_{0,1}$ to give an object in $\tilde\CC_{1,1}$. $\tilde\CC$ can also be thought of as a 2-category made out of two objects `0' and `1'.

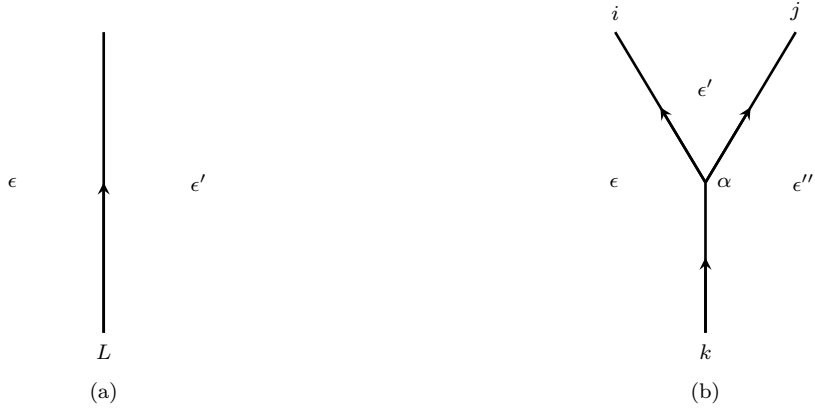
\begin{figure}
\centering
\begin{tikzpicture}[line width=1pt]
\begin{scope}[every node/.style={sloped,allow upside down}]
\coordinate (lowA) at (0,0);
\coordinate (highA) at (0,4);
\arrowpath{(lowA)}{(highA)}{0.5};
\node[below] at (lowA) {\scriptsize{$L$}};
\node[left] at ($(lowA)+(-1,2)$) {\scriptsize{$\epsilon$}};
\node[right] at ($(lowA)+(1,2)$) {\scriptsize{$\epsilon'$}};
\node[below] at ($(lowA) - (0, 0.5)$) {\scriptsize{(a)}};

\coordinate (C) at (8,0);
\coordinate (mid) at ($(C)+(0,2)$);
\coordinate (A) at ($(mid)+(-1.2,2)$);
\coordinate (B) at ($(mid)+(1.2,2)$);

\arrowpath{(C)}{(mid)}{0.5};
\arrowpath{(mid)}{(A)}{0.5};
\arrowpath{(mid)}{(B)}{0.5};

\node[below] at (C) {\scriptsize{$k$}};
\node[above] at (A) {\scriptsize{$i$}};
\node[above] at (B) {\scriptsize{$j$}};
\node[right] at (mid) {\scriptsize{$\alpha$}};
\node[left] at ($(mid)+(-1,0)$) {\scriptsize{$\epsilon$}};
\node[right] at ($(mid)+(1,0)$) {\scriptsize{$\epsilon''$}};
\node[above] at ($(mid)+(0,1)$) {\scriptsize{$\epsilon'$}};
\node[below] at ($(C)-(0,0.5)$) {\scriptsize{(b)}};
\end{scope}
\end{tikzpicture}
\caption{(a) A line in $\tilde\CC_{\epsilon,\epsilon'}$. (b) A sample graph in $\tilde\CC$ showing a morphism $\alpha$ in the morphism space $(V^{ij}_k)_\epsilon$.} \label{fig:sample}
\end{figure}

$\tilde\CC_{0,0}$ is the same as the spherical fusion category $\CC$. $\tilde\CC_{1,1}$ has same objects as that of $\CC$. Similarly, $\tilde\CC_{1,0}$ is a copy of $\tilde\CC_{0,1}$ at the level of objects. Graphically, we describe simple objects of $\tilde\CC_{\epsilon,\epsilon'}$ as lines with the left plaquette labeled by $\epsilon$ and right plaquette labeled by $\epsilon'$. In general, we draw graphs $\Gamma$ in $\tilde\CC$ in which we label the each plaquette by some $\epsilon$. See Figure \ref{fig:sample}. 

In our notation, the labels $i$, $j$, $k$ etc. tell us that we have a line of $\tilde\CC_{\epsilon,\epsilon'}$ with a specific value of $\epsilon+\epsilon'$ but do not determine the individual $\epsilon$, $\epsilon'$. The data of individual $\epsilon$ is captured in the labeling of plaquettes by $\epsilon$, $\epsilon'$, $\epsilon''$ etc. Thus the labeling of plaquettes is slightly redundant. We need only specify the label of a single plaquette and the labels for the other plaquettes can be determined from the labels $i$, $j$, $k$ etc. In what follows, we will often just specify the label of the left-most plaquette.

\begin{figure}
\centering
\begin{tikzpicture}[line width=1pt]
\begin{scope}[every node/.style={sloped,allow upside down}]
\coordinate (C) at (0,0);
\coordinate (mid) at ($(C)+(0,2)$);
\coordinate (A) at ($(mid)+(-1.2,2)$);
\coordinate (B) at ($(mid)+(1.2,2)$);

\arrowpath{(C)}{(mid)}{0.5};
\arrowpath{(mid)}{(A)}{0.5};
\arrowpath{(mid)}{(B)}{0.5};

\node[below] at (C) {\scriptsize{$k$}};
\node[above] at (A) {\scriptsize{$i$}};
\node[above] at (B) {\scriptsize{$j$}};
\node[right] at (mid) {\scriptsize{$\alpha$}};
\node[left] at ($(mid)+(-1,0)$) {\scriptsize{$\epsilon$}};
\node[right] at ($(mid)+(1,0)$) {\scriptsize{$\epsilon''$}};
\node[above] at ($(mid)+(0,1)$) {\scriptsize{$\epsilon'$}};
\node[below] at ($(C)-(0,0.5)$) {\scriptsize{(a)}};

\coordinate (to) at ($(mid)+(4,0)$);
\node at (to) {$\longrightarrow$};
\node[above] at (to) {\scriptsize{$I$}};

\coordinate (C') at (8,0);
\coordinate (mid') at ($(C')+(0,2)$);
\coordinate (A') at ($(mid')+(-1.2,2)$);
\coordinate (B') at ($(mid')+(1.2,2)$);

\arrowpath{(C')}{(mid')}{0.5};
\arrowpath{(mid')}{(A')}{0.5};
\arrowpath{(mid')}{(B')}{0.5};

\node[below] at (C') {\scriptsize{$k$}};
\node[above] at (A') {\scriptsize{$i$}};
\node[above] at (B') {\scriptsize{$j$}};
\node[right] at (mid') {\scriptsize{$\alpha$}};
\node[left] at ($(mid')+(-1,0)$) {\scriptsize{$\epsilon+1$}};
\node[right] at ($(mid')+(1,0)$) {\scriptsize{$\epsilon''+1$}};
\node[above] at ($(mid')+(0,1)$) {\scriptsize{$\epsilon'+1$}};
\node[below] at ($(C')-(0,0.5)$) {\scriptsize{(b)}};
\end{scope}
\end{tikzpicture}
\caption{The action of anti-linear isomorphism $I$.} \label{fig:I}
\end{figure}
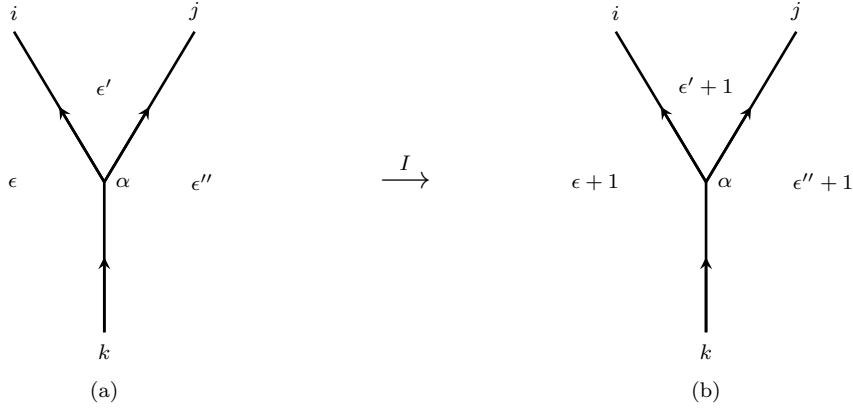

$\tilde\CC$ comes equipped with the data of an anti-linear isomorphism $I$ between various morphism spaces. This map is easy to describe in terms of simple objects. It takes the morphism space $(V^{ij}_k)_\epsilon$ from $L_k$ to $L_i\otimes L_j$ with some labeling of plaquettes (such that the left-most plaquette is $\epsilon$) to the morphism space $(V^{ij}_k)_{\epsilon+1}$ from $L_k$ to $L_i\otimes L_j$ but with the labeling on plaquettes flipped. Thus, we just have to pick a basis $\alpha$ of morphism spaces $(V^{ij}_k)_0$. The basis $(V^{ij}_k)_1$ is determined by the action of $I$ and we label the resulting basis by the same labels $\alpha$. See Figure \ref{fig:I}. Thus, under a change of basis $(U^{ij}_k)_{\alpha'\alpha}$ of $(V^{ij}_k)_0$, the basis of $(V^{ij}_k)_1$ transforms by $(U^{ij}_k)^*_{\alpha'\alpha}$.

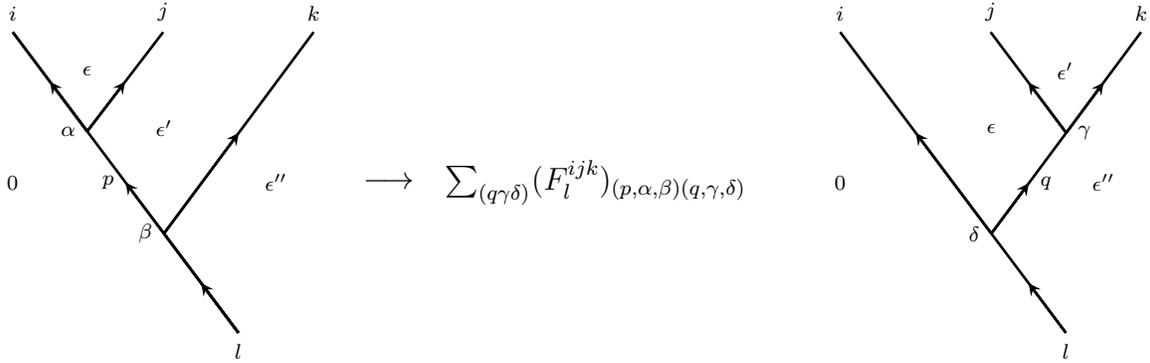
\begin{figure}
\centering
\begin{tikzpicture}[line width=1pt]
\begin{scope}[every node/.style={sloped,allow upside down}]
\coordinate (l) at (3,0);
\coordinate (i) at (0,4);
\coordinate (j) at (2,4);
\coordinate (k) at (4,4);
\coordinate (ij) at ($(l)!0.67!(i)$);
\coordinate (jk) at ($(l)!0.33!(i)$);

\arrowpath{(l)}{(i)}{0.5};
\arrowpath{(l)}{(jk)}{0.5};
\arrowpath{(ij)}{(i)}{0.5};
\arrowpath{(jk)}{(k)}{0.5};
\arrowpath{(ij)}{(j)}{0.5};

\node[above] at (i) {\scriptsize{$i$}};
\node[above] at (j) {\scriptsize{$j$}};
\node[above] at (k) {\scriptsize{$k$}};
\node[below] at (l) {\scriptsize{$l$}};
\node[left] at (ij) {\scriptsize{$\alpha$}};
\node[left] at (jk) {\scriptsize{$\beta$}};
\node[left] at ($(ij)!0.5!(jk)$) {\scriptsize{$p$}};

\coordinate (0) at ($(ij)!0.5!(jk)-(1.5,0)$);
\node at ($(ij)!0.5!(jk)-(1.5,0)$) {\scriptsize{$0$}};
\node at ($(ij)+(0,0.8)$) {\scriptsize{$\epsilon$}};
\node at ($(jk)+(0,1.4)$) {\scriptsize{$\epsilon'$}};
\node at ($(0)+(3.5,0)$) {\scriptsize{$\epsilon''$}};

\coordinate (to) at ($(l)+(2,2)$);
\node at (to) {$\longrightarrow$};
\node at ($(to)+(2.75,0)$) {$\sum_{(q\gamma\delta)}(F^{ijk}_l)_{(p,\alpha,\beta)(q,\gamma,\delta)}$};

\coordinate (L) at ($(l)+(11,0)$);
\coordinate (I) at ($(i)+(11,0)$);
\coordinate (J) at ($(j)+(11,0)$);
\coordinate (K) at ($(k)+(11,0)$);
\coordinate (JK) at ($(jk)+(11,0)$);
\coordinate (IJ) at ($(JK)!0.5!(K)$);

\arrowpath{(JK)}{(IJ)}{0.5};
\arrowpath{(L)}{(JK)}{0.5};
\arrowpath{(IJ)}{(K)}{0.5};
\arrowpath{(JK)}{(I)}{0.5};
\arrowpath{(IJ)}{(J)}{0.5};

\node[above] at (I) {\scriptsize{$i$}};
\node[above] at (J) {\scriptsize{$j$}};
\node[above] at (K) {\scriptsize{$k$}};
\node[below] at (L) {\scriptsize{$l$}};
\node[right] at (IJ) {\scriptsize{$\gamma$}};
\node[left] at (JK) {\scriptsize{$\delta$}};
\node[right] at ($(IJ)!0.5!(JK)$) {\scriptsize{$q$}};

\coordinate (0') at ($(0)+(11,0)$);
\node at ($(0')$) {\scriptsize{$0$}};
\node at ($(ij)+(0,0.8)+(11,0)+(2,0)$) {\scriptsize{$\epsilon'$}};
\node at ($(jk)+(0,1.4)+(11,0)$) {\scriptsize{$\epsilon$}};
\node at ($(0)+(3.5,0)+(11,0)$) {\scriptsize{$\epsilon''$}};

\end{scope}
\end{tikzpicture}
\caption{Definition of $F$-symbols for $\tilde\CC$.}\label{fig:tF}
\end{figure}

The associators are compatible with $I$. Let's denote $F$-symbol associated to a graph having the left-most plaquette 0 as $(F^{ijk}_l)_{(p,\alpha,\beta)(q,\gamma,\delta)}$ as shown in Figure \ref{fig:tF}. Then, compatibility of associator and $I$ implies that the $F$-symbol associated to the same graph but with the labels of all plaquettes flipped is $(F^{ijk}_l)^*_{(p,\alpha,\beta)(q,\gamma,\delta)}$. Thus, the pentagon equation for the associator becomes
\be
\sum_{r,\delta,\epsilon,\mu} (F^{ijk}_q)_{(p,\alpha,\beta)(q,\gamma,\delta)}(F^{irl}_m)_{(q,\epsilon,\gamma)(s,\mu,\nu)}(F^{jkl}_s)^{s(i)}_{(r,\delta,\mu)(t,\rho,\sigma)}=\sum_\tau(F^{pkl}_m)_{(q,\beta,\gamma)(t,\rho,\tau)}(F^{ijt}_m)_{(p,\alpha,\tau)(s,\sigma,\nu)} \label{tpentagon}
\ee
where $s(i)=*$ if $i$ labels a simple object of $\tilde\CC_{0,1}$ or $\tilde\CC_{1,0}$ and $s(i)=1$ otherwise. As the equation in terms of $F$-symbols looks different from (\ref{pentagon}), we refer to this equation as the twisted pentagon equation even though it still descends from the pentagon equation on the associators. This equation also appeared in \cite{cheng2016exactly}.

Notice that the gauge transformations on $F$-symbols now take the following form
\be
(F^{ijk}_l)_{(p,\alpha,\beta)(q,\gamma,\delta)}\to(F^{ijk}_l)_{(p,\alpha,\beta)(q,\gamma,\delta)}(U^{jk}_q)^{*s(i)}_{\gamma'\gamma}(U^{iq}_l)^*_{\delta'\delta}(U^{ij}_p)_{\alpha'\alpha}(U^{pk}_l)_{\beta'\beta} \label{tgauge}
\ee
and one can check that (\ref{tpentagon}) is invariant under this gauge transformation.

The identity line of $\CC=\tilde\CC_{0,0}$ can be inserted anywhere in plaquettes labeled by $0$ without changing any answers. Similarly, the identity line of $\tilde\CC_{1,1}$ can be inserted anywhere in plaquettes labeled by $1$ without changing any answers. For each object in $\tilde\CC_{0,1}$, we define a dual object in $\tilde\CC_{1,0}$ and vice-versa. These duals are slightly different from the usual duals in a spherical fusion category $\CC$. That is, given a line $L$ in $\tilde\CC_{0,1}$, the evaluation maps take $L\otimes L^*$ to identity in $\tilde\CC_{0,0}$ and take $L^*\otimes L$ to identity in $\tilde\CC_{1,1}$. Similar statements hold true if we flip 0 and 1 or replace evaluation with co-evaluation.

Given a graph $\Gamma$ in $\tilde\CC$ drawn on the sphere, different projections to planar graphs $\Gamma_p$ must be equivalent. In other words, we demand that there are conditions on $\tilde\CC$ similar to that of a spherical structure on a spherical fusion category $\CC$.

\begin{figure}
\centering
\begin{tikzpicture}[line width=1pt]
\begin{scope}[every node/.style={sloped,allow upside down}]
\coordinate (C) at (0,0);
\coordinate (mid) at ($(C)+(0,2)$);
\coordinate (A) at ($(mid)+(-1.2,2)$);
\coordinate (B) at ($(mid)+(1.2,2)$);

\arrowpath{(C)}{(mid)}{0.5};
\arrowpath{(mid)}{(A)}{0.5};
\arrowpath{(mid)}{(B)}{0.5};

\node[below] at (C) {\scriptsize{$k$}};
\node[above] at (A) {\scriptsize{$i$}};
\node[above] at (B) {\scriptsize{$j$}};
\node[right] at (mid) {\scriptsize{$\alpha$}};
\node[left] at ($(mid)+(-1,0)$) {\scriptsize{$0$}};
\node[right] at ($(mid)+(1,0)$) {\scriptsize{$0$}};
\node[above] at ($(mid)+(0,1)$) {\scriptsize{$0$}};
\node[below] at ($(C)-(0,0.5)$) {\scriptsize{(a)}};

\coordinate (to) at ($(mid)+(3,0)$);
\node at (to) {$\longrightarrow$};
\node[above] at (to) {\scriptsize{$U_R$}};

\coordinate (C') at (6,0);
\coordinate (mid') at ($(C')+(0,2)$);
\coordinate (A') at ($(mid')+(-1.2,2)$);
\coordinate (B') at ($(mid')+(1.2,2)$);

\arrowpath{(C')}{(mid')}{0.5};
\arrowpath{(mid')}{(A')}{0.5};
\arrowpath{(mid')}{(B')}{0.5};

\node[below] at (C') {\scriptsize{$k^*$}};
\node[above] at (A') {\scriptsize{$i^*$}};
\node[above] at (B') {\scriptsize{$j^*$}};
\node[left] at ($(mid')+(-1,0)$) {\scriptsize{$1$}};
\node[right] at ($(mid')+(1,0)$) {\scriptsize{$1$}};
\node[above] at ($(mid')+(0,1)$) {\scriptsize{$1$}};
\node[below] at ($(C')-(0,0.5)$) {\scriptsize{(b)}};

\coordinate (to') at ($(mid')+(3,0)$);
\node at (to') {$\longrightarrow$};
\node[above] at (to') {\scriptsize{Mirror}};

\coordinate (C'') at (12,0);
\coordinate (mid'') at ($(C'')+(0,2)$);
\coordinate (A'') at ($(mid'')+(-1.2,2)$);
\coordinate (B'') at ($(mid'')+(1.2,2)$);

\arrowpath{(C'')}{(mid'')}{0.5};
\arrowpath{(mid'')}{(A'')}{0.5};
\arrowpath{(mid'')}{(B'')}{0.5};

\node[below] at (C'') {\scriptsize{$k^*$}};
\node[above] at (A'') {\scriptsize{$j^*$}};
\node[above] at (B'') {\scriptsize{$i^*$}};
\node[right] at (mid'') {\scriptsize{$\alpha$}};
\node[left] at ($(mid'')+(-1,0)$) {\scriptsize{$1$}};
\node[right] at ($(mid'')+(1,0)$) {\scriptsize{$1$}};
\node[above] at ($(mid'')+(0,1)$) {\scriptsize{$1$}};
\node[below] at ($(C'')-(0,0.5)$) {\scriptsize{(c)}};
\end{scope}
\end{tikzpicture}
\caption{The action of $U_R$ sends a graph (a) on $\fB$ to a graph (b) on $\fB'$ but in the ``wrong" orientation. This means that the tensor product in graph (b) is taken from right to left. In order to get back to our convention of tensor product from left to right, we take a mirror of graph (b) and obtain graph (c). The resulting graph (c) is read in $\CC'$.} \label{fig:mirror}
\end{figure}
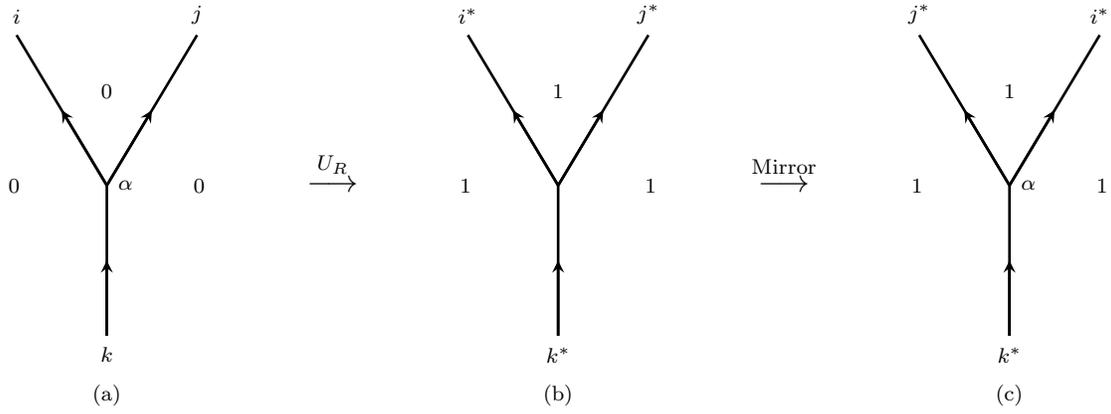


We now turn our attention to the physical interpretation of the structure we have just described. An unoriented 3d TFT $\tilde\fT$ has an orientation reversing defect $U_R$ implementing a reflection transformation. This defect can fuse with other orientation preserving defects $U_g$ to form more orientation reversing defects $U_{Rg}$. The fusion of these defects froms a group $\tilde G=G\times\Z_2$ and there is a canonical homomorphism $\rho_1$ from $\tilde G$ to $\Z_2$ whose kernel is $G$.

$U_R$ can be fused with the boundary $\fB$ to give a new boundary $\fB'$. Under such a fusion, the orientation of the boundary flips as well. There is a spherical fusion category $\CC'$ associated to $\fB'$ which is identified as $\tilde\CC_{1,1}$. If there is a line $L$ on $\fB$, then fusion of $\fB$ with $U_R$ flips its orientation and we obtain the line $L^*$ on $\fB'$. Consider a morphism from $L_k$ to $L_i\otimes L_j$ on $\fB$. Slapping $U_R$ on top of it, we send each line to its dual and $\fB$ to $\fB'$. However, since this process flips the orientation of the boundary, we have to take a mirror of the resulting configuration of lines to read it in terms of $\CC'$. See Figure \ref{fig:mirror}. Thus, fusion with $U_R$ provides a linear isomorphism from $V^{ij}_k$ in $\CC$ to $V^{j^*i^*}_{k^*}$ in $\CC'$. This is the origin of the anti-linear isomorphism $I$ in $\tilde\CC$ described above.

\begin{figure}
\centering
\begin{tikzpicture}[line width=1pt]
\begin{scope}[every node/.style={sloped,allow upside down}]
\coordinate (023) at (0,0);
\coordinate (012) at ($(023)+(-1.5,1.2)$);
\coordinate (123) at ($(023)+(1.5,2.4)$);
\coordinate (013) at ($(023)+(0,3.6)$);
\coordinate (controlu) at ($(013)+(3,3)$);
\coordinate (controld) at ($(023)+(3,-3)$);
\coordinate (mid) at ($(controlu)!0.5!(controld)-(0.75,0)$);

\arrowpath{(023)}{(012)}{0.5};
\arrowpath{(023)}{(123)}{0.5};
\arrowpath{(012)}{(123)}{0.5};
\arrowpath{(012)}{(013)}{0.5};
\arrowpath{(123)}{(013)}{0.5};
\draw (013)..controls(controlu) and (controld)..(023);
\arrowpath{($(mid)+(0,0.1)$)}{($(mid)-(0,0.1)$)}{0.5};

\node[left] at (012) {\scriptsize{$\alpha$}};
\node[below left] at (023) {\scriptsize{$\beta$}};
\node[above left] at (013) {\scriptsize{$\delta$}};
\node[right] at (123) {\scriptsize{$\gamma$}};

\node[below left] at ($(023)!0.5!(012)$) {\scriptsize{$p$}};
\node[right] at ($(023)!0.5!(123)$) {\scriptsize{$k$}};
\node[above] at ($(012)!0.5!(123)$) {\scriptsize{$j$}};
\node[left] at ($(012)!0.5!(013)$) {\scriptsize{$i$}};
\node[above right] at ($(123)!0.5!(013)$) {\scriptsize{$q$}};
\node[right] at (mid) {\scriptsize{$l$}};

\node[left] at ($(012)+(0,1.5)$) {\scriptsize{$\epsilon$}};
\node at ($(012)+(0,1.5)+(1.5,0)$) {\scriptsize{$\epsilon'$}};
\node at ($(012)+(0,1.5)+(1.5,0)+(0,-1.5)$) {\scriptsize{$\epsilon''$}};
\node[below] at ($(012)+(0,1.5)+(1.5,0)+(0,-1.5)+(1.5,0)$) {\scriptsize{$\epsilon'''$}};


\node[below] at ($(023) + (0.5,-1)$) {\scriptsize{(a)}};

\coordinate (023') at ($(023)+(8,0)$);
\coordinate (012') at ($(023')+(-1.5,2.4)$);
\coordinate (123') at ($(023')+(1.5,1.2)$);
\coordinate (013') at ($(023')+(0,3.6)$);
\coordinate (controlu') at ($(013')+(3,3)$);
\coordinate (controld') at ($(023')+(3,-3)$);
\coordinate (mid') at ($(controlu')!0.5!(controld')-(0.75,0)$);

\arrowpath{(023')}{(012')}{0.5};
\arrowpath{(023')}{(123')}{0.5};
\arrowpath{(123')}{(012')}{0.5};
\arrowpath{(012')}{(013')}{0.5};
\arrowpath{(123')}{(013')}{0.5};
\draw (013')..controls(controlu') and (controld')..(023');
\arrowpath{($(mid')+(0,0.1)$)}{($(mid')-(0,0.1)$)}{0.5};

\node[left] at (012') {\scriptsize{$\alpha$}};
\node[below left] at (023') {\scriptsize{$\delta$}};
\node[above left] at (013') {\scriptsize{$\beta$}};
\node[right] at (123') {\scriptsize{$\gamma$}};

\node[below left] at ($(023')!0.5!(012')$) {\scriptsize{$i$}};
\node[below right] at ($(023')!0.5!(123')$) {\scriptsize{$q$}};
\node[above] at ($(012')!0.5!(123')$) {\scriptsize{$j$}};
\node[above left] at ($(012')!0.5!(013')$) {\scriptsize{$p$}};
\node[above right] at ($(123')!0.5!(013')$) {\scriptsize{$k$}};
\node[right] at (mid') {\scriptsize{$l$}};

\node[left] at ($(012')-(0,1.5)$) {\scriptsize{$\epsilon+1$}};
\node[above] at ($(012')-(0,1.5)+(1.5,0)$) {\scriptsize{$\epsilon'+1$}};
\node[above] at ($(012')-(0,1.5)+(1.5,0)-(0,-1.5)$) {\scriptsize{$\epsilon''+1$}};
\node[above left] at ($(012')-(0,1.5)+(1.5,0)-(0,-1.5)+(1.5,0.75)$) {\scriptsize{$\epsilon'''+1$}};


\node[below] at ($(023') + (0.5,-1)$) {\scriptsize{(b)}};
\end{scope}
\end{tikzpicture}
\caption{The symmetry of the theory under a reflection guarantees that the above two graphs evaluate to the same number.} \label{fig:reflection}
\end{figure}
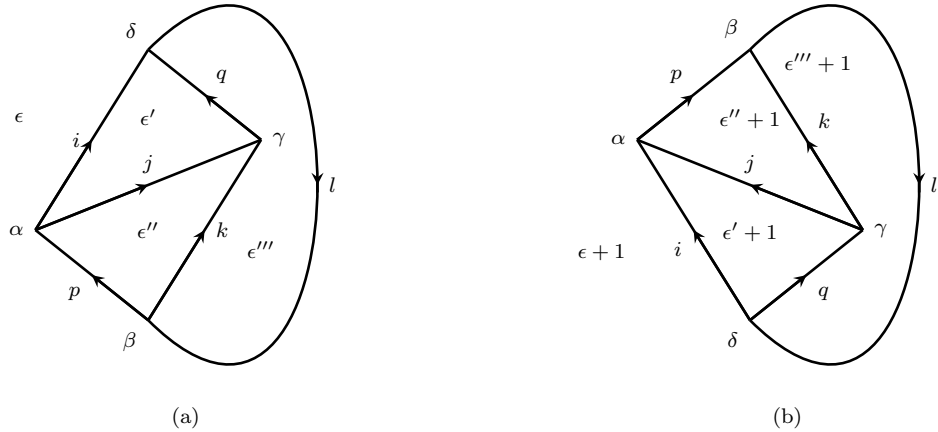

$U_{Rg}$ can end on the boundary giving an interface between $\fB$ and $\fB'$ and an interface between $\fB'$ and $\fB$. The lines living on these interfaces give rise to $\tilde\CC_{0,1}$ and $\tilde\CC_{1,0}$ respectively. Together they form a ``twisted" spherical fusion category $\tilde\CC$ described above.

The label 0 and 1 of plaquettes in a graph in $\tilde\CC$ corresponds respectively to the boundary $\fB$ and $\fB'$ in the physical setting. A graph $\Gamma$ in $\tilde\CC$ drawn on a sphere computes the partition function of $\tilde\fT$ on a 3-ball with a network of boundary lines specified by $\Gamma$. The bulk of the 3-ball contains orientation reversing defects which end on the boundary at the location of lines living in $\tilde\CC_{0,1}$ or $\tilde\CC_{1,0}$.

Given such a 3-ball with $\Gamma$ on the boundary, we can bubble a $U_R$ in the bulk of the 3-ball and bring it to the boundary. This sends $\Gamma$ to $\Gamma'$ (after taking the mirror) and both of these graphs must evaluate to the same number. This is the origin of the compatibility between associators and $I$. See Figure \ref{fig:reflection}.

\subsection{Unoriented Turaev-Viro} \label{2.4}
In this subsection, we generalize the Turaev-Viro prescription to compute the partition function of an unoriented theory $\tilde\fT$ on an unoriented 3-manifold $M$. We will assume that the reader has read subsection \ref{2.2} before reading this subsection and so we will sometimes cut corners in what follows.

Fix an orientation $\CO$ on $\R^3$. An unoriented 3-manifold $M$ can be constructed as follows. We pick open sets of $\R^3$ and glue them along codimension one loci using piecewise-linear maps. This gives us a locus $\CL$ in $M$ which is defined by the property that the transition functions are orientation reversing. The Poincare dual of this locus is a representative of first Stiefel-Whitney class and we call it $w_1$. We assign a local orientation $\CO_t$ to any small tetrahedron $t$ in $M-\CL$ by first using the local chart to map it to a tetrahedron in $\R^3$ where we have already picked an orientation $\CO$. $\CO_t$ remains invariant under deformations of $t$ inside $M-\CL$.

Pick a branched triangulation $T$ of $M$. $w_1$ assigns a number $p_e$ valued in $\{0,1\}$ to every edge $e$. And the $G$-connection $\alpha_1$ on $M$ assigns an element $g_e$ of the group $G$ to each directed edge $e$. 

Let's extract a set of labels $S_{0,0}$ such that each label in the set corresponds to a simple object of $\tilde\CC_{0,0}$. Similarly, extract a set of labels $S_{0,1}$ such that each label in the set corresponds to a simple object of $\tilde\CC_{0,1}$. Define an involution $*$ on $S_{0,0}$ induced by taking the dual of simple objects. Similarly, define an involution $*$ on $S_{0,1}$ under which $i$ is sent to $j$ if the $*$ operation of $\tilde\CC$ sends the object $L_i$ in $\tilde\CC_{0,1}$ to the object $L_j$ in $\tilde\CC_{1,0}$. There is also a $G$-grading on both of these sets descending from the $G$-grading of simple objects. 

We now label each directed edge $e$ by a label in $(S_{0,p_e})_{g_e}$. Pick a face $f$ of $T$. We can label $f$ by some label $\alpha$ just as in the oriented case. Thus we obtain a labeling of edges and faces of a branched triangulation. Call one such labeling as $\tilde l$.

\begin{figure}
\centering
\begin{tikzpicture}[line width=1pt]
\begin{scope}[every node/.style={sloped,allow upside down}]
\coordinate (023) at (0,0);
\coordinate (012) at ($(023)+(-1.5,1.2)$);
\coordinate (123) at ($(023)+(1.5,2.4)$);
\coordinate (013) at ($(023)+(0,3.6)$);
\coordinate (controlu) at ($(013)+(3,3)$);
\coordinate (controld) at ($(023)+(3,-3)$);
\coordinate (mid) at ($(controlu)!0.5!(controld)-(0.75,0)$);

\arrowpath{(023)}{(012)}{0.5};
\arrowpath{(023)}{(123)}{0.5};
\arrowpath{(012)}{(123)}{0.5};
\arrowpath{(012)}{(013)}{0.5};
\arrowpath{(123)}{(013)}{0.5};
\draw (013)..controls(controlu) and (controld)..(023);
\arrowpath{($(mid)+(0,0.1)$)}{($(mid)-(0,0.1)$)}{0.5};

\node[left] at (012) {\scriptsize{$\alpha$}};
\node[below left] at (023) {\scriptsize{$\beta$}};
\node[above left] at (013) {\scriptsize{$\delta$}};
\node[right] at (123) {\scriptsize{$\gamma$}};

\node[below left] at ($(023)!0.5!(012)$) {\scriptsize{$p$}};
\node[right] at ($(023)!0.5!(123)$) {\scriptsize{$k$}};
\node[above] at ($(012)!0.5!(123)$) {\scriptsize{$j$}};
\node[left] at ($(012)!0.5!(013)$) {\scriptsize{$i$}};
\node[above right] at ($(123)!0.5!(013)$) {\scriptsize{$q$}};
\node[right] at (mid) {\scriptsize{$l$}};

\node[left] at ($(012)+(0,1.5)$) {\scriptsize{$\epsilon$}};
\node at ($(012)+(0,1.5)+(1.5,0)$) {\scriptsize{$\epsilon'$}};
\node at ($(012)+(0,1.5)+(1.5,0)+(0,-1.5)$) {\scriptsize{$\epsilon''$}};
\node[below] at ($(012)+(0,1.5)+(1.5,0)+(0,-1.5)+(1.5,0)$) {\scriptsize{$\epsilon'''$}};


\node[below] at ($(023) + (0.5,-1)$) {\scriptsize{(a)}};

\coordinate (023') at ($(023)+(8,0)$);
\coordinate (012') at ($(023')+(-1.5,2.4)$);
\coordinate (123') at ($(023')+(1.5,1.2)$);
\coordinate (013') at ($(023')+(0,3.6)$);
\coordinate (controlu') at ($(013')+(3,3)$);
\coordinate (controld') at ($(023')+(3,-3)$);
\coordinate (mid') at ($(controlu')!0.5!(controld')-(0.75,0)$);

\arrowpath{(023')}{(012')}{0.5};
\arrowpath{(023')}{(123')}{0.5};
\arrowpath{(123')}{(012')}{0.5};
\arrowpath{(012')}{(013')}{0.5};
\arrowpath{(123')}{(013')}{0.5};
\draw (013')..controls(controlu') and (controld')..(023');
\arrowpath{($(mid')+(0,0.1)$)}{($(mid')-(0,0.1)$)}{0.5};

\node[left] at (012') {\scriptsize{$\alpha$}};
\node[below left] at (023') {\scriptsize{$\delta$}};
\node[above left] at (013') {\scriptsize{$\beta$}};
\node[right] at (123') {\scriptsize{$\gamma$}};

\node[below left] at ($(023')!0.5!(012')$) {\scriptsize{$i$}};
\node[below right] at ($(023')!0.5!(123')$) {\scriptsize{$q$}};
\node[above] at ($(012')!0.5!(123')$) {\scriptsize{$j$}};
\node[above left] at ($(012')!0.5!(013')$) {\scriptsize{$p$}};
\node[above right] at ($(123')!0.5!(013')$) {\scriptsize{$k$}};
\node[right] at (mid') {\scriptsize{$l$}};

\node[left] at ($(012')-(0,1.5)$) {\scriptsize{$\epsilon$}};
\node[above] at ($(012')-(0,1.5)+(1.5,0)$) {\scriptsize{$\epsilon'$}};
\node[above] at ($(012')-(0,1.5)+(1.5,0)-(0,-1.5)$) {\scriptsize{$\epsilon''$}};
\node[above left] at ($(012')-(0,1.5)+(1.5,0)-(0,-1.5)+(1.5,0.75)$) {\scriptsize{$\epsilon'''$}};


\node[below] at ($(023') + (0.5,-1)$) {\scriptsize{(b)}};
\end{scope}
\end{tikzpicture}
\caption{The possible graphs that we attach to a tetrahedron.} \label{fig:TV}
\end{figure}
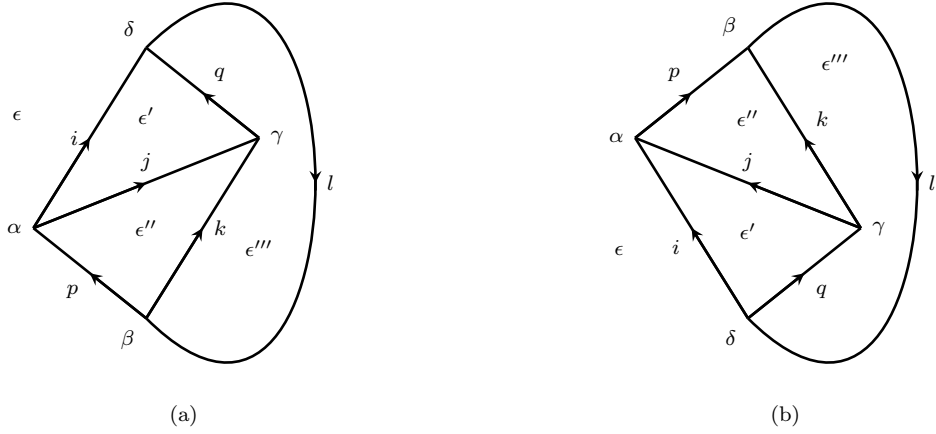

Now pick a tetrahedron $t$ in $M-\CL$ in the labeling $\tilde l$. To each such $t$ we will assign a planar graph $\Gamma_t$ in $\tilde\CC$. If the chirality of $t$ matches the local orientation $\CO_t$, we assign the graph shown in Figure \ref{fig:TV}(a) with $\epsilon=\epsilon'=\epsilon''=\epsilon'''=0$ and if the chirality doesn't match the local orientation we assign the graph shown in Figure \ref{fig:TV}(b) with $\epsilon=\epsilon'=\epsilon''=\epsilon'''=0$. To define $\Gamma_t$ for a $t$ intersecting $\CL$, we choose a small neighborhood $U_t$ of $t$ such that $\CL$ looks like a wall cutting $U_t$ into two parts. On one side of the wall, we assign $0$ to every vertex and on the other side we assign $1$. We assign a global orientation $\CO_{U_t}$ to $U_t$ given by local orientation $\CO_{t'}$ of any tetrahedron $t'$ lying completely on one side of the wall where vertices are labeled by $0$. We now assign the graph shown in Figure \ref{fig:TV}(a) with $\epsilon=0$ and arbitrary $\epsilon'$, $\epsilon''$, $\epsilon'''$ if chirality of $t$ matches $\CO_{U_t}$ and the graph shown in Figure \ref{fig:TV}(b) with $\epsilon=0$ and arbitrary $\epsilon'$, $\epsilon''$, $\epsilon'''$ if it does not.

Notice that if we flip the choice of $0$ and $1$ that we assigned to the sides of the wall and apply the above presciption, then $\Gamma_t$ is flipped to the ``reflected" graph $\Gamma'_t$ which is the graph obtained by acting $U_R$ on $\Gamma_t$. See Figure \ref{fig:reflection}. $\Gamma_t$ and $\Gamma'_t$ evaluate to the same number. 

Also notice that if we take a tetrahedron $t''$ in $U_t$ whose vertices are all labeled by $1$ and assign to it a new graph $\Gamma'_{t''}$ by matching its chirality with $\CO_{U_t}$ instead, then $\Gamma'_{t''}$ will be the ``reflected" version of the old graph $\Gamma_{t''}$ that we assigned in the starting of last paragraph by matching its chirality with the local orientation, and hence $\Gamma'_{t''}$ will evaluate to the same number as $\Gamma_{t''}$. Thus, we see that we could have also given the prescription to compute the partition function in various patches $U_i$ by using the local orientations and assigning $\{0,1\}$ to the two sides produced by an intersection of $U_i$ with the $w_1$ wall. The tetrahedra lying the intersection of $U_i$ and $U_j$ would give the same contribution in each patch. Thus, we would just have to make sure that we ``glue" the tetrahedra in various intersections properly.

Returning back to our original prescription, we just repeat what we already said for the oriented case. Let's call the evaluation of $\Gamma_t$ as $n_t(\tilde l)$ and define $N(l)=\prod_t n_t(\tilde l)$. To each edge $e$ of $T$, we can associate a number $d_e(\tilde l)$ which is the quantum dimension of the simple line assigned to $e$ in $\tilde l$. Define $d(\tilde l)=\prod_e d_e(\tilde l)$. The partition function $Z(M)$ is then given by
\be
Z(M)=D^{-2v}\sum_{\tilde l} N(\tilde l)d(\tilde l)
\ee
where $D=\sqrt{\sum_i d_i^2}$ is what we dub as the total quantum dimension of $\tilde\CC$ (where $d_i$ is the quantum dimension of simple line $L_i$) and $v$ is the number of vertices in $T$. We would like to emphasize that we are picking labels $i$ only in ``half" of $\tilde\CC$ {\ie} $\tilde\CC_{0,0}$ and $\tilde\CC_{0,1}$. Hence, the total quantum dimension only involves square of quantum dimensions of half of the simple lines.

The invariance of $Z(M)$ under Pachner moves and under change of representative of $w_1$ is guaranteed by the twisted pentagon equation (\ref{tpentagon}) satisfied by the $F$-symbols in $\tilde\CC$. In the rest of the paper, by ``twisted spherical fusion category" we will mean the data of $\tilde\CC_{0,0}\oplus\tilde\CC_{0,1}$ and we will often repackage this data as $\CC'=\oplus_{\tilde g} \CC'_{\tilde g}=\oplus_g\CC'_{g}\oplus_g\CC'_{Rg}$ where $\tilde g\in\tilde G=G\times\Z_2$ and $R$ is the generator of $\Z_2$ in $\tilde G$. $\CC'_{g}=(\tilde\CC_{0,0})_g$ and $\CC'_{Rg}=(\tilde\CC_{0,1})_g$. We also define a homomorphism $\rho$ (also called $\rho_1$) from $\tilde G$ to $\Z_2$ which sends $G\times\{e\}$ to $0$ and $G\times\{R\}$ to $1$.

\subsection{Example: Bosonic SPT phases}
Bosonic SPT phases protected by $G=G_0\times\Z_2^T$ are invertible unoriented TFTs with global symmetry $G_0$. Such a phase is constructed by a twisted spherical fusion category $\CC$ having a single simple object $L_g$ in each subcategory $\CC_g$. The fusion rules are $L_g\otimes L_{g'}\simeq L_{gg'}$. 

F-matrices define a $U(1)$ valued function of three group elements $\alpha_3(g,g',g'')=F_{g,g',g'';gg'g''}$. The twisted pentagon equation (\ref{tpentagon}) translates to
\be
\frac{\alpha_3^{s(g)}(g',g'',g''')\alpha_3(g,g'g'',g''')\alpha_3(g,g',g'')}{\alpha_3(gg',g'',g''')\alpha_3(g,g',g''g''')}=1
\ee
where $s(g)=(-1)^{\rho(g)}$. This means that $\alpha_3$ is a $\rho$-twisted group cocycle. On the other hand, gauge transformations (\ref{tgauge}) become
\be
\alpha_3(g,g',g'')\to\alpha_3(g,g',g'')\frac{\beta_2^{s(g)}(g',g'')\beta_2(g,g'g'')}{\beta_2(g,g')\beta_2(gg',g'')}
\ee
which corresponds to adding an exact $\rho$-twisted cocycle $\delta\beta_2$ to $\alpha_3$.

This means that the bosonic SPT phases are classified by the $\rho$-twisted group cohomology $H^3(BG,U(1)_\rho)$ \cite{Chen:2011pg}. A background connection $\alpha_1$ for $G_0$ on $M$ combines with $w_1$ to give a background connection for $G$ which is represented as a map from $M$ to $BG$. An element of $H^3(BG,U(1)_\rho)$ is then pulled back to a density on $M$ which can be integerated on $M$ to produce the partition function $Z(M,\alpha_1)$.

\section{\PP-TFTs} \label{3}
We start this section by reviewing the construction of Spin-TFTs from their shadows \cite{Bhardwaj:2016clt}. We will argue that the \PP-shadows must have an additional kind of anomaly which was not present in the case of Spin-shadows. Incorporating this addtional anomaly will allow us to generalize the shadow construction to \PP-TFTs. We finish the section by showing how to take a product of \PP-TFTs at the level of shadows.
\subsection{Review of Spin case}
\begin{figure}
\centering
\begin{tikzpicture}[line width=1pt]
\begin{scope}[every node/.style={sloped,allow upside down}]
\coordinate (lowb1) at (0,0);
\coordinate (lowb2) at ($(lowb1)+(1,0)$);
\coordinate (midb1) at ($(lowb1)+(0,1.5)$);
\coordinate (midb2) at ($(lowb2)+(0,1.5)$);
\coordinate (lowb3) at ($(lowb1)+(2,0)$);
\coordinate (highb3) at ($(lowb3)+(0,4)$);

\draw[double] (lowb1)--(midb1);
\draw[double] (lowb2)--(midb2);
\draw[double] (lowb3)--(highb3);
\draw[double] (midb1) arc[radius=0.5, start angle=180, end angle=0];

\node[below] at (lowb1) {\scriptsize{$\Pi$}};
\node[below] at (lowb2) {\scriptsize{$\Pi$}};
\node[below] at (lowb3) {\scriptsize{$\Pi$}};

\coordinate (eq) at ($(lowb3)+(1,2)$);
\node at (eq) {$=$};

\coordinate (lowbb1) at ($(lowb3)+(2,0)$);
\coordinate (highbb1) at ($(lowbb1)+(0,4)$);
\coordinate (lowbb2) at ($(lowbb1)+(1,0)$);
\coordinate (lowbb3) at ($(lowbb2)+(1,0)$);
\coordinate (midbb2) at ($(lowbb2)+(0,1.5)$);
\coordinate (midbb3) at ($(lowbb3)+(0,1.5)$);

\draw[double] (lowbb2)--(midbb2);
\draw[double] (lowbb3)--(midbb3);
\draw[double] (lowbb1)--(highbb1);
\draw[double] (midbb2) arc[radius=0.5, start angle=180, end angle=0];

\node[below] at (lowbb1) {\scriptsize{$\Pi$}};
\node[below] at (lowbb2) {\scriptsize{$\Pi$}};
\node[below] at (lowbb3) {\scriptsize{$\Pi$}};

\node[below] at ($(eq) + (0,-2.5)$) {\scriptsize{(a)}};

\coordinate (llowb1) at ($(lowbb3)+(1.75,0)$);
\coordinate (llowb2) at ($(llowb1)+(1.5,0)$);
\coordinate (hhighb2) at ($(llowb1)+(0,4)$);
\coordinate (hhighb1) at ($(llowb2)+(0,4)$);

\draw[double] (llowb1)--(hhighb1);
\paddedline{(llowb2)}{(hhighb2)}{(0.1,0)};
\draw[double] (llowb2)--(hhighb2);

\node[below] at (llowb1) {\scriptsize{$\Pi$}};
\node[below] at (llowb2) {\scriptsize{$\Pi$}};

\coordinate (eq2) at ($(llowb2)+(0.75,2)$);
\node at (eq2) {$=$};

\coordinate (pm) at ($(eq2)+(0.75,0)$);
\coordinate (llowbb1) at ($(llowb2)+(2,0)$);
\coordinate (hhighbb1) at ($(llowbb1)+(0,4)$);
\coordinate (llowbb2) at ($(llowbb1)+(1,0)$);
\coordinate (hhighbb2) at ($(llowbb2)+(0,4)$);

\draw[double] (llowbb2)--(hhighbb2);
\draw[double] (llowbb1)--(hhighbb1);

\node[below] at (llowbb1) {\scriptsize{$\Pi$}};
\node[below] at (llowbb2) {\scriptsize{$\Pi$}};
\node at (pm) {$-$};

\node[below] at ($(eq2) + (0,-2.5)$) {\scriptsize{(b)}};
\end{scope}
\end{tikzpicture}
\caption{(a) A property of bulk line $\Pi$ generating an anomalous $\mathbb{Z}_2$ 1-form symmetry. (b) Half-braiding $\Pi$ lines across each other gives a factor of $- 1$ when compared to $\Pi$ lines without braiding.}
\label{fig:fermion}
\end{figure}
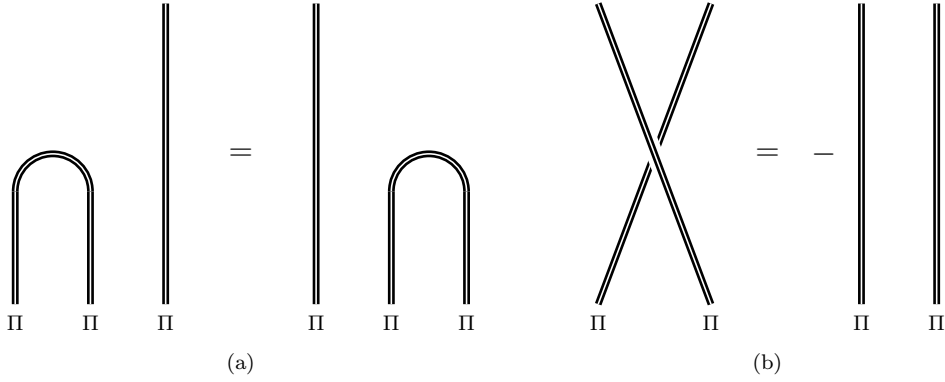

\cite{Bhardwaj:2016clt} provided a recipe to construct a 3d Spin-TFT $\fT_s$ from its shadow $\fT_f$. The shadow is an ordinary TFT with an anomalous $\Z_2$ 1-form symmetry. This manifests itself in the existence of a bulk line $\Pi$ which fuses with itself to the identity and has certain properties. See Figure \ref{fig:fermion}. 

We want to couple $\fT_f$ to a background 2-connection $\beta_2$ for the 1-form symmetry. We can do so by inserting $\Pi$ lines inside a triangulated manifold such that an even number of $\Pi$ lines cross a face having $\beta_2=0$ and an odd number of $\Pi$ lines cross a face having $\beta_2=1$. Since $\Pi$ has a non-trivial crossing with itself, topologically different ways of gluing $\Pi$ lines inside the tetrahedron will differ by signs. Hence, we need to pick a convention of how we will glue the $\Pi$ lines crossing these faces inside each tetrahedron when we say that $\fT_f$ is coupled to a background 2-connection $\beta_2$. Once we have picked this convention, the partition function will not be invariant under gauge transformations of $\beta_2$.

After fixing the convention, the change in the partition function under gauge transformations is independent of the theory, however. To see this, consider the product $\fT=\fT_1\times\fT_2$ of two shadow theories $\fT_1$ and $\fT_2$, and couple it to a background 2-connection for the diagonal $\Z_2$ 1-form symmetry. The partition function would then be the product
\be
Z(M,\beta_2)=Z_1(M,\beta_2)Z_2(M,\beta_2)
\ee
and a gauge transformation would leave $Z$ invariant. This is because resolving a crossing of the product line $\Pi_1\Pi_2$ gives no minus sign as the signs from crossing of $\Pi_1$ and crossing of $\Pi_2$ cancel each other.

The strategy of \cite{Bhardwaj:2016clt} was to compute this anomaly for a simple shadow theory, namely the shadow of Gu-Wen fermionic SPT phases. The anomaly under $\beta_2\to\beta_2+\delta\lambda_1$ turns out to be
\be
Z_f(M,\beta_2)\to (-1)^{\int_M\lambda_1\cup\beta_2+\beta_2\cup\lambda_1+\lambda_1\cup\delta\lambda_1}Z_f(M,\beta_2)
\ee
This transformation is the same as the transformation of a spin-structure $\eta_1$ dependent sign $z(M,\eta_1,\beta_2)$. This sign can be written as \cite{Gaiotto:2015zta}
\be
z(M,\eta_1,\beta_2)=(-1)^{\int_M\eta_1\cup\beta_2+\int_N\beta_2\cup\beta_2+w_2\cup\beta_2}
\ee
where $N$ is a 4-manifold whose boundary is $M$ and $w_2$ is a representative of second Stiefel-Whitney class. For oriented manifolds, this sign is independent of $N$ because $\beta_2\cup\beta_2+w_2\cup\beta_2$ is exact if $\beta_2$ is a cocycle. It is easy to see from this expression that
\be
z(M,\eta_1,\beta_2+\delta\lambda_1)=(-1)^{\int_M\lambda_1\cup\beta_2+\beta_2\cup\lambda_1+\lambda_1\cup\delta\lambda_1}z(M,\eta_1,\beta_2)
\ee
Here we have used a representation of spin structure as an equivalence class of 1-cochains $\eta_1$ satisfying $\delta\eta_1=w_2$ under the equivalence relation given by addition of exact 1-cochains to $\eta_1$ \cite{Gaiotto:2015zta}.

\begin{figure}
\centering
\begin{tikzpicture}[line width=1pt]
\begin{scope}[every node/.style={sloped,allow upside down}]
\coordinate (lowO) at (0,0);
\coordinate (highO) at ($(lowO)+(1.5,4)$);
\coordinate (lowX) at ($(lowO)+(1.5,0)$);
\coordinate (highX) at ($(lowO)+(0,4)$);
\coordinate (OX) at (intersection of lowO--highO and lowX--highX);
\node [below] at (lowO) {\scriptsize{$P$}};
\node [below] at (lowX) {\scriptsize{$X$}};
\node [above] at (highO) {\scriptsize{$P$}};
\node [above] at (highX) {\scriptsize{$X$}};
\arrowpathdouble{(lowO)}{(highO)}{0.3};
\paddedline{(lowX)}{(highX)}{(0.1,0)};
\arrowpath{(lowX)}{(highX)}{0.3};

\coordinate (eq) at ( $(OX) + (1.5,0)$);
\node at (eq) {$=$};

\coordinate (lowOs) at ($(lowO) + (3,0)$);
\coordinate (highXs) at ($(lowOs) + (0,4)$);
\coordinate (lowXs) at ($(lowOs) + (1.5,0)$);
\coordinate (highOs) at ($(lowXs) + (0,4)$);
\coordinate (OXs) at (intersection of lowOs--highOs and lowXs--highXs);

\node [below] at (lowOs) {\scriptsize{$P$}};
\node [below] at (lowXs) {\scriptsize{$X$}};
\node [above] at (highOs) {\scriptsize{$P$}};
\node [above] at (highXs) {\scriptsize{$X$}};
\arrowpathdouble {(lowOs)}{(highOs)}{0.3};
\draw (lowXs)--(highXs);
\arrowpath{(lowXs)}{(highXs)}{0.3};
\coordinate (halfbox) at (0.3, 0.2);
\draw[fill=white] ($(OXs) - (halfbox)$) rectangle ($(OXs) + (halfbox)$);
\node at (OXs) {\scriptsize{$\beta_X$}};

\node[below] at ($(lowO)!0.5!(lowXs) - (0,0.5)$) {\scriptsize{(a)}};

\coordinate (lowO) at (9,0);
\coordinate (highO) at ($(lowO)+(1.5,4)$);
\coordinate (lowX) at ($(lowO)+(0.5,0)$);
\coordinate (highX) at ($(lowO)+(0.5,4)$);
\node [below] at (lowO) {\scriptsize{$P$}};
\node [below] at (lowX) {\scriptsize{$X$}};
\node [above] at (highO) {\scriptsize{$P$}};
\node [above] at (highX) {\scriptsize{$Y$}};
\arrowpathdouble{(lowO)}{(highO)}{0.7};
\paddedline{(lowX)}{(highX)}{(0.05,0)};
\draw[fill=black] ($(lowX)!0.5!(highX)$) circle(2pt);
\arrowpath{(lowX)}{(highX)}{0.2};
\arrowpath{(lowX)}{(highX)}{0.75};

\coordinate (eq) at ( $(lowX)!0.5!(highX) + (1.75,0)$);
\node at (eq) {$=$};

\coordinate (lowOs) at ($(lowO) + (3,0)$);
\coordinate (lowXs) at ($(lowOs) + (1,0)$);
\coordinate (highXs) at ($(lowXs) + (0,4)$);
\coordinate (highOs) at ($(highXs) + (0.5,0)$);
\coordinate (OXs) at (intersection of lowOs--highOs and lowXs--highXs);

\node [below] at (lowOs) {\scriptsize{$P$}};
\node [below] at (lowXs) {\scriptsize{$X$}};
\node [above] at (highOs) {\scriptsize{$P$}};
\node [above] at (highXs) {\scriptsize{$Y$}};
\arrowpathdouble {(lowOs)}{(highOs)}{0.3};
\paddedline{(lowXs)}{(highXs)}{(0.05,0)};
\draw[fill=black] ($(lowXs)!0.5!(highXs)$) circle(2pt);
\arrowpath{(lowXs)}{(highXs)}{0.25};
\arrowpath{(lowXs)}{(highXs)}{0.85};

\node[below] at ($(lowXs)!0.5!(lowX) - (0,0.5)$) {\scriptsize{(b)}};
\end{scope}
\end{tikzpicture}
\caption{Properties of boundary image of bulk lines. These properties imply that the bulk lines are elements of the Drinfeld center of the spherical fusion category formed by boundary lines.}\label{fig:bulk}
\end{figure}
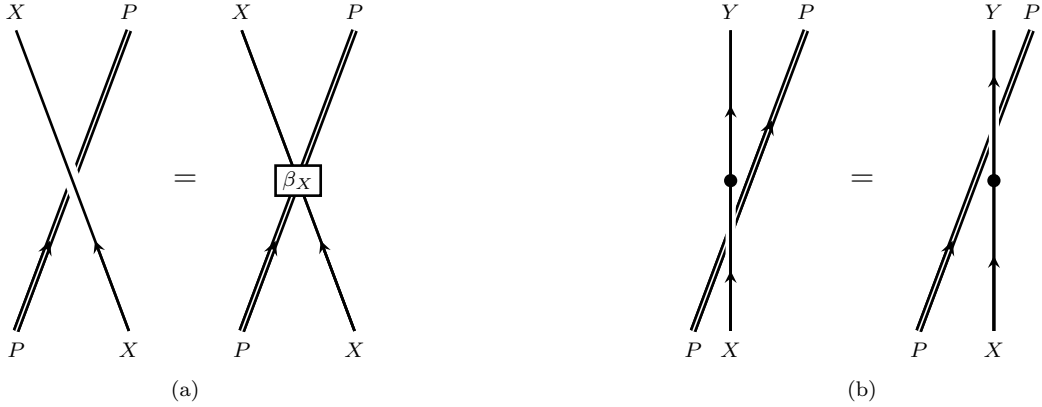

Thus combining the shadow theory with this sign gives a theory with a non-anomalous $\Z_2$ 1-form symmetry. The spin theory $\fT_s$ is obtained from this by gauging this 1-form symmetry
\be
Z_s(M,\eta_1)=\frac{|H^0(M,\Z_2)|}{|H^1(M,\Z_2)|}\sum_{[\beta_2]\in H^2(M,\Z_2)}z(M,\eta_1,\beta_2)Z_f(M,\beta_2)
\ee
So, we have a one-to-one correspondence $[\fT_s]\leftrightarrow(\fT_f,\Pi)$ where $[\fT_s]$ denotes the equivalence class of Spin-TFTs under permutations of spin structures $\eta_1\to\eta_1+\alpha_1$ where $[\alpha_1]\in H^1(M,\Z_2)$.

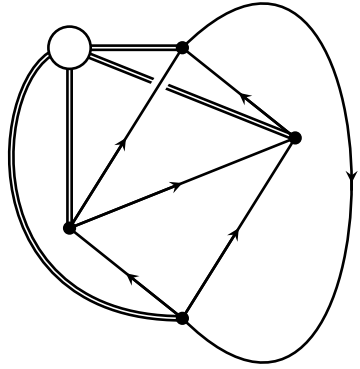
\begin{figure}
\begin{tikzpicture}[line width=1pt]
\begin{scope}[every node/.style={sloped,allow upside down}]
\node at (-8,0) {};
\coordinate (023) at (0,0);
\coordinate (012) at ($(023)+(-1.5,1.2)$);
\coordinate (123) at ($(023)+(1.5,2.4)$);
\coordinate (013) at ($(023)+(0,3.6)$);
\coordinate (controlu) at ($(013)+(3,3)$);
\coordinate (controld) at ($(023)+(3,-3)$);
\coordinate (mid) at ($(controlu)!0.5!(controld)-(0.75,0)$);
\coordinate (blob) at ($(012)+(0,2.4)$);
\coordinate (controllu) at ($(blob)-(1,0)$);
\coordinate (controlld) at ($(023)-(3,0)$);

\draw[double] (123) to (blob);
\paddedline{(012)}{(013)}{(0.1,0)};
\draw[double] (013) to (blob);
\draw[double] (012) to (blob);
\arrowpath{(023)}{(012)}{0.5};
\arrowpath{(023)}{(123)}{0.5};
\arrowpath{(012)}{(123)}{0.5};
\arrowpath{(012)}{(013)}{0.5};
\arrowpath{(123)}{(013)}{0.5};
\draw (013)..controls(controlu) and (controld)..(023);
\arrowpath{($(mid)+(0,0.1)$)}{($(mid)-(0,0.1)$)}{0.5};
\draw[double] (023)..controls(controlld) and (controllu)..(blob);

\draw[fill=black] (012) circle(2pt);
\draw[fill=black] (013) circle(2pt);
\draw[fill=black] (023) circle(2pt);
\draw[fill=black] (123) circle(2pt);
\draw[fill=white] (blob) circle(8pt);
\end{scope}
\end{tikzpicture}
\caption{The basic tetrahedron graph in the Turaev-Viro construction of the partition function $Z_f(M,\beta_2)$ of the shadow theory in the presence of a background 2-connection $\beta_2$. A $\Pi$ line (shown as double line in the figure) leaves the vertex if the dual face has $\beta_2=1$. We let such lines meet without crossing each other in the region denoted by a disk in the graph. Different ways of joining the lines in the disk are equivalent because of the property of $\Pi$ lines shown in Figure \ref{fig:fermion}(a).} \label{fig:shadow}
\end{figure}

We would like to have a Turaev-Viro construction for $Z_f(M,\beta_2)$. To this end, we should understand how to encode the $\Pi$ line in terms of the spherical fusion category $\CC$. Notice that $\Pi$ is mapped to a boundary line $P$ by bringing it to the boundary. If we bring $\Pi$ to the boundary such that it crosses a boundary line $X$, we obtain a canonical isomorphism $\beta_X:X\otimes P\to P\otimes X$. Bringing $\Pi$ to the boundary in topologically equivalent ways should lead to same answers. Hence, $(P,\beta)$ can be moved across other morphisms. See Figure \ref{fig:bulk}.

Mathematically, this means that $\Pi$ is an element $(P,\beta)$ of Drinfeld center of $\CC$. This element fuses with itself to identity and $\beta_P=-1$. The Turaev-Viro construction for $Z_f(M, \beta_2)$ is achieved by inserting a $\Pi$ line emanating from every vertex whose dual face has $\beta_2=1$. See Figure \ref{fig:shadow}.

\subsection{Fermion in \PP-theories}
\PP-TFTs are a generalization of Spin-TFTs to the unoriented case. Spinors can be defined on an $n$-dimensional non-orientable manifold by using transition functions valued in \PP(n) group or \PM(n) group, both of which are double covers of $O(n)$. They are distinguished by the value of $R^2$ acting on spinors where $R$ is a spatial reflection. $R^2=+1$ for \PP-group and $R^2=-1$ for \PM-group. In terms of time reversal symmetry $T$, the action on spinors is $T^2=-1$ for the {\PP} case and $T^2=+1$ for the {\PM} case. A \PP-structure exists only if the second Stiefel-Whitney class $[w_2]$ vanishes. On the other hand, a \PM-structure exists only if $[w_2+w_1^2]$ vanishes where $[w_1]$ is the first Stiefel-Whitney class. Two {\PP} or \PM-structures differ by an element of $H^1(M,\Z_2)$.


In the {\PP} case, there is a choice in defining the action of reflection in $i$-th spatial direction on spinors. We can either multiply the spinor by the gamma matrix $\gamma_i$ or by $-\gamma_i$. This suggests that in a \PP-shadow there are two canonical choices $m_R$ and $n_R=-m_R$ of local operators at the junction of a $\Pi$ line and $R$-defect. These operators square to 1. The orientation preserving defects always have a single canonical local operator at the junction.

\begin{figure}
\begin{tikzpicture}[line width=1pt]
\begin{scope}[every node/.style={sloped,allow upside down}]
\draw (0,0) circle(2cm);
\draw[double] (0,2) circle(0.25cm);
\draw[fill=black] ($(0,2)-(0.25,0)$) circle(2pt);
\draw[fill=black] ($(0,2)+(0.25,-0)$) circle(2pt);

\node[below] at ($(0,0) - (0,2.5)$) {\scriptsize{(a)}};

\draw[double] (6,0) circle(2.5cm);
\draw[double] (6,0) circle(2cm);
\draw[fill=black] (6,2.25) circle(0.5cm);
\node[below] at ($(6,0) - (0,2.5)$) {\scriptsize{(b)}};

\draw[fill=white] (11,0) circle(1.5cm);
\draw[double] ($(11,0)+({1.5*cos(135)},{1.5*sin(135)})$)--($(11,0)+({1.5*cos(-45)},{1.5*sin(-45)})$);
\paddedline{($(11,0)+({1.5*cos(225)},{1.5*sin(225)})$)}{($(11,0)+({1.5*cos(45)},{1.5*sin(45)})$)}{(0.1,0)};
\draw[double] ($(11,0)+({1.5*cos(225)},{1.5*sin(225)})$)--($(11,0)+({1.5*cos(45)},{1.5*sin(45)})$);
\draw (11,0) circle(1.5cm);
\node[below] at ($(11,0) - (0,2.5)$) {\scriptsize{(c)}};
\end{scope}
\end{tikzpicture}
\caption{(a) The big circle is a cartoon representing the locus $\CM$ dual to $w_1$. A $\Pi$ line is bubbled nearby and dipped into this locus. The operators at the two marked junctions are inverses of each other. (b) Taking the left half of $\Pi$ line around a loop $C$ in $\CM$ which intersects once the locus $\CL'$ dual to $w_1^2$, brings us to the configuration shown in the figure. We omit $\CM$ in this figure for brevity. The region denoted by a black disk is shown in (c). That is, the $\Pi$ lines are glued inside this black region in such a way that there is a non-trivial half-braiding ({\ie} crossing) of the $\Pi$ lines. The reason for the appearance of this crossing is that the normal direction to $\CM$ is reflected across $\CL'$ and hence the top and bottom parts of the (left half of) $\Pi$ loop are exchanged as $C$ crosses $\CL'$.} \label{fig:cross}
\end{figure}
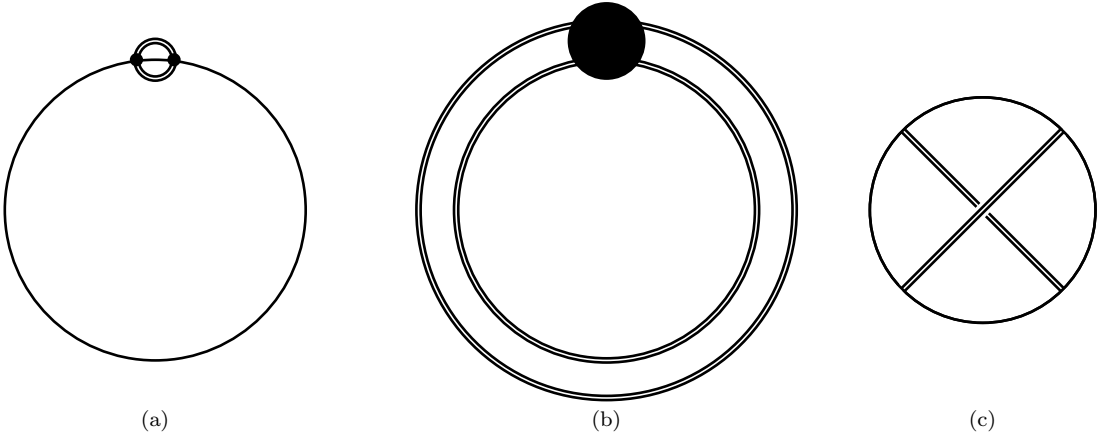

Now we will argue that, in general, there must be a locus $\CL$ embedded inside the locus $\CM$ of orientation reversing defects which implements the transformation $m\leftrightarrow n$. Moreover, the homology class of $\CL$ must be the Poincare dual of $[w_1^2]$. Choose a locus $\CL'$ embedded inside $\CM$ whose homology class is the dual of $[w_1^2]$. Now bubble a fermion line near $\CM$ and move it such that it intersects $\CM$ in two junctions. See Figure \ref{fig:cross}(a). The local opeartors at the two junctions must be inverses of each other. Take one of these junctions around a cycle $C$ in $\CM$. If the cycle intersects $\CL'$, then fusing the fermion line with itself at the end of this process gives a crossing of fermion line which provides a factor of $-1$. See Figure \ref{fig:cross}(b). Topological invariance demands that $C$ must intersect $\CL$ as well so that the fusion of the local operators at the end of the process provides a factor of $-1$ which cancels the sign from the crossing. Similarly, if $C$ doesn't intersect $\CL$, it doesn't intersect $\CL'$ either. Hence, $\CL$ and $\CL'$ are in the same homology class. Thus, we can choose to identify $\CL$ with the representative $w_1^2$.

We will see in the next section that this flip $m\leftrightarrow n$ as $\Pi$ crosses $\CL'$ is responsible for the presence of mixed anomaly between time reversal symmetry and $\Z_2$ 1-form symmetry in \PP-shadows.

\subsection{Shadows of \PP-TFTs}

Just as in the spin case, to define what we mean by a \PP-shadow $\fT_f$ coupled to a background $\beta_2$, we need to pick a convention for configuring $\Pi$ lines. In addition to this, we also need to choose whether we will put $m$ or $n$ on the junctions when $\Pi$ crosses $R$-defect. The \PP-TFT $\fT_+$ is obtained as
\be
Z_+(M,\eta_1)=\frac{|H^0(M,\Z_2)|}{|H^1(M,\Z_2)|}\sum_{[\beta_2]\in H^2(M,\Z_2)}z_+(M,\eta_1,\beta_2)Z_f(M,\beta_2) \label{pres}
\ee
where the sign which cancels the anomaly for $\Z_2$ 1-form symmetry of $\fT_f$ can be defined as
\be
z_+(M,\eta_1,\beta_2)=(-1)^{\int_M\eta_1\cup\beta_2+\int_N\beta_2\cup\beta_2+(w_1^2+w_2)\cup\beta_2} \label{sign}
\ee
where $\partial N=M$ and $\eta_1$ parametrizes \PP-structures. The expression is independent of $N$ as $\beta_2\cup\beta_2+(w_1^2+w_2)\cup\beta_2$ is exact if $\beta_2$ is a cocycle. Fliping the choice of local operator changes the partition function as $Z_f(M,\beta_2)\to (-1)^{\int_M w_1\cup\beta_2} Z_f(M,\beta_2)$. This can be absorbed into a permutation of \PP-structures $\eta_1\to\eta_1+w_1$. Thus, as in the spin case, we have a one-to-one correspondence $[\fT_+]\leftrightarrow(\fT_f,\Pi)$ where $[\fT_+]$ denotes the equivalence class of \PP-TFTs under permutations of \PP-structures.

Now, notice that \PP-shadows have a time reversal anomaly in the presence of a background 2-connection $\beta_2$. As we add $\delta v_0$ to $w_1$, we add $\delta u_1$ to $w_1^2$ where $u_1=w_1\cup v_0+v_0\cup w_1+ v_0\cup\delta v_0$. This corresponds to moving $\CM$ and $\CL'$. But during such movements, $\CL'$ will cross some $\Pi$ lines encoding $\beta_2$ and the partition function will change as
\be
Z_f(M,\beta_2)\to (-1)^{\int_M u_1\cup\beta_2}Z_f(M,\beta_2) \label{tanomaly}
\ee
Under this transformation, the sign also transforms in the same way
\be
z_+(M,\eta_1,\beta_2)\to (-1)^{\int_M u_1\cup\beta_2}z_+(M,\eta_1,\beta_2)
\ee
and the corresponding \PP-TFTs have no time-reversal anomaly.

The signs $z_+$ written above implies the following anomaly under $\beta_2\to\beta_2+\delta\lambda_1$ 
\be
Z_f(M,\beta_2)\to (-1)^{\int_M\lambda_1\cup\beta_2+\beta_2\cup\lambda_1+\lambda_1\cup\delta\lambda_1+w_1^2\cup\lambda_1}Z_f(M,\beta_2) \label{anomaly}
\ee
where $w_1$ is a representative of first Stiefel-Whitney class. As the anomaly is universal, we will verify that this is the correct anomaly by computing the anomaly directly for shadows of {\PP} generalization of Gu-Wen fermionic SPT phases in the next section.

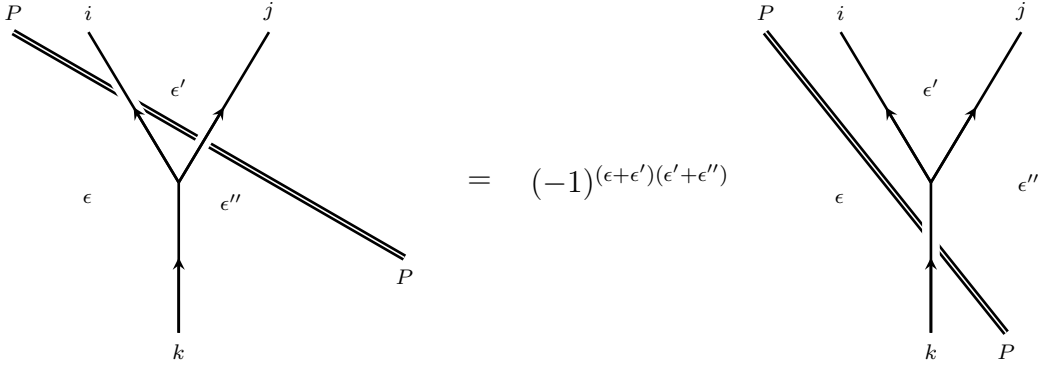
\begin{figure}
\begin{tikzpicture}[line width=1pt]
\begin{scope}[every node/.style={sloped,allow upside down}]
\coordinate (C) at (0,0);
\coordinate (mid) at ($(C)+(0,2)$);
\coordinate (A) at ($(mid)+(-1.2,2)$);
\coordinate (B) at ($(mid)+(1.2,2)$);
\coordinate (P) at ($(mid)+(3,-1)$);
\coordinate (P') at ($(A)-(1,0)$);

\draw[double] (P)--(P');
\paddedline{(mid)}{(A)}{(0.1,0)};
\paddedline{(mid)}{(B)}{(0.1,0)};
\arrowpath{(C)}{(mid)}{0.5};
\arrowpath{(mid)}{(A)}{0.5};
\arrowpath{(mid)}{(B)}{0.5};

\node[below] at (C) {\scriptsize{$k$}};
\node[below] at (P) {\scriptsize{$P$}};
\node[above] at (A) {\scriptsize{$i$}};
\node[above] at (P') {\scriptsize{$P$}};
\node[above] at (B) {\scriptsize{$j$}};
\node[below left] at ($(mid)+(-1,0)$) {\scriptsize{$\epsilon$}};
\node[below left] at ($(mid)+(1,0)$) {\scriptsize{$\epsilon''$}};
\node[above] at ($(mid)+(0,1)$) {\scriptsize{$\epsilon'$}};

\coordinate (eq) at ( $(P) + (1,1)$);
\node at (eq) {$=$};

\node at ($(eq)+(2,0)$) {$(-1)^{(\epsilon+\epsilon')(\epsilon'+\epsilon'')}$};


\coordinate (C1) at ($(C)+(10,0)$);
\coordinate (mid1) at ($(C1)+(0,2)$);
\coordinate (A1) at ($(mid1)+(-1.2,2)$);
\coordinate (B1) at ($(mid1)+(1.2,2)$);
\coordinate (P1) at ($(C1)+(1,0)$);
\coordinate (P1') at ($(A1)-(1,0)$);

\draw[double] (P1)--(P1');
\paddedline{(C1)}{(mid1)}{(0.1,0)};
\arrowpath{(C1)}{(mid1)}{0.5};
\arrowpath{(mid1)}{(A1)}{0.5};
\arrowpath{(mid1)}{(B1)}{0.5};

\node[below] at (C1) {\scriptsize{$k$}};
\node[below] at (P1) {\scriptsize{$P$}};
\node[above] at (A1) {\scriptsize{$i$}};
\node[above] at (P1') {\scriptsize{$P$}};
\node[above] at (B1) {\scriptsize{$j$}};
\node[below left] at ($(mid1)+(-1,0)$) {\scriptsize{$\epsilon$}};
\node[right] at ($(mid1)+(1,0)$) {\scriptsize{$\epsilon''$}};
\node[above] at ($(mid1)+(0,1)$) {\scriptsize{$\epsilon'$}};

\end{scope}
\end{tikzpicture}
\caption{The equations defining twisted Drinfeld center.} \label{fig:tdrin}
\end{figure}

To obtain the Turaev-Viro construction for $Z_f(M,\beta_2)$, we need to know how to encode the $\Pi$ line in terms of the data of $\CC$. As in the spin case, $\Pi$ is mapped to some boundary line $P$ with canonical isomorphisms $\beta_X:X\otimes P\to P\otimes X$. However, unlike the spin case, $\Pi$ is not an element of Drinfeld center of $\CC$. Rather, we need to insert extra signs whenever we move $\Pi$ across $\CL'$. This descends to the statement that $(P,\beta)$ is an element of a twisted Drinfeld center which is defined in Figure \ref{fig:tdrin}.

\subsection{Product of \PP-TFTs}
In this subsection, we want to figure out the shadow of the product of two \PP-TFTs. This will lead to the definition of a product on the shadow theories which we will call the \emph{shadow product}.

First, notice that \footnote{See Appendix B of \cite{Gaiotto:2015zta} for an introduction to higher cup products like $\cup_1$ used in the following equation.}
\be
z_+(M,\eta_1,\beta_2)z_+(M,\eta_1,\beta_2')=z_+(M,\eta_1,\beta_2+\beta_2')(-1)^{\int_M \beta_2\cup_1\beta_2'}
\ee
Now consider two \PP-TFTs $\fT_+$ and $\fT_+'$ with their corresponding shadows $\fT_f$ and $\fT_f'$. Using the above, we can write the partition function of the product theory as
\be
Z_+(M)Z_+'(M)=\frac{|H^0(M,\Z_2)|^2}{|H^1(M,\Z_2)|^2}\sum_{[\beta_2],[\beta_2']}z_+(M,\beta_2+\beta_2')Z_f(M,\beta_2)Z_f'(M,\beta_2')(-1)^{\int_M \beta_2\cup_1\beta_2'}
\ee
which can be massaged as
\be
Z_+(M,\eta_1)Z_+'(M,\eta_1)=\frac{|H^0(M,\Z_2)|}{|H^1(M,\Z_2)|}\sum_{[\beta_2]\in H^2(M,\Z_2)}z_+(M,\eta_1,\beta_2)\tilde Z_f(M,\beta_2)
\ee
with
\be
\tilde Z_f(M,\beta_2)=\frac{|H^0(M,\Z_2)|}{|H^1(M,\Z_2)|}\sum_{[\beta_2']\in H^2(M,\Z_2)}(-1)^{\int_M (\beta_2+\beta_2')\cup_1\beta_2'}Z_f(M,\beta_2+\beta_2')Z_f'(M,\beta_2')
\ee
being the partition function of the shadow corresponding to the product theory. We denote this shadow theory as the shadow product $\fT_f\times_f\fT_f'$.

Physically, we are constructing the shadow of the product by gauging the diagonal $\Z_2$ 1-form symmetry in the product of the shadow theories. Notice that this 1-form symmetry is non-anomalous and hence gauging it makes sense.

\begin{figure}
\centering
\begin{tikzpicture}[line width=1pt]
\begin{scope}[every node/.style={sloped,allow upside down}]
\coordinate (lowb1) at (0,0);
\coordinate (lowb2) at ($(lowb1)+(1,0)$);
\coordinate (midb1) at ($(lowb1)+(0,1.5)$);
\coordinate (midb2) at ($(lowb2)+(0,1.5)$);
\coordinate (lowb3) at ($(lowb1)+(2,0)$);
\coordinate (highb3) at ($(lowb3)+(0,4)$);

\draw[double] (lowb1)--(midb1);
\draw[double] (lowb2)--(midb2);
\draw[double] (lowb3)--(highb3);
\draw[double] (midb1) arc[radius=0.5, start angle=180, end angle=0];

\node[below] at (lowb1) {\scriptsize{$b$}};
\node[below] at (lowb2) {\scriptsize{$b$}};
\node[below] at (lowb3) {\scriptsize{$b$}};

\coordinate (eq) at ($(lowb3)+(1,2)$);
\node at (eq) {$=$};

\coordinate (lowbb1) at ($(lowb3)+(2,0)$);
\coordinate (highbb1) at ($(lowbb1)+(0,4)$);
\coordinate (lowbb2) at ($(lowbb1)+(1,0)$);
\coordinate (lowbb3) at ($(lowbb2)+(1,0)$);
\coordinate (midbb2) at ($(lowbb2)+(0,1.5)$);
\coordinate (midbb3) at ($(lowbb3)+(0,1.5)$);

\draw[double] (lowbb2)--(midbb2);
\draw[double] (lowbb3)--(midbb3);
\draw[double] (lowbb1)--(highbb1);
\draw[double] (midbb2) arc[radius=0.5, start angle=180, end angle=0];

\node[below] at (lowbb1) {\scriptsize{$b$}};
\node[below] at (lowbb2) {\scriptsize{$b$}};
\node[below] at (lowbb3) {\scriptsize{$b$}};

\node[below] at ($(eq) + (0,-2.5)$) {\scriptsize{(a)}};

\coordinate (llowb1) at ($(lowbb3)+(1.75,0)$);
\coordinate (llowb2) at ($(llowb1)+(1.5,0)$);
\coordinate (hhighb2) at ($(llowb1)+(0,4)$);
\coordinate (hhighb1) at ($(llowb2)+(0,4)$);

\draw[double] (llowb1)--(hhighb1);
\paddedline{(llowb2)}{(hhighb2)}{(0.1,0)};
\draw[double] (llowb2)--(hhighb2);

\node[below] at (llowb1) {\scriptsize{$b$}};
\node[below] at (llowb2) {\scriptsize{$b$}};

\coordinate (eq2) at ($(llowb2)+(0.75,2)$);
\node at (eq2) {$=$};

\coordinate (pm) at ($(eq2)+(0.75,0)$);
\coordinate (llowbb1) at ($(llowb2)+(2,0)$);
\coordinate (hhighbb1) at ($(llowbb1)+(0,4)$);
\coordinate (llowbb2) at ($(llowbb1)+(1,0)$);
\coordinate (hhighbb2) at ($(llowbb2)+(0,4)$);

\draw[double] (llowbb2)--(hhighbb2);
\draw[double] (llowbb1)--(hhighbb1);

\node[below] at (llowbb1) {\scriptsize{$b$}};
\node[below] at (llowbb2) {\scriptsize{$b$}};

\node[below] at ($(eq2) + (0,-2.5)$) {\scriptsize{(b)}};
\end{scope}
\end{tikzpicture}
\caption{The properties of a bulk line $b$ generating a non-anomalous $\Z_2$ 1-form symmetry are very similar to that of $\Pi$. The only difference is that crossing $b$ lines doesn't lead to a sign.}
\label{fig:1-form}
\end{figure}
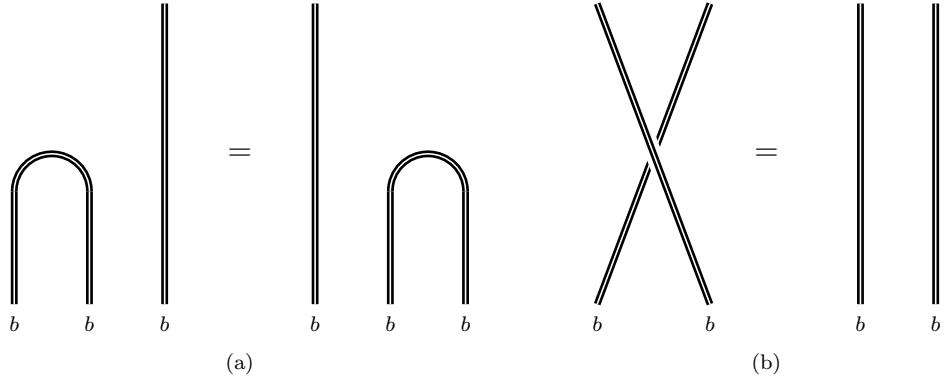

To implement the shadow product in the Turaev-Viro description, we first take a graded product of $\CC\times_G\CC'$ of $\CC$ and $\CC'$. This means that $(\CC\times_G\CC')_g=\CC_g\times\CC'_g$. Now we need a notion of gauging the line $b=\Pi\Pi'$ in the Drinfeld center of $\CC\times_G\CC'$. In general, we can consider the following problem. Take a theory $\fT_b$ specified by a twisted spherical fusion category $\CC_b$ having a non-anomalous $\Z_2$ 1-form symmetry. This means that there exists a line $b$ in the Drinfeld center of $\CC_b$ which fuses with itself to identity and has the properties shown in Figure \ref{fig:1-form}. We want to construct the twisted spherical fusion category for the theory $\fT_{\Z_2}$ obtained after gauging the 1-form symmetry generated by $b$.

\begin{figure}
\begin{tikzpicture}[line width=1pt]
\begin{scope}[every node/.style={sloped,allow upside down}]
\coordinate (lowA1) at (0,0);
\coordinate (highA1) at ($(lowA1) + (0,4)$);
\coordinate (midA1) at ($(lowA1)!0.5!(highA1)$);
\arrowpath{(lowA1)}{(midA1)}{0.5};
\arrowpath{(midA1)}{(highA1)}{0.5};
\draw[fill=black] (midA1) circle(2pt);
\node[below] at (lowA1) {\scriptsize{$A$}};
\node[above] at (highA1) {\scriptsize{$B$}};

\coordinate (lowA2) at ($(lowA1)+(3,0)$);
\coordinate (highA2) at ($(lowA2) + (0,4)$);
\coordinate (midA2) at ($(lowA2)!0.5!(highA2)$);
\coordinate (b) at ($(highA2)-(1,0)$);
\arrowpath{(lowA2)}{(midA2)}{0.5};
\arrowpath{(midA2)}{(highA2)}{0.5};
\draw[double] (midA2)--(b);
\draw[fill=black] (midA2) circle(2pt);
\node[above] at (b) {\scriptsize{$b$}};
\node[below] at (lowA2) {\scriptsize{$A$}};
\node[above] at (highA2) {\scriptsize{$B$}};

\node[below] at ($(lowA1)!0.5!(lowA2) - (0, 0.5)$) {\scriptsize{(a)}};

\coordinate (lowB) at (6,0);
\coordinate (highB) at ($(lowB) + (0,4)$);
\coordinate (midB) at ($(lowB)!0.5!(highB)$);
\arrowpath{(lowB)}{(highB)}{0.17};
\arrowpath{(lowB)}{(highB)}{0.5};
\arrowpath{(lowB)}{(highB)}{0.83};
\draw[double] ($(lowB)!0.67!(highB)$) arc[radius=0.67, start angle=90, end angle=270];
\draw[fill=black] ($(lowB)!0.33!(highB)$) circle(2pt);
\draw[fill=black] ($(lowB)!0.67!(highB)$) circle(2pt);
\node[below] at (lowB) {\scriptsize{$A$}};
\node[above] at (highB) {\scriptsize{$B$}};

\node[below] at ($(lowB) - (0, 0.5)$) {\scriptsize{(b)}};

\coordinate (lowC1) at (8.5,0);
\coordinate (highC1) at ($(lowC1) + (0,4)$);
\coordinate (midC1) at ($(lowC1)!0.5!(highC1)$);
\coordinate (lowC2) at ($(lowC1)+(1,0)$);
\coordinate (highC2) at ($(lowC2) + (0,4)$);
\coordinate (midC2) at ($(lowC2)!0.5!(highC2)$);

\draw[double] (midC1) .. controls ($(midC1) + (-1,1.5)$) .. (midC2);

\paddedline{($(lowC1)!0.6!(highC1)$)}{(highC1)}{(0.05,0)};
\arrowpath{(lowC1)}{(highC1)}{0.25};
\arrowpath{(lowC1)}{(highC1)}{0.8};
\draw[fill=black] (midC1) circle(2pt);
\node[below] at (lowC1) {\scriptsize{$A_1$}};
\node[above] at (highC1) {\scriptsize{$B_1$}};

\arrowpath{(lowC2)}{(highC2)}{0.25};
\arrowpath{(lowC2)}{(highC2)}{0.8};
\draw[fill=black] (midC2) circle(2pt);
\node[below] at (lowC2) {\scriptsize{$A_2$}};
\node[above] at (highC2) {\scriptsize{$B_2$}};
\node[below] at ($(lowC1)!0.5!(lowC2) - (0, 0.5)$) {\scriptsize{(c)}};

\coordinate (lowC1') at (12,0);
\coordinate (highC1') at ($(lowC1') + (0,4)$);
\coordinate (midC1') at ($(lowC1')!0.5!(highC1')$);
\coordinate (lowC2') at ($(lowC1')+(1,0)$);
\coordinate (highC2') at ($(lowC2') + (0,4)$);
\coordinate (midC2') at ($(lowC2')!0.5!(highC2')$);
\coordinate (left) at ($(highC1') - (1,0)$);

\draw[double] (midC2')--(left);

\paddedline{($(lowC1')!0.6!(highC1')$)}{(highC1')}{(0.05,0)};
\arrowpath{(lowC1')}{(highC1')}{0.25};
\arrowpath{(lowC1')}{(highC1')}{0.8};
\draw[fill=black] (midC1') circle(2pt);
\node[below] at (lowC1') {\scriptsize{$A_1$}};
\node[above] at (highC1') {\scriptsize{$B_1$}};

\arrowpath{(lowC2')}{(highC2')}{0.25};
\arrowpath{(lowC2')}{(highC2')}{0.8};
\draw[fill=black] (midC2') circle(2pt);
\node[below] at (lowC2') {\scriptsize{$A_2$}};
\node[above] at (highC2') {\scriptsize{$B_2$}};
\node[below] at ($(lowC1')!0.5!(lowC2') - (0, 0.5)$) {\scriptsize{(d)}};
\end{scope}
\end{tikzpicture}
\caption{Construction of $\CC_{\Z_2}$: (a) Two types of morphisms from $A$ to $B$. (b) Composition of two morphisms of the second type. (c) Tensor product of two morphisms of the second type. (d) Tensor product of a morphism of first type on the left and of second type on the right will similarly involve a crossing of $b$ line. And the tensor product of second type on left and first type on right doesn't involve any crossing.}
\label{fig:condensed}
\end{figure}
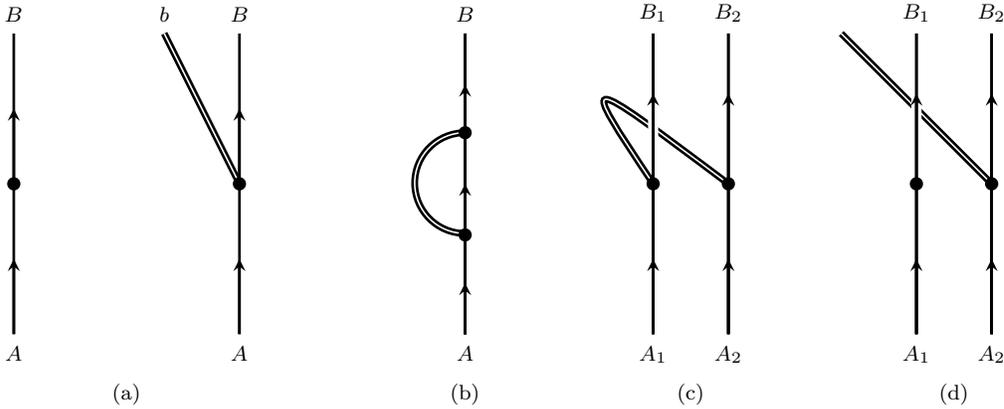

$b$ is invisble in the gauge theory. This means that a morphism from $A$ to $b\otimes B$ in $\CC_b$ has to be regarded as a morphism from $A$ to $B$ in $\CC_{\Z_2}$. And the morphisms from $A$ to $B$ in $\CC_b$ are also morphisms from $A$ to $B$ in $\CC_{\Z_2}$. The composition and tensor product of new morphisms are defined as shown in the Figure \ref{fig:condensed}.

Let's try to understand what happens to the simple objects under this operation. If $L$ is a simple object in $\CC_b$, $M=b\otimes L$ is simple as well. If $M$ is not isomorphic to $L$, then the morphism from $L$ to $b\otimes M$ in $\CC_b$ provides an isomorphism from $L$ to $M$ in $\CC_{\Z_2}$ combining them into a single simple object in $\CC_{\Z_2}$. If $M$ is isomorphic to $L$, then the morphism from $L$ to $b\otimes M$ in $\CC_b$ provides an additional endomorphism $\xi_L$ of $L$ in $\CC_{\Z_2}$. Since there are two independent morphisms from $L$ to itself in $\CC_{\Z_2}$, it must split into two simple objects $L^+$ and $L^-$ which can be constructed by inserting a projector
\be
\pi_L^\pm=\frac{1}{2}(1\pm\xi_L)
\ee
on $L$.

\section{Fermionic SPT phases} \label{4}
In this section, we discuss \PP-SPT phases. We also explicitly compute the partition function on an arbitrary manifold $M$ of a certain \PP-shadow which gives rise to the {\PP} Gu-Wen phases. We can read the anomaly of \PP-shadows from the expression for the partition function. The anomaly matches the expectation of the previous section. We finish the section by reproducing $\Z_2$ group of \PP-SPT phases without any global symmetry.
\subsection{Gu-Wen phases}
In this subsection, we will discuss {\PP} Gu-Wen SPT (f-SPT) phases with global symmetry $G$. Gu-Wen fermionic SPT phases were first described in \cite{Gu:2012ib} and explored further in \cite{Gaiotto:2015zta}.

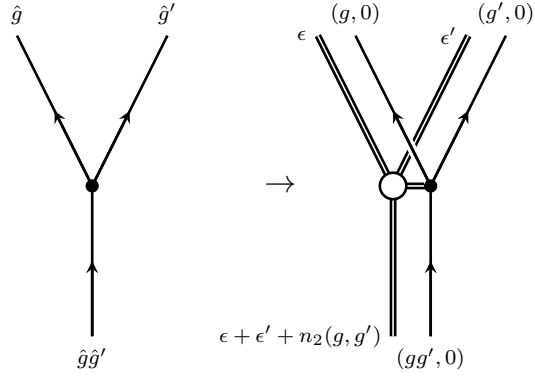
\begin{figure}
\centering
\begin{tikzpicture}[line width=1pt]

\begin{scope}[every node/.style={sloped,allow upside down}]
\coordinate (00) at (0,0);

\coordinate (C) at ($(00)+(1,0)$);
\coordinate (mid) at ($(C)+(0,2)$);
\coordinate (A) at ($(mid)+(-1,2)$);
\coordinate (B) at ($(mid)+(1,2)$);

\arrowpath{(C)}{(mid)}{0.5};
\arrowpath{(mid)}{(A)}{0.5};
\arrowpath{(mid)}{(B)}{0.5};
\draw[fill=black] (mid) circle(2pt);

\node[below] at (C) {\scriptsize{${\hat g \hat g'}$}};
\node[above] at (A) {\scriptsize{${\hat g }$}};
\node[above] at (B) {\scriptsize{${\hat g'}$}};

\coordinate (eq2) at ($(mid)+(2.5,0)$);
\node at (eq2) {$\to$};

\coordinate (C) at ($(00)+(5,0)$);
\coordinate (mid) at ($(C)+(0,2)$);
\coordinate (A) at ($(mid)+(-1,2)$);
\coordinate (B) at ($(mid)+(1,2)$);

\draw[double] (C)--(mid);
\draw[double] (mid)--(A);
\draw[double] (mid)--(B);

\coordinate (mid2) at (mid);

\node[left] at (C) {\scriptsize{$\epsilon + \epsilon' + n_2(g,g')$}};
\node[left] at (A) {\scriptsize{$\epsilon$}};
\node[left] at (B) {\scriptsize{$\epsilon'$}};

\coordinate (C) at ($(00)+(5.5,0)$);
\coordinate (mid) at ($(C)+(0,2)$);
\coordinate (A) at ($(mid)+(-1,2)$);
\coordinate (B) at ($(mid)+(1,2)$);

\draw[double] (mid)--(mid2);
\draw[fill=white] (mid2) circle(5pt);

\arrowpath{(C)}{(mid)}{0.5};
\paddedline{(mid)}{(A)}{(0.05,0)};
\arrowpath{(mid)}{(A)}{0.5};
\arrowpath{(mid)}{(B)}{0.5};
\draw[fill=black] (mid) circle(2pt);

\node[below] at (C) {\scriptsize{${(g g',0)}$}};
\node[above] at (A) {\scriptsize{${(g,0)}$}};
\node[above] at (B) {\scriptsize{${(g',0)}$}};

\end{scope}
\end{tikzpicture}
\caption{We choose our basis for morphism space $L_{gg,\epsilon+\epsilon'+n_2(g,g')}\to L_{g,\epsilon}\otimes L_{g',\epsilon'}$ such that the basis for different $(\epsilon,\epsilon')$ are related by crossing of a $\Pi$ line as shown in the figure. Here a label $\epsilon$ adjacent to double line denotes that the double line is $\Pi$ if $\epsilon=1$ and the double line is the identity line if $\epsilon=0$.}
\label{fig:gaugefix}
\end{figure}

\begin{figure}
\centering
\begin{tikzpicture}[line width=1pt]
\begin{scope}[every node/.style={sloped,allow upside down}]
\coordinate (00) at (0,0);
\coordinate (023) at (00);
\coordinate (012) at ($(023)+(-1.5,1.2)$);
\coordinate (123) at ($(023)+(1.5,2.4)$);
\coordinate (013) at ($(023)+(0,3.6)$);
\coordinate (controlu) at ($(013)+(3,3)$);
\coordinate (controld) at ($(023)+(3,-3)$);
\coordinate (mid) at ($(controlu)!0.5!(controld)-(0.75,0)$);

\coordinate (023b) at ($(023)+ (-.7,0)$);
\coordinate (012b) at ($(012)+(-.7,0)$);
\coordinate (123b) at ($(123)+(-.7,0)$);
\coordinate (013b) at ($(013)+(-.7,0)$);
\coordinate (controlub) at ($(controlu)+(.7,.5)$);
\coordinate (controldb) at ($(controld)+(.7,-.5)$);

\draw[double] (023b) to (012b);
\draw[double] (023b) to (123b);
\draw[double] (012b) to (123b);
\draw[double] (012b) to (013b);
\draw[double] (123b) to (013b);
\draw[double] (013b)..controls(controlub) and (controldb)..(023b);

\draw[double] (023b) to (023);
\draw[double] (123b) to (123);
\draw[double] (012b) to (012);
\draw[double] (013b) to (013);

\draw[fill=white] (012b) circle(4pt);
\draw[fill=white] (013b) circle(4pt);
\draw[fill=white] (023b) circle(4pt);
\draw[fill=white] (123b) circle(4pt);

\paddedline{(023)}{(012)}{(.1,0)};
\paddedline{(023)}{(123)}{(.1,0)};
\paddedline{(012)}{(123)}{(.1,0)};
\paddedline{(012)}{(013)}{(.1,0)};
\paddedline{(123)}{(013)}{(.1,0)};

\arrowpath{(023)}{(012)}{0.5};
\arrowpath{(023)}{(123)}{0.5};
\arrowpath{(012)}{(123)}{0.5};
\arrowpath{(012)}{(013)}{0.5};
\arrowpath{(123)}{(013)}{0.5};
\draw (013)..controls(controlu) and (controld)..(023);
\arrowpath{($(mid)+(0,0.1)$)}{($(mid)-(0,0.1)$)}{0.5};

\draw[fill=black] (012) circle(2pt);
\draw[fill=black] (013) circle(2pt);
\draw[fill=black] (023) circle(2pt);
\draw[fill=black] (123) circle(2pt);

\node (a) at ($(00)+(0.25,-2)$) {\scriptsize{(a)}};

\node (eq) at ($(00)+(4,2)$) {$= (-1)^{\rho_1^2\cup\epsilon_1}$};

\coordinate (00) at (11.5,0);
\coordinate (023) at (00);
\coordinate (012) at ($(023)+(-1.5,1.2)$);
\coordinate (123) at ($(023)+(1.5,2.4)$);
\coordinate (013) at ($(023)+(0,3.6)$);
\coordinate (controlu) at ($(013)+(3,3)$);
\coordinate (controld) at ($(023)+(3,-3)$);
\coordinate (mid) at ($(controlu)!0.5!(controld)-(0.75,0)$);

\coordinate (023b) at ($(023)+ (-3,0)$);
\coordinate (012b) at ($(012)+(-3,0)$);
\coordinate (123b) at ($(123)+(-3,0)$);
\coordinate (013b) at ($(013)+(-3,0)$);
\coordinate (controlub) at ($(controlu)+(-10,0)$);
\coordinate (controldb) at ($(controld)+(-10,0)$);

\draw[double] (023b) to (023);
\draw[double] (123b) to (123);
\draw[double] (012b) to (012);
\draw[double] (013b) to (013);

\draw[double] (023b) to (012b);
\paddedline{(023b)}{(123b)}{(.1,0)};
\draw[double] (023b) to (123b);
\draw[double] (012b) to (123b);
\draw[double] (012b) to (013b);
\draw[double] (123b) to (013b);
\draw[double] (013b)..controls(controlub) and (controldb)..(023b);

\draw[fill=white] (012b) circle(4pt);
\draw[fill=white] (013b) circle(4pt);
\draw[fill=white] (023b) circle(4pt);
\draw[fill=white] (123b) circle(4pt);

\paddedline{(023)}{(012)}{(.1,0)};
\paddedline{(023)}{(123)}{(.1,0)};
\paddedline{(012)}{(123)}{(.1,0)};
\paddedline{(012)}{(013)}{(.1,0)};
\paddedline{(123)}{(013)}{(.1,0)};

\arrowpath{(023)}{(012)}{0.5};
\arrowpath{(023)}{(123)}{0.5};
\arrowpath{(012)}{(123)}{0.5};
\arrowpath{(012)}{(013)}{0.5};
\arrowpath{(123)}{(013)}{0.5};
\draw (013)..controls(controlu) and (controld)..(023);
\arrowpath{($(mid)+(0,0.1)$)}{($(mid)-(0,0.1)$)}{0.5};

\draw[fill=black] (012) circle(2pt);
\draw[fill=black] (013) circle(2pt);
\draw[fill=black] (023) circle(2pt);
\draw[fill=black] (123) circle(2pt);

\node at ($(a)+(7,0)$) {\scriptsize{(b)}};
\end{scope}
\end{tikzpicture}
\caption{Intermediate computational steps relating $\hat \alpha_3$ and $\nu_3$. The sign arises from dragging a $\Pi$ line over a vertex. We can further resolve the crossing on the right hand side to make contact with $\nu_3$ which is defined in Figure \ref{fig:nu}.}
\label{fig:bis}
\end{figure}
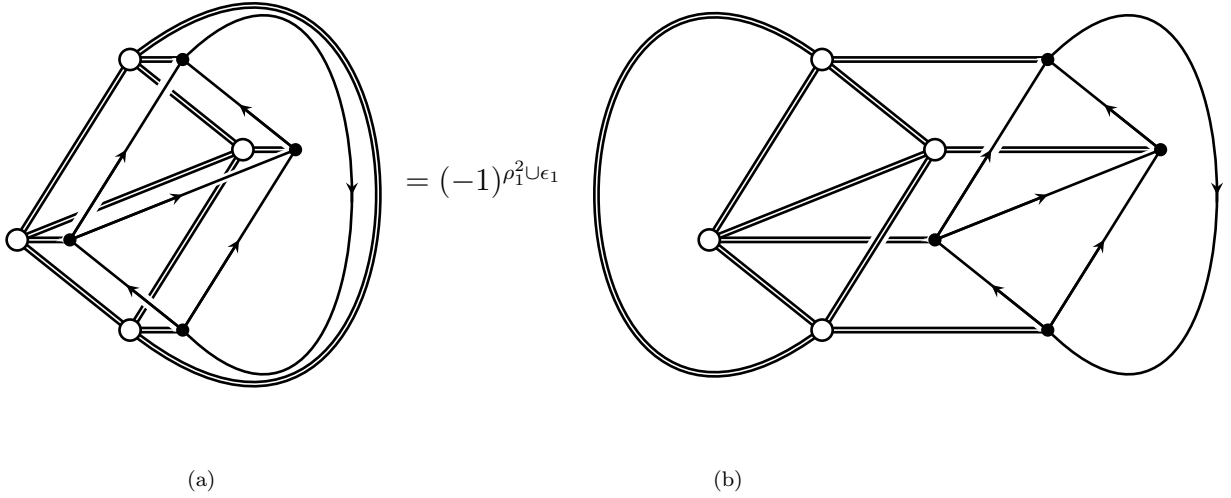

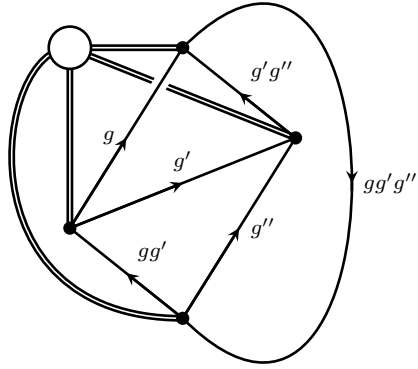
\begin{figure}
\centering
\begin{tikzpicture}[line width=1pt]
\begin{scope}[every node/.style={sloped,allow upside down}]
\coordinate (023) at (0,0);
\coordinate (012) at ($(023)+(-1.5,1.2)$);
\coordinate (123) at ($(023)+(1.5,2.4)$);
\coordinate (013) at ($(023)+(0,3.6)$);
\coordinate (controlu) at ($(013)+(3,3)$);
\coordinate (controld) at ($(023)+(3,-3)$);
\coordinate (mid) at ($(controlu)!0.5!(controld)-(0.75,0)$);
\coordinate (blob) at ($(012)+(0,2.4)$);
\coordinate (controllu) at ($(blob)-(1,0)$);
\coordinate (controlld) at ($(023)-(3,0)$);

\draw[double] (123) to (blob);
\paddedline{(012)}{(013)}{(0.1,0)};
\draw[double] (013) to (blob);
\draw[double] (012) to (blob);
\arrowpath{(023)}{(012)}{0.5};
\arrowpath{(023)}{(123)}{0.5};
\arrowpath{(012)}{(123)}{0.5};
\arrowpath{(012)}{(013)}{0.5};
\arrowpath{(123)}{(013)}{0.5};
\draw (013)..controls(controlu) and (controld)..(023);
\arrowpath{($(mid)+(0,0.1)$)}{($(mid)-(0,0.1)$)}{0.5};
\draw[double] (023)..controls(controlld) and (controllu)..(blob);

\node[above right] at ($(023)!0.5!(012)$) {\scriptsize{${gg'}$}};
\node[right] at ($(023)!0.5!(123)$) {\scriptsize{${g''}$}};
\node[above] at ($(012)!0.5!(123)$) {\scriptsize{${g'}$}};
\node[left] at ($(012)!0.5!(013)$) {\scriptsize{${g}$}};
\node[above right] at ($(123)!0.5!(013)$) {\scriptsize{${g'g''}$}};
\node[right] at (mid) {\scriptsize{${gg'g''}$}};

\draw[fill=black] (012) circle(2pt);
\draw[fill=black] (013) circle(2pt);
\draw[fill=black] (023) circle(2pt);
\draw[fill=black] (123) circle(2pt);
\draw[fill=white] (blob) circle(8pt);

\end{scope}
\end{tikzpicture}
\caption{The definition of $\nu_3(g,g',g'')$.}
\label{fig:nu}
\end{figure}

The twisted spherical fusion category for these phases is such that $\CC_g$ has two simple objects $L_{g,0}$ and $L_{g,1}$ for any $g$ in $G\times\Z_2^R$. The fusion rule is
\be
L_{g,\epsilon}\otimes L_{g',\epsilon'}\simeq L_{gg',\epsilon+\epsilon'+n_2(g,g')}
\ee
where $n_2$ is a $\Z_2$-valued group cocycle, {\ie} it is an element of $H^2(B(G\times\Z_2^R),\Z_2)$. $H^2(B(G\times\Z_2^R),\Z_2)$ is also the group of central extensions of the form
\be
0\to\Z_2\to\hat G\to G\times\Z_2^R\to 0
\ee
Thus, we can view $\CC$ as descending from $\hat\CC$ which is a $\hat G$-graded category with a single simple object in each grade. One obtains $\CC$ by forgetting the sub-grading corresponding to the $\Z_2$ subgroup appearing in the above central extension. More physically, $\hat\CC$ can be viewed as generalizing the notion of unoriented bosonic SPT phases to bosonic SPT phases with more complicated structure group. Forgetting the $\Z_2$ grading corresponds to gauging the $\Z_2$ symmetry. The associator of elements in $\CC$ can be read from the associator in $\hat\CC$ which we denote as $\hat\alpha_3$. It is an element of $H^3(B\hat G,U(1)_\rho)$. As a note, we will denote an arbitrary element of $G\times\Z_2^R$ by $g$ in what follows.

We demand the existence a fermionic line $\Pi$ in the twisted Drinfeld center of $\CC$ which fuses with itself to the identity. For the 1-form symmetry generated by this line to be compatible with $G$, the line must be of the form $(L_{e,0},\beta)$ or $(L_{e,1},\beta)$. The former case cannot lead to a fermionic line. Hence, $\Pi$ must be of the form $(L_{e,1},\beta)$. The existence of such a line will put some constraints on the form of $\CC$ which we now explore. First, we choose our basis of morphisms as shown in the Figure \ref{fig:gaugefix}. Consider the basic graph dual to the tetrahedron. Using our basis, it can be written as in Figure \ref{fig:bis}(a). This, in turn, can be manipulated to the final graph shown in Figure \ref{fig:nu} which we define to be $\nu_3(g,g',g'')$. During this manipulation we obtain a sign from resolving a crossing and another sign from moving a $\Pi$ line across a vertex. See Figure \ref{fig:bis}(b). Thus, we see that
\be
\hat\alpha_3=\nu_3(-1)^{(n_2+\rho_1^2)\cup\epsilon_1} \label{nu}
\ee
where $\epsilon_1$ is a $\Z_2$-valued co-chain which sends $(g,\epsilon)$ to $\epsilon$.


We find that a {\PP} Gu-Wen phase is specified by a double $(\nu_3,n_2)$ where $\nu_3$ satisfies
\be
\delta\nu_3=(-1)^{n_2\cup n_2+\rho_1^2\cup n_2}
\ee
However, there is a redundancy in such a description. We will see in subsection \ref{4.3} that the phase defined by $\nu_3=1$ and $n_2=\rho_1^2$ is the same as the trivial phase specified by $\nu_3=1$ and $n_2=0$.

To completely specify the {\PP} Gu-Wen phase, we also need to pick a specific $\Pi$ line. The twisted Drinfeld center equations (see Figure \ref{fig:tdrin} with $i=(g,\epsilon)$ and $j=(g',\epsilon')$) for such a line tell us that
\bea
\hat\alpha_3(g,\epsilon;g',\epsilon';e,1)\beta(g',\epsilon')\hat\alpha_3^{-1}(g,\epsilon;e,1;g',\epsilon')\beta(g,\epsilon)\hat\alpha_3(e,1;g,\epsilon;g',\epsilon')\\ \nonumber
=(-1)^{\rho_1(g)\rho_1(g')}\beta(gg',\epsilon+\epsilon'+n_2(g,g'))
\eea
Using (\ref{nu}), we see that it reduces to
\be
\beta(gg',\epsilon+\epsilon'+n_2(g,g'))=(-1)^{n_2(g,g')}\beta(g,\epsilon)\beta(g',\epsilon')
\ee
Using the fact that $\Pi$ is fermion tells us that
\be
\beta(g,\epsilon+1)=-\beta(g,\epsilon)
\ee
Feeding it back, we obtain that
\be
\beta(gg',0)=\beta(g,0)\beta(g',0)
\ee
The only solution to this equation that works uniformly for any group $G$ is
\bea
\beta(g,\epsilon)=(\pm 1)^{\rho_1(g)}(-1)^{\epsilon}
\eea
Since flipping the sign of all the $\beta$ in the orientation reversing sector doesn't change the resulting \PP-TFT, we can choose $L_{e,1}$ equipped with
\bea
\beta(g,\epsilon)=(-1)^\epsilon
\eea
as the fermion.

For the rest of this subsection, we note that we can write $H^2(B(G\times\Z_2^R),\Z_2)$ in terms of group cohomology of $G$. Let's denote an arbitrary element of $G\times\Z_2^R$ as $g_1$, $g_2$ etc. We also denote an arbitrary element of $G$ as $g$ and $R$ as the generator of $\Z_2^R$. We have the gauge transofrmations
\be
n_2(g_1,g_2)\to n_2(g_1,g_2)+n_1(g_1)+n_1(g_1g_2)+n_1(g_2)
\ee
Pick $n_1$ such that $n_1(g)=0$ for all $g$ and $n_1(R)+n_1(gR)=n_2(g,R)$. Thus we have fixed a gauge such that $n_2(g,R)=0$ for all $g$. Then using the cocycle condition
\be
n_2(g_2,g_3)+n_2(g_1g_2,g_3)+n_2(g_1,g_2g_3)+n_2(g_1,g_2)=0
\ee
we find that we can express $n_2$ as
\be
n_2=\tilde m_2+\rho_1\cup\tilde m_1+\rho_1\cup\rho_1 \label{break}
\ee
where $m_2$ parametrizes an element of $H^2(BG,\Z_2)$, $m_1$ parametrizes an element of $H_1(BG,\Z_2)$ and $\tilde m_{1,2}$ denotes the pullback of $m_{1,2}$ from $G$ to $G\times\Z_2^R$. This analysis establishes that
\be
H^2(B(G\times\Z_2^R),\Z_2)=H^2(BG,\Z_2)\times H^1(BG,\Z_2)\times\Z_2 \label{gcoh}
\ee




\subsection{Anomaly for \PP-shadows}
In this subsection we will compute the partition function $Z_f(M,\beta_2)$ for a {\PP} Gu-Wen phase. The explicit expression will allow us to compute the anomaly under a gauge transformation $\beta_2\to\beta_2+\delta\lambda_1$. As the anomaly is universal, this will justify our prescription (\ref{pres}) for constructing \PP-TFTs in terms of their shadows.

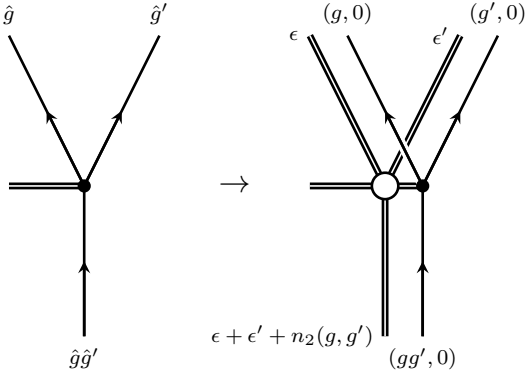
\begin{figure}
\centering
\begin{tikzpicture}[line width=1pt]

\begin{scope}[every node/.style={sloped,allow upside down}]
\coordinate (00) at (0,0);

\coordinate (C) at ($(00)+(1,0)$);
\coordinate (mid) at ($(C)+(0,2)$);
\coordinate (A) at ($(mid)+(-1,2)$);
\coordinate (B) at ($(mid)+(1,2)$);
\coordinate (D) at ($(mid)+(-1,0)$);

\arrowpath{(C)}{(mid)}{0.5};
\arrowpath{(mid)}{(A)}{0.5};
\arrowpath{(mid)}{(B)}{0.5};
\draw[double] (mid)--(D);
\draw[fill=black] (mid) circle(2pt);

\node[below] at (C) {\scriptsize{${\hat g \hat g'}$}};
\node[above] at (A) {\scriptsize{${\hat g }$}};
\node[above] at (B) {\scriptsize{${\hat g'}$}};

\coordinate (eq2) at ($(mid)+(2,0)$);
\node at (eq2) {$\to$};

\coordinate (C) at ($(00)+(5,0)$);
\coordinate (mid) at ($(C)+(0,2)$);
\coordinate (A) at ($(mid)+(-1,2)$);
\coordinate (B) at ($(mid)+(1,2)$);
\coordinate (D) at ($(mid)+(-1,0)$);

\draw[double] (C)--(mid);
\draw[double] (mid)--(A);
\draw[double] (mid)--(B);
\draw[double] (mid)--(D);

\coordinate (mid2) at (mid);

\node[left] at (C) {\scriptsize{$\epsilon + \epsilon' + n_2(g,g')$}};
\node[left] at (A) {\scriptsize{$\epsilon$}};
\node[left] at (B) {\scriptsize{$\epsilon'$}};

\coordinate (C) at ($(00)+(5.5,0)$);
\coordinate (mid) at ($(C)+(0,2)$);
\coordinate (A) at ($(mid)+(-1,2)$);
\coordinate (B) at ($(mid)+(1,2)$);

\draw[double] (mid)--(mid2);
\draw[fill=white] (mid2) circle(5pt);

\arrowpath{(C)}{(mid)}{0.5};
\paddedline{(mid)}{(A)}{(0.05,0)};
\arrowpath{(mid)}{(A)}{0.5};
\arrowpath{(mid)}{(B)}{0.5};
\draw[fill=black] (mid) circle(2pt);

\node[below] at (C) {\scriptsize{${(g g',0)}$}};
\node[above] at (A) {\scriptsize{${(g,0)}$}};
\node[above] at (B) {\scriptsize{${(g',0)}$}};

\end{scope}
\end{tikzpicture}
\caption{Gauge fixing: We choose basis for various morphisms related in the way shown in the figure.}
\label{fig:gfix}
\end{figure}

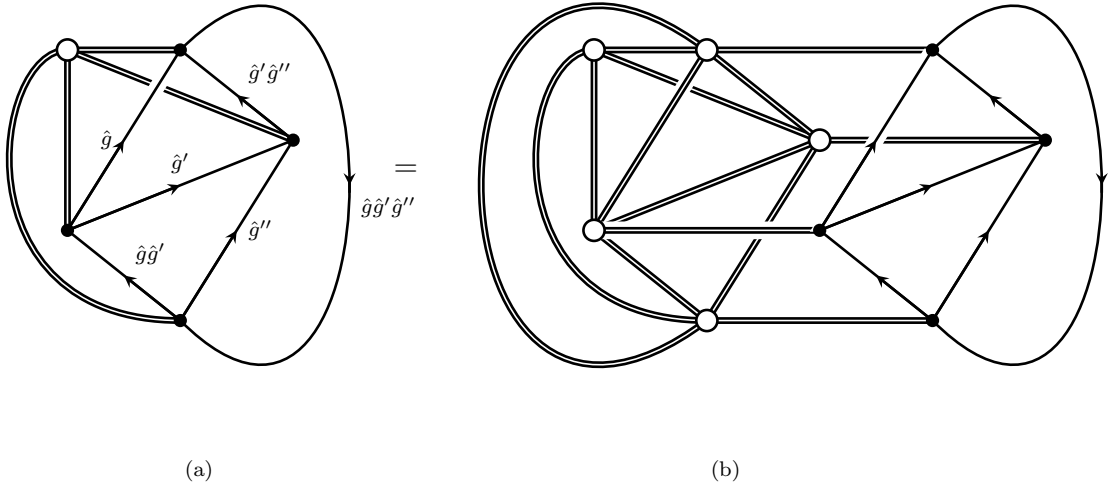
\begin{figure}
\centering
\begin{tikzpicture}[line width=1pt]
\begin{scope}[every node/.style={sloped,allow upside down}]
\coordinate (00) at (0,0);
\coordinate (023) at (00);

\coordinate (012) at ($(023)+(-1.5,1.2)$);
\coordinate (123) at ($(023)+(1.5,2.4)$);
\coordinate (013) at ($(023)+(0,3.6)$);
\coordinate (controlu) at ($(013)+(3,3)$);
\coordinate (controld) at ($(023)+(3,-3)$);
\coordinate (mid) at ($(controlu)!0.5!(controld)-(0.75,0)$);
\coordinate (blob) at ($(012)+(0,2.4)$);
\coordinate (controllu) at ($(blob)-(1,0)$);
\coordinate (controlld) at ($(023)-(3,0)$);

\draw[double] (123) to (blob);
\paddedline{(012)}{(013)}{(0.1,0)};
\draw[double] (013) to (blob);
\draw[double] (012) to (blob);
\arrowpath{(023)}{(012)}{0.5};
\arrowpath{(023)}{(123)}{0.5};
\arrowpath{(012)}{(123)}{0.5};
\arrowpath{(012)}{(013)}{0.5};
\arrowpath{(123)}{(013)}{0.5};
\draw (013)..controls(controlu) and (controld)..(023);
\arrowpath{($(mid)+(0,0.1)$)}{($(mid)-(0,0.1)$)}{0.5};
\draw[double] (023)..controls(controlld) and (controllu)..(blob);

\node[above right] at ($(023)!0.5!(012)$) {\scriptsize{${\hat g \hat g'}$}};
\node[right] at ($(023)!0.5!(123)$) {\scriptsize{${\hat g''}$}};
\node[above] at ($(012)!0.5!(123)$) {\scriptsize{${\hat g'}$}};
\node[left] at ($(012)!0.5!(013)$) {\scriptsize{${\hat g}$}};
\node[above right] at ($(123)!0.5!(013)$) {\scriptsize{${\hat g' \hat g''}$}};
\node[below right] at (mid) {\scriptsize{${\hat g \hat g' \hat g''}$}};

\draw[fill=black] (012) circle(2pt);
\draw[fill=black] (013) circle(2pt);
\draw[fill=black] (023) circle(2pt);
\draw[fill=black] (123) circle(2pt);
\draw[fill=white] (blob) circle(4pt);

\node (a) at ($(00)+(0.25,-2)$) {\scriptsize{(a)}};

\node (eq) at ($(00)+(3,2)$) {=};

\coordinate (023) at ($(00)+(7,0)$);
\coordinate (012) at ($(023)+(-1.5,1.2)$);
\coordinate (123) at ($(023)+(1.5,2.4)$);
\coordinate (013) at ($(023)+(0,3.6)$);
\coordinate (controlu) at ($(013)+(3,3)$);
\coordinate (controld) at ($(023)+(3,-3)$);
\coordinate (mid) at ($(controlu)!0.5!(controld)-(0.75,0)$);
\coordinate (blob) at ($(012)+(0,2.4)$);
\coordinate (controllu) at ($(blob)-(1,0)$);
\coordinate (controlld) at ($(023)-(3,0)$);
\coordinate (controlu1) at ($(controlu)+(-7,0)$);
\coordinate (controld1) at ($(controld)+(-7,0)$);
\coordinate (mid1) at ($(controlu1)!0.5!(controld1)-(0.75,0)$);

\draw[double] (123) to (blob);
\paddedline{(012)}{(013)}{(0.1,0)};
\draw[double] (013) to (blob);
\draw[double] (012) to (blob);
\draw[double] (023) to (012);
\draw[double] (023) to (123);
\draw[double] (012) to (123);
\draw[double] (012) to (013);
\draw[double] (123)to (013);
\draw[double] (013)..controls(controlu1) and (controld1)..(023);
\draw[double] (023)..controls(controlld) and (controllu)..(blob);

\coordinate (023b) at ($(023)+ (3,0)$);
\coordinate (012b) at ($(012)+ (3,0)$);
\coordinate (123b) at ($(123)+ (3,0)$);
\coordinate (013b) at ($(013)+ (3,0)$);
\coordinate (controlub) at ($(controlu)+ (3,0)$);
\coordinate (controldb) at ($(controld)+ (3,0)$);
\coordinate (midb) at ($(mid)+ (3,0)$);
\coordinate (blobb) at ($(blob)+ (3,0)$);
\coordinate (controllub) at ($(controllu)+ (3,0)$);
\coordinate (controlldb) at ($(controlld)+ (3,0)$);

\paddedline{(012b)}{(012)}{(0,0.1)};
\draw[double] (023b) to (023);
\draw[double] (012b) to (012);
\draw[double] (123b) to (123);
\draw[double] (013b) to (013);

\draw[fill=white] (012) circle(4pt);
\draw[fill=white] (013) circle(4pt);
\draw[fill=white] (023) circle(4pt);
\draw[fill=white] (123) circle(4pt);
\draw[fill=white] (blob) circle(4pt);

\paddedline{(012b)}{(013b)}{(0.1,0)};

\arrowpath{(023b)}{(012b)}{0.5};
\arrowpath{(023b)}{(123b)}{0.5};
\arrowpath{(012b)}{(123b)}{0.5};
\arrowpath{(012b)}{(013b)}{0.5};
\arrowpath{(123b)}{(013b)}{0.5};
\draw (013b)..controls(controlub) and (controldb)..(023b);
\arrowpath{($(midb)+(0,0.1)$)}{($(midb)-(0,0.1)$)}{0.5};

\draw[fill=black] (012b) circle(2pt);
\draw[fill=black] (013b) circle(2pt);
\draw[fill=black] (023b) circle(2pt);
\draw[fill=black] (123b) circle(2pt);

\node at ($(a)+(7,0)$) {\scriptsize{(b)}};
\end{scope}
\end{tikzpicture}
\caption{The graph on the right can be obtained from graph on the left after gauge fixing and deforming the $\Pi$ lines.}
\label{fig:compute}
\end{figure}

In the presence of a background $\beta_2$, the basic tetrahedron graph is as shown in Figure \ref{fig:compute}(a). This can be gauge fixed as shown in Figure \ref{fig:gfix}. After the gauge fixing, we can move the $\Pi$ lines to the position shown in Figure \ref{fig:compute}(b). This implies that the partition function can be written as
\be
\frac{1}{2^v}\prod\nu_3\sum_{\epsilon_1|\delta\epsilon_1=n_2+\beta_2}(-1)^{\int_M n_2\cup\epsilon_1+w_1^2\cup\epsilon_1+\epsilon_1\cup\beta_2}
\ee
This expression is non-zero only when the $G$-connection is such that $n_2=\beta_2+\delta\alpha_1$. By shifting $\epsilon_1\to\epsilon_1+\alpha_1$, the above expression can be re-written as
\be
\frac{1}{2^v}(-1)^{\int_M n_2\cup\alpha_1+\alpha_1\cup n_2+\alpha_1\cup\delta\alpha_1+w_1^2\cup\alpha_1}\prod\nu_3\sum_{\epsilon_1|\delta\epsilon_1=0}(-1)^{\int_M n_2\cup\epsilon_1+\epsilon_1\cup n_2+w_1^2\cup\epsilon_1}
\ee
The sign inside the sum is exact and hence we obtain
\be
Z_f(M,\beta_2)=\frac{1}{2^v}(-1)^{\int_M n_2\cup\alpha_1+\alpha_1\cup n_2+\alpha_1\cup\delta\alpha_1+w_1^2\cup\alpha_1}\prod\nu_3\sum_{\epsilon_1|\delta\epsilon_1=0}1 \label{GW}
\ee

Shifting $\beta_2\to\beta_2+\delta\lambda_1$ is the same as shifting $\alpha_1\to\alpha_1+\lambda_1$ under which the partition function changes as
\be
Z_f(M,\beta_2+\delta\lambda_1)=(-1)^{\int_M\lambda_1\cup\beta_2+\beta_2\cup\lambda_1+\lambda_1\cup\delta\lambda_1+w_1^2\cup\lambda_1}Z_f(M,\beta_2)
\ee
which matches the expectation in (\ref{anomaly}) exactly.

\subsection{Group structure of Gu-Wen phases} \label{4.3}
Now we would like to compute the product of two Gu-Wen phases labeled by $(\nu_3,n_2)$ and $(\nu_3',n_2')$. The $G$-graded product of corresponding categories has 4 simple objects in each grade $L_{g,\epsilon,\epsilon'}$ which fuse according to the cocycle $(n_2,n_2')$ and have associators $\hat\alpha_3\hat\alpha_3'$. The non-anomalous $\Z_2$ 1-form symmetry is generated by $L_{e,1,1}$ which has crossing $(-1)^{\epsilon+\epsilon'}$.

Gauging the symmetry identifies $L_{g,\epsilon,\epsilon'}$ with $L_{g,\epsilon+1,\epsilon'+1}$. We pick representative objects $L_{g,\epsilon,0}$ in each grade and compute the associator of $L_{g,\epsilon,0}$, $L_{g',\epsilon',0}$ and $L_{g'',\epsilon'',0}$ via the tetrahedron graph. Multiplying two representative objects $L_{g,\epsilon,0}$ and $L_{g',\epsilon',0}$, we obtain $L_{gg',\epsilon+\epsilon'+n_2(g,g'),n_2'(g,g')}$ which can be mapped back to the representative object $L_{gg',\epsilon+\epsilon'+n_2(g,g')+n_2'(g,g'),0}$ by inserting $n_2'(g,g')$ number of $\Pi\Pi'$ lines emanating from the corresponding vertex. The representative objects thus fuse according to the cocycle $n_2+n_2'$. Now, we gauge fix as in the previous subsection. Then, doing same manipulations as in the previous subsection, we find that the tetrahedron graph evaluates to
\be
\nu_3\nu_3'(-1)^{n_2\cup\epsilon_1+w_1^2\cup\epsilon_1+\epsilon_1\cup n_2'}
\ee
Upto a gauge redefintion, it can be written as
\be
(\nu_3\nu_3'(-1)^{n_2\cup_1n_2'})(-1)^{(n_2+n_2'+w_1^2)\cup\epsilon_1}
\ee
Thus the product is a Gu-Wen phase with $\tilde\nu_3=\nu_3\nu_3'(-1)^{n_2\cup_1n_2'}$ and $\tilde n_2=n_2+n_2'$.

However, notice that substituting $\nu_3=1, n_2=0$ in (\ref{GW}) and writing $w_1^2=\delta\sigma_1$ gives
\be
Z_f(M,\beta_2)=\frac{1}{2^v}(-1)^{\int_M \alpha_1\cup\delta\alpha_1+\delta\sigma_1\cup\alpha_1}\sum_{\epsilon_1|\delta\epsilon_1=0}1 \label{trivial}
\ee
and substituting $\nu_3=1, n_2=\rho_1^2$ gives
\be
Z_f(M,\beta_2)=\frac{1}{2^v}(-1)^{\int_M (\alpha_1+\sigma_1)\cup\delta\alpha_1}\sum_{\epsilon_1|\delta\epsilon_1=0}1
\ee
which are the same expressions! Thus, the Gu-Wen phase labeled by $(\nu_3=1,n_2=\rho_1^2)$ is the trivial phase. The reader might complain that (\ref{trivial}) does not seem to describe a \emph{trivial} phase. We would like to stress that this is the partition function of the shadow theory describing the trivial \PP-TFT. The trivial \PP-TFT is obtained by combining a non-trivial shadow with a non-trivial sign.

Thus, the group $\CG\CW(G)$ of {\PP} Gu-Wen phases with global symmetry $G$ can be described as follows. Consider the set parametrized by $(\nu_3,n_2)$ with $\nu_3$ parametrizing elements of $H^3(B(G\times\Z_2^R),U(1)_\rho)$ and $n_2$ parametrizing elements of $H^2(B(G\times\Z_2^R),\Z_2)$. Provide it a group structure given by
\be
(\nu_3,n_2)(\nu_3',n_2')=(\nu_3\nu_3'(-1)^{n_2\cup_1n_2'},n_2+n_2')
\ee
Finally, quotient it by the $\Z_2$ subgroup generated by $(\nu_3,n_2)=(1,\rho_1^2)$.

An alternative description can be given by first defining a group $H(G)=H^3(B(G\times\Z_2^R),U(1)_\rho)/\Z_2$ where the $\Z_2$ is generated by the cocycle $(-1)^{\rho_1^2\cup_1\rho_1^2}$. Then, $\CG\CW(G)$ is a central extension
\be
0\to H(G)\to\CG\CW(G)\to H^2(BG,\Z_2)\times H^1(BG,\Z_2)\to 0
\ee
with cocycle valued in $H(G)$ being $(-1)^{n_2\cup_1n_2'}\in H^3(B(G\times\Z_2^R),U(1)_\rho)$ where $n_2$ and $n_2'$ are valued in $H^2(BG,\Z_2)\times H^1(BG,\Z_2)$ as in (\ref{break}) but without the $\rho_1^2$ summand.

\subsection{$\Z_2^R$ version of Ising}
As an application of our formalism, we would like to construct all {\PP}-SPT phases wth global symmetry group $G$ being the trivial group $\{id\}$. There is only one Gu-Wen phase in this class, which is the trivial phase. There is a non-trivial phase in this class which is given by the $\Z_2^R$ analogue of a $\Z_2$ graded spherical fusion category $\CI$ which is known as the Ising fusion category. Below we recall the construction of $\CI$ and its $\Z_2^R$ cousin. It turns out that the analysis for both the cases is similar and we treat both of them together.

We are looking for a $\Z_2$ graded (twisted) spherical fusion category such that $\CC_0=\{I,P\}$ and $\CC_1=\{S\}$ are the simple objects. The fusion rules are
\begin{align}
P\otimes P&\simeq I\\
S\otimes P&\simeq S\\
P\otimes S&\simeq S\\
S\otimes S&\simeq I\oplus P
\end{align}
The $F$-symbols can be bootstrapped from these fusion rules by using (twisted) pentagon equation and taking advantage of the gauge freedom.

When the $\Z_2$ grading corresponds to a $\Z_2$ global symmetry, the non-trivial $F$-symbols are determined to be
\begin{align}
(F^{PSP}_S)_{(S)(S)}&=-1\\
(F^{SPS}_{P})_{(S)(S)}&=-1\\
(F^{SSS}_{S})_{(I)(I)}=(F^{SSS}_{S})_{(P)(I)}=(F^{SSS}_{S})_{(I)(P)}&=\pm\frac{1}{\sqrt{2}}\\
(F^{SSS}_{S})_{(P)(P)}&=\mp\frac{1}{\sqrt{2}}
\end{align}
When the $\Z_2$ grading corresponds to $\Z_2^R$ orientation reversing symmetry, the non-trivial $F$-symbols are determined to be
\begin{align}
(F^{PSP}_S)_{(S)(S)}&=-1\\
(F^{SPS}_{P})_{(S)(S)}&=-1\\
(F^{SSS}_{S})_{(I)(I)}=(F^{SSS}_{S})_{(P)(I)}=(F^{SSS}_{S})_{(I)(P)}&=\frac{1}{\sqrt{2}}\\
(F^{SSS}_{S})_{(P)(P)}&=-\frac{1}{\sqrt{2}}
\end{align}
That is, the choice of sign becomes a gauge freedom in the $\Z_2^R$ case.

The fermion line is given by an element of (twisted) Drinfeld center of the form $(P,\beta)$. Solving the Drinfeld center equations for the $\Z_2$ case, we obtain
\bea
\beta_P=-1\\
\beta_S=\pm i
\eea
Thus, there are two choices for the fermion line $\Pi$. Given the choice in picking the associator and the choice in $\Pi$, we can construct four Spin-TFTs with global symmetry $\Z_2$.

On the other hand, solving the twisted Drinfeld center equations for the $\Z_2^R$ case, we obtain
\bea
\beta_P=-1\\
\beta_S=\pm 1
\eea
However, as we know from before, flipping the sign of all the $\beta$ in the orientation reversing sector doesn't change the resulting {\PP}-TFT and we can fix $\beta_S=+1$. Hence, in the $\Z_2^R$ case, there are no choices and we obtain only one {\PP}-TFT which we call $\CI_+$.

\subsection{\PP-SPT phases with no global symmetry}
Cobordism hypothesis predicts a $\Z_2$ group of \PP-SPT phases \cite{Kapustin:2014dxa}. We have already found the trivial phase as a Gu-Wen phase. We claim that the non-trivial phase corresponds to the \PP-TFT $\CI_+$ that we encountered in last subsection. To justify this, we will show that the square of $\CI_+$ is the trivial Gu-Wen phase. This will prove that $\CI_+$ is indeed an SPT phase and provide an explicit construction of \PP-SPT phases without global symmetry. The existence of this non-trivial phase was also discussed in \cite{2015arXiv150101313C}.

\begin{figure}
\centering
\begin{tikzpicture}[line width=1pt]
\begin{scope}[every node/.style={sloped,allow upside down}]
\coordinate (lowA) at (0,0);
\coordinate (highA) at ($(lowA)+(0,4)$);
\arrowpath{(lowA)}{(highA)}{0.5};
\node[below] at (lowA) {\scriptsize{$SS^\pm$}};

\coordinate (eq) at ($(lowA)+(2,2)$);
\node at (eq) {$=$};
\node at ($(eq)+(1,0)$) {$\frac{1}{2}$};

\coordinate (lowS1) at ($(lowA)+(4,0)$);
\coordinate (highS1) at ($(lowS1)+(0,4)$);
\node[below] at (lowS1) {\scriptsize{$SS$}};
\arrowpath{(lowS1)}{(highS1)}{0.5};

\coordinate (p1) at ($(lowS1)+(1.5,2)$);
\node at (p1) {$\pm$};
\node at ($(p1)+(1,0)$) {$\frac{1}{2}$};

\coordinate (lowS3) at ($(lowS1)+(4,0)$);
\coordinate (highS3) at ($(lowS3)+(0,4)$);
\coordinate (midS3) at ($(lowS3)!0.5!(highS3)$);
\arrowpath{(lowS3)}{(midS3)}{0.5};
\arrowpath{(midS3)}{(highS3)}{0.5};
\coordinate (PP1) at ($(highS3)-(1,0)$);
\draw[double] (midS3)--(PP1);
\draw[fill=black] (midS3) circle(2pt);
\node[above] at (PP1) {\scriptsize{$PP$}};
\node[below] at (lowS3) {\scriptsize{$SS$}};
\end{scope}
\end{tikzpicture}
\caption{Definition of $SS^\pm$.} \label{fig:def}
\end{figure}
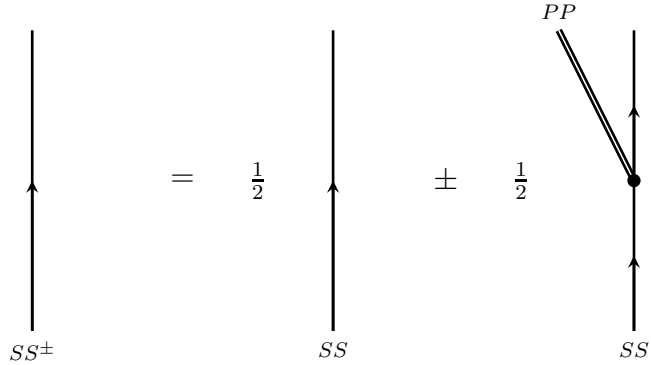

The graded product of $\CI_+$ with itself has simple objects $II$, $PI$, $IP$, $PP$ in the trivial grade and a simple object $SS$ in the non-trivial grade. Gauging the 1-form symmetry generated by $\Pi\Pi$, we obtain a category $\CC$ with $\CC_0$ having simple objects $II$, $PI$ and $\CC_1$ having simple objects $SS^+$, $SS^-$. $SS^+$ and $SS^-$ are constructed by using projectors obtained by using the non-trivial endomorphism of $SS$. See Figure \ref{fig:def}.

$SS^+\otimes PI$ involves the $F$-symbol $F^{PSP}$ which flips the sign of $\xi_S$ and hence $SS^+\otimes PI\simeq SS^-$. On the other hand, $PI\otimes SS^+$ involves $\beta_P$ and hence $PI\otimes SS^+\simeq SS^-$. The computation of $SS^+\otimes SS^+$ can be done in a similar but more involved manner which we explain in Figure \ref{fig:Isquared}. We find that $SS^+\otimes SS^+\simeq II$. All the statements above hold true if we replace $SS^+$ with $SS^-$. Thus, $\CC$ has the fusion rules of the Gu-Wen phase which is trivial.

\begin{figure}
\centering
\begin{tikzpicture}[line width=1pt]
\begin{scope}[every node/.style={sloped,allow upside down}]
\coordinate (lowA) at (0,0);
\coordinate (lowB) at ($(lowA)+(1,0)$);
\coordinate (highA) at ($(lowA)+(0,4)$);
\coordinate (highB) at ($(lowB)+(0,4)$);

\arrowpath{(lowA)}{(highA)}{0.5};
\arrowpath{(lowB)}{(highB)}{0.5};

\node[below] at (lowA) {\scriptsize{$SS^+$}};
\node[below] at (lowB) {\scriptsize{$SS^+$}};

\coordinate (eq) at ($(lowB)+(1,2)$);
\node at (eq) {$=$};

\coordinate (lowS1) at ($(lowB)+(2,0)$);
\coordinate (lowS2) at ($(lowS1)+(0.5,0)$);
\coordinate (highS1) at ($(lowS1)+(0,4)$);
\coordinate (highS2) at ($(lowS2)+(0,4)$);

\arrowpath{(lowS1)}{(highS1)}{0.5};
\arrowpath{(lowS2)}{(highS2)}{0.5};

\node[below] at (lowS1) {\scriptsize{$SS$}};
\node[below] at (lowS2) {\scriptsize{$SS$}};

\coordinate (p1) at ($(lowS2)+(1,2)$);
\node at (p1) {$+$};

\coordinate (lowS3) at ($(lowS2)+(2.5,0)$);
\coordinate (lowS4) at ($(lowS3)+(0.5,0)$);
\coordinate (highS3) at ($(lowS3)+(0,4)$);
\coordinate (highS4) at ($(lowS4)+(0,4)$);

\coordinate (midS3) at ($(lowS3)!0.5!(highS3)$);
\coordinate (PP1) at ($(highS3)-(1,0)$);
\arrowpath{(lowS3)}{(midS3)}{0.5};
\arrowpath{(midS3)}{(highS3)}{0.5};
\draw[double] (midS3)--(PP1);
\draw[fill=black] (midS3) circle(2pt);
\node[above] at (PP1) {\scriptsize{$PP$}};

\arrowpath{(lowS4)}{(highS4)}{0.5};

\node[below] at (lowS3) {\scriptsize{$SS$}};
\node[below] at (lowS4) {\scriptsize{$SS$}};

\coordinate (p2) at ($(lowS4)+(1,2)$);
\node at (p2) {$+$};

\coordinate (lowS5) at ($(lowS4)+(2,0)$);
\coordinate (lowS6) at ($(lowS5)+(0.5,0)$);
\coordinate (highS5) at ($(lowS5)+(0,4)$);
\coordinate (highS6) at ($(lowS6)+(0,4)$);

\coordinate (midS6) at ($(lowS6)!0.5!(highS6)$);
\coordinate (PP2) at ($(highS6)-(1,0)$);
\arrowpath{(lowS6)}{(midS6)}{0.5};
\arrowpath{(midS6)}{(highS6)}{0.5};
\draw[double] (midS6)--(PP2);
\paddedline{($(lowS5)$)}{(highS5)}{(0.05,0)};
\draw[fill=black] (midS6) circle(2pt);
\node[above] at (PP2) {\scriptsize{$PP$}};

\arrowpath{(lowS5)}{(highS5)}{0.5};

\node[below] at (lowS5) {\scriptsize{$SS$}};
\node[below] at (lowS6) {\scriptsize{$SS$}};

\coordinate (p3) at ($(lowS6)+(1,2)$);
\node at (p3) {$+$};

\coordinate (lowS7) at ($(lowS6)+(2,0)$);
\coordinate (highS7) at ($(lowS7) + (0,4)$);
\coordinate (midS7) at ($(lowS7)!0.5!(highS7)$);
\coordinate (lowS8) at ($(lowS7)+(1,0)$);
\coordinate (highS8) at ($(lowS8) + (0,4)$);
\coordinate (midS8) at ($(lowS8)!0.5!(highS8)$);

\draw[double] (midS7) .. controls ($(midS7) + (-1,1.5)$) .. (midS8);

\paddedline{($(lowS7)!0.6!(highS7)$)}{(highS7)}{(0.05,0)};
\arrowpath{(lowS7)}{(highS7)}{0.25};
\arrowpath{(lowS7)}{(highS7)}{0.8};
\draw[fill=black] (midS7) circle(2pt);
\node[below] at (lowS7) {\scriptsize{$SS$}};

\arrowpath{(lowS8)}{(highS8)}{0.25};
\arrowpath{(lowS8)}{(highS8)}{0.8};
\draw[fill=black] (midS8) circle(2pt);
\node[below] at (lowS8) {\scriptsize{$SS$}};

\node at ($(eq)+(0.5,0)$) {$\frac{1}{4}$};
\node at ($(p1)+(0.5,0)$) {$\frac{1}{4}$};
\node at ($(p2)+(0.5,0)$) {$\frac{1}{4}$};
\node at ($(p3)+(0.5,0)$) {$\frac{1}{4}$};

\end{scope}
\end{tikzpicture}
\caption{Computation of $SS^+\otimes SS^+$ is by definition a sum of four terms which involve associators and crossings. $II$ inside $SS\otimes SS$ is mapped to $II$ by first and fourth terms and to $PP$ (which is isomorphic to $II$ in the new category) by the second and third terms. Similarly, $PI$ is mapped to the zero object as the four terms cancel in pairs. Hence, $SS^+\otimes SS^+\simeq II$.} \label{fig:Isquared}
\end{figure}

For a general $G$, we can consider the pullback of $\CI_+$ along $\rho_1$ which we denote as $\CI_+(G)$. $\CI_+(G)_g$ has two simple elements $I_g$, $P_g$ if $\rho_1(g)=0$ and has a single simple object $S_g$ if $\rho_1(g)=1$. The fusion rules and associators are just pulled back from $\CI_+$. Clearly, $\CI_+(G)$ will also square to 0 as our argument above is independent of $G$-grading. 

This allows us to construct $\CG\CW(G)\times\Z_2$ worth of \PP-SPT phases with global symmetry $G$. We suspect that this is not the full classification and comment on how to complete the classification in the next section.

\section{Conclusion and future directions} \label{5}
In this paper we discussed the generalization of Turaev-Viro construction of oriented 3d TFTs to unoriented 3d TFTs. We proposed that the input data of this construction in the unoriented case should be a ``twisted" spherical fusion category in which the pentagon equation for the $F$-symbols is modified.

As a generalization of the construction of \cite{Bhardwaj:2016clt}, we also proposed a construction for \PP-TFTs in terms of their shadows. The shadows are ordinary unoriented TFTs with a $\Z_2$ 1-form symmetry which is anomalous and has a mixed anomaly with time-reversal symmetry.

Combining the above two ingredients, we were able to give explicit constructions of a large class of invertible \PP-TFTs with global symmetry $G$. Such theories are known as \PP-SPT phases. We also reproduced the $\Z_2$ group of \PP-SPT phases without any global symmetry.

There are plenty of interesting directions in which this work can be extended in the future and we make some very speculative comments about them in what follows. Perhaps the most immediate future direction is to use the machinery developed in this paper to provide a classification of \PP-SPT phases for an arbitrary group $G$ which admit a topological boundary condition. The author suggests to look at a spherical fusion category graded by $\Z_2\times\Z_2^R$ with simple elements $I,P$ in the $(0,0)$ grade, $I_1,P_1$ in the $(0,1)$ grade, $S$ in the $(1,0)$ grade and $S_1$ in the $(1,1)$ grade. The fusion rules mimic the Ising category. Is it possible to find a consistent set of $F$-symbols? If yes, then the class of \PP-SPT phases we presented in this paper is not the full answer. It should then be possible to finish the classification, in a spirit similar to the one in \cite{Bhardwaj:2016clt}, by pulling back this $\Z_2\times\Z_2^R$ phase and combining it with the class of phases presented in this paper.

It would be very interesting to provide a construction (Turaev-Viro-like or some other construction) for TFTs with more general struture groups. For instance, one could mix $O(n)$ and $G$ or mix \PP$(n)$ and $G$ in the fermionic case. It seems that a proper treatment of these generalizations should involve a rich interplay of symmetry defects along with higher codimension defects living in the worldvolume of symmetry defects.

Let us comment about the {\PM}$(n)\times G$ case. It seems natural that the kernel for \PM-TFTs would be the sign
\be
z_-(M,\eta_1,\beta_2)=(-1)^{\int_M\eta_1\cup\beta_2+\int_N\beta_2\cup\beta_2+(w_1^2+w_2)\cup\beta_2}
\ee
which seems to be the same expression as (\ref{sign}) but this time we take $\eta_1$ to parametrize \PM-structures. This would suggest that the corresponding shadow theory has no mixed anomaly between time reversal and $\Z_2$ 1-form symmetry. Also, the anomaly for the 1-form symmetry should now be
\be
Z_f(M,\beta_2)\to (-1)^{\int_M\lambda_1\cup\beta_2+\beta_2\cup\lambda_1+\lambda_1\cup\delta\lambda_1}Z_f(M,\beta_2) \label{anomaly-}
\ee
However, for {\PP} case, we saw in Figure \ref{fig:cross} that moving the fermion $\Pi$ across the locus dual to $w_1^2$ should change the operator at the junction of $\Pi$ line and the orientation reversing defects. The argument given there was that this sign was needed to cancel the sign coming from the crossing of $\Pi$ lines. This lead to different anomalies than the ones we want for the {\PM} case. So, in the {\PM} case, we do not want such a change in the sign of the corresponding local operator. The author suspects that in this case the sign coming from crossing of $\Pi$ lines will be canceled by factors coming from \emph{patching} of $\Pi$ with $R\Pi$ where $R\Pi$ is $\Pi$ line with a reflected framing. This would make sure that $\Pi$ is an element of Drinfeld center rather than a twisted Drinfeld center, which would in turn imply the anomalies given above. It would be interesting to work out the details and provide a Turaev-Viro construction for {\PM} shadows. 

Of course, this means that one will have to first understand how to compute (in terms of the twisted spherical fusion category) the extra data attached to a bulk line which corresponds to patching the line with itself but with reflected framing. In other words, this corresponds to a generalization of Moore-Seiberg data \cite{moore1989classical,moore1990lectures} to the unoriented case. A step towards this was recently taken in \cite{Barkeshli:2016mew}.

A puzzle here is that there should be no non-trivial \PM-SPT phase according to \cite{Kapustin:2014dxa}. So, somehow the \PM-TFTs produced by the potential \PM-shadows having $\Z_2^R$ version of Ising as their twisted spherical fusion category should be trivial.

Another interesting direction to pursue would be to see if it is possible to find a generalization of Turaev-Viro construction which could construct anomalous 3d TFTs. Such TFTs live at the boundary of a 4d SPT phase. Hence, such TFTs should not admit topological boundaries of their own but they can admit interfaces to other 3d TFTs with the same anomaly. Perhaps it is possible to choose a simple TFT in each anomaly class and build a Turaev-Viro construction using a topological interface between the TFT we want to construct and the simple TFT. See \cite{Witten:2016cio,Tachikawa:2016cha,Wang:2016qkb,Tachikawa:2016nmo} for recent interesting work on anomalous unoriented 3d TFTs.

Finally, it would be interesting to concretely construct a time-reversal invariant commuting projector Hamiltonian using the data of twisted spherical fusion category. This Hamiltonian goes into the string-net construction of fermionic phases of matter. See \cite{Levin:2004mi}, \cite{Bhardwaj:2016clt} for more details.

\section*{Acknowledgements}
The author is grateful to Davide Gaiotto, Theo Johnson-Freyd and Yuji Tachikawa for reading the draft of this paper and giving helpful suggestions for its improvement. The author thanks Kevin Costello, Davide Gaiotto, Theo Johnson-Freyd, Anton Kapustin and Yuji Tachikawa for useful comments and discussions.

Part of this work was completed at IPMU, Japan and the author is thankful for the hospitality there.

This research was supported by the Perimeter Institute for Theoretical Physics. Research at Perimeter Institute is supported by the Government of Canada through Industry Canada and by the Province of Ontario through the Ministry of Economic Development and Innovation.

\bibliographystyle{ytphys}
\let\bbb\bibitem\def\bibitem{\itemsep4pt\bbb}
\bibliography{ref}


\appendix

\end{document}